\documentclass{aa}
\usepackage{multirow}
\usepackage{graphicx} 
\usepackage[usenames,dvipsnames]{xcolor}
\usepackage{float}
\usepackage{enumitem}
\usepackage{txfonts}
\usepackage{natbib}
\usepackage[utf8]{inputenc}
\usepackage{tikz}
\usepackage{units}
\usepackage{url}
\usepackage{comment}
\usepackage{pdflscape}
\usepackage{xcolor}
\usepackage{supertabular,booktabs}
\usepackage{rotating}
\usepackage{soul}
\usepackage{ulem}
\usepackage{hyperref} 
\usepackage{nccmath}
\usepackage{xspace}

\newcommand{\hii}{\mbox{\ion{H}{\small II}}\xspace}
\newcommand{\uchii}{\mbox{UC-\ion{H}{\small II}}\ }

\newcommand{\blobcat}{\texttt{BLOBCAT}}

\newcommand{\firstblobs}{{ \color{black}13\,001}} 
\newcommand{\includeLarge}{{\color{black}839}}  
\newcommand{\includeSNR}{{\color{black}1\,126}} 
\newcommand{\includeLSS}{{\color{black}1\,965}} 
\newcommand{\rejectedUnclear}{{\color{black}1\,790}}
\newcommand{\NumUnknown}{{\color{black}19}}
\newcommand{\realsVI}{{\color{black}701}} 
\newcommand{\WITHtenarcsec}{{\color{black}8 235}} 
\newcommand{\YfMthree}{{\color{black}297}}
\newcommand{\FinCatNum}{{\color{black}11\,211}}
\newcommand{\FinDiscNum}{{\color{black}9\,227}}

\newcommand{\largestructure}{{ \color{black} 106 }} 
\newcommand{\visualSources}{{\color{black} 2\,807 }}

\newcommand{\Gatlasgal}{{ \color{black} 198}}
\newcommand{\Grms}{{ \color{black} 150}}
\newcommand{\Gcornish}{{ \color{black} 1\,231}}
\newcommand{\GBconf}{{ \color{black} 2\,496}}
\newcommand{\Gthor}{{\color{black} 5\,179}}
\newcommand{\Ghigal}{{\color{black} 1\,096}}
\newcommand{\Gwise}{{\color{black} 5\,473}}
\newcommand{\Gglimpse}{{\color{black} 7\,326}}

\newcommand{\sGatlasgal}{{ \color{black} 109}}
\newcommand{\sGrms}{{ \color{black} 145}}
\newcommand{\sGcornish}{{ \color{black} 1\,221}}
\newcommand{\sGBconf}{{ \color{black} 2\,473}}
\newcommand{\sGthor}{{ \color{black} 5\,092}}
\newcommand{\sGhigal}{{ \color{black} 918}}
\newcommand{\sGwise}{{ \color{black} 3\,034}}
\newcommand{\sGglimpse}{{\color{black} 5\,525}}

\newcommand{\obit}{\texttt{Obit}}

\setstcolor{red}
\usepackage[switch]{lineno}

\begin{document}

\title{A global view on star formation: The GLOSTAR Galactic plane survey. XI. Radio source catalog IV: $2\degr < \ell < 28\degr$, $36\degr < \ell < 60\degr$  and $|b| < 1\degr$
\thanks{The full version of Table\,7, and Fig.\,1 are only available in electronic form at the CDS via anonymous ftp to cdsarc.u-strasbg.fr (130.79.125.5) or via \url{https://glostar.mpifr-bonn.mpg.de}}}
\author{S.-N.\,X.\,Medina \inst{1,2}\thanks{E-mail: smedina@mpifr-bonn.mpg.de}, S.\,A.\,Dzib\inst{2},  \thanks{E-mail: sdzib@mpifr-bonn.mpg.de}, 
J.\,S.\,Urquhart\inst{3},  A.\,Y. Yang\inst{4, 5}, A. Brunthaler\inst{2},
K. M. Menten \inst{2}, 
F. Wyrowski \inst{2}, 
W.~D.~Cotton\inst{6},
A. Cheema\inst{2},
R. Dokara \inst{2},
Y. Gong \inst{2},
S. Khan\inst{2},
H. Nguyen\inst{2},
G. N.~Ortiz-Le\'{o}n\inst{7,2},
M. R.~Rugel\inst{2,6,8}, 
V.~S.~Veena\inst{2,9},
H.~Beuther\inst{10},
T.~Csengeri\inst{11},
J.~D.~Pandian\inst{12},
and N.~Roy\inst{13} }

\institute{German Aerospace Center, Scientific Information, 51147 Cologne, Germany
\and 
Max-Planck-Institut f\"ur Radioastronomie, Auf dem H\"ugel 69, D-53121 Bonn, Germany  
\and
Centre for Astrophysics and Planetary Science, University of Kent, Canterbury, CT2\,7NH, UK
\and
National Astronomical Observatories, Chinese Academy of Sciences, A20 Datun Road, Chaoyang District, Beijing,  100101, P.~R.~China
\and Key Laboratory of Radio Astronomy and Technology, Chinese Academy of Sciences, A20 Datun Road, Chaoyang District, Beijing, 100101, P.~R.~China 
\and 
National Radio Astronomy Observatory,  520 Edgemont Road, Charlottesville, VA 22903, USA
\and
Instituto Nacional de Astrof\'isica, \'Optica y Electr\'onica, Apartado Postal 51 y 216, 72000 Puebla, Mexico
\and Center for Astrophysics | Harvard $\&$ Smithsonian, 60 Garden St., Cambridge, MA 02138, USA
\and
 I. Physikalisches Institut, Universit\"at zu Köln, Z\"ulpicher Str. 77, 50937 K\"oln, Germany
\and Max Planck Institute for Astronomy, Koenigstuhl 17, 69117 Heidelberg, Germany 
\and Laboratoire d'astrophysique de Bordeaux, Univ. Bordeaux, CNRS, B18N, all\'ee Geoffroy Saint-Hilaire, 33615 Pessac, France 
\and Department of Earth \& Space Sciences, Indian Institute of Space Science and Technology, Trivandrum 695547, India
\and Department of Physics, Indian Institute of Science, Bangalore 560012, India
}

\date{Received 2020; }

\date{Received February 2024; Accepted July 2024}

\abstract
{ The GLObal view on STAR formation in the Milky Way (GLOSTAR) survey studies star formation with the Very Large Array (VLA) and the Effelsberg 100 meter radio telescope in the Galactic plane between $-2\degr < \ell < 60\degr$  and $|b| < 1\degr$, and the Cygnus\,X region ($76\degr < \ell < 83\degr$  and $-1\degr < b< 2\degr$), with unprecedented sensitivity in both flux density ($1\sigma\sim$50\,$\mu$Jy\,beam$^{-1}$) and the capability of detecting emission with angular scales in the range from 1\rlap{.}$''$0 to the largest radio structures in the Galaxy on the order of a few degrees in size.}
{Here, we provide a complete GLOSTAR-VLA D-configuration radio source catalog for the part of the Galactic disk covered. A catalog for the ``pilot region'' between  $28\degr < \ell < 36\degr$ has been published in a previous paper and here we present the complementary catalog for the area within $2\degr < \ell < 28\degr$, $36\degr < \ell < 60\degr$ and $|b| < 1\degr$.}
{Observations were taken with the VLA in a 4 -- 8 GHz band to image 100 square degrees of the inner Galactic disk at a reference frequency of 5.8 GHz, using a total of 260 hours of telescope time. We determined spectral indices ($\alpha$; $S_\nu\propto\nu^\alpha$) inside the observed band and in the frequency range of 1.4 -- 5.8 GHz by complementing our results with those from The HI/OH/Recombination line survey of the inner Milky Way (THOR), which covers 1 -- 2 GHz.  }
{The final images have an angular resolution of $18''$ and an average sensitivity of 123\,$\mu$Jy\,beam$^{-1}$. The sensitivity is better ($\sim 60\,\mu$Jy\,beam$^{-1}$) in areas free of extended emission. The complementary Galactic disk catalog presented in this work consists of \FinCatNum\ radio sources.  Of these, \includeLSS\ are  known large-scale structure sources such as star-forming region complexes, well-known supernova remnants (SNRs),
SNR candidates, or parts thereof. The remaining \FinDiscNum\ are discrete individual sources. Source parameters --- namely flux densities, sizes, spectral indices, and classifications --- are reported. We identify 769 \hii\ region candidates, 359 of which have been newly classified as such. 
The mean value of spectral indices of 225  \hii\ regions is $+0.14\pm0.02$, consistent with most of them emitting optically thin thermal radio emission.  Combining our results with the previously published catalog of the pilot region, the final GLOSTAR-VLA D-configuration catalog contains 12\,981 radio sources.
} 
{}
\titlerunning{GLOSTAR Catalog IV}
\authorrunning{Medina, S.-N. X. et al.}
\maketitle

\newpage

\section{Introduction}
The wide-field radio continuum surveys have been fundamental in significantly increasing the number of known radio objects, particularly sources representing high-mass young stellar objects still embedded in or close to dense gas. These objects, when far away in the Galactic plane, are typically obscured by large volumes of dust, making observations of stars in the earliest evolutionary stages difficult (see a review in high-mass star formation by \citealt{Motte2018}). Therefore, the GLObal view on STAR formation in the Milky Way (GLOSTAR)\footnote{https://glostar.mpifr-bonn.mpg.de/glostar/} survey \citep{Medina2019,brunthaler2021} focuses on finding and characterizing star-forming regions in the Galactic plane at radio frequencies with unprecedented sensitivity using the {\it Karl G. Jansky} Very Large Array (VLA) and the Effelsberg 100 meter telescope. The GLOSTAR survey detects tell-tale tracers of star formation: compact, ultra- and hyper-compact \hii regions and class II methanol masers that trace different stages of early stellar evolution. The capabilities of the GLOSTAR survey help to locate the center of early stages of star-forming activity and complement previous radio surveys like MAGPIS \citep[the Multi-Array Galactic Plane Imaging Survey;][]{helfand2006}, CORNISH \citep[Co-Ordinated Radio 'N' Infrared Survey for High-mass star formation;][]{hoare2012}, and THOR \citep[The HI/OH/Recombination line survey of the inner Milky Way;][]{beuther2016} that have also largely contributed to star formation studies. Furthermore, the sensitivity achieved by GLOSTAR will highly increase and confirm the number of known sources within the Galactic plane.\\

To achieve its goal, the GLOSTAR survey covers a large area of the Galactic plane within 145 $\deg^2$, covering $-2\degr < \ell < 60\degr$; $|b| < 1\degr$, which includes the Galactic center, and, in addition, the Cygnus X region ($76\degr < \ell < 83\degr$; $-1\degr < b < 2\degr$). It uses the results of a complementary set of observations with the VLA of the National Radio Astronomy Observatory (NRAO)\footnote{The NRAO is operated by Associated Universities Inc. under cooperative agreement with the National Science Foundation.} and the Effelsberg 100\,m telescope. With this combination, the GLOSTAR survey can detect compact radio sources and recover extended emission. The GLOSTAR-VLA observations used the wideband C-band (4--8 GHz) receivers. They were performed using both the compact D-configuration (as well as complementary DnC- and C-configurations) and the extended B-configuration (as well as complementary BnA- and A-configurations) to obtain good surface brightness sensitivity and a range of resolutions to map the radio emission over a large range of angular scales. The C-band also comprises the frequency of the prominent class II CH$_3$OH (methanol)  maser line  (6.668\,GHz), the 4.829 GHz transition line H$_2$CO (formaldehyde), and several radio recombination lines (RRLs). Spectrally resolved data have also been taken within the GLOSTAR survey.\\

The GLOSTAR survey results can be found in a series of publications, starting with the first catalog of radio continuum sources by \citet{Medina2019} based on the   VLA D-configuration data of the GLOSTAR ``pilot region'' (defined within the following limits: $28\degr < \ell < 36\degr$ and $|b| < 1\degr$). An overview of the GLOSTAR survey is presented by \citet{brunthaler2021}, who additionally provided examples of the combination of the VLA and the Effelsberg observations. At the same time, \citet{Dokara2021} presented a study of Galactic supernova remnants (SNRs), \citet{Nguyen2021} presented radio continuum detections of young stellar objects (YSOs) in the Galactic center, and \citet{Ortiz2021} reported 6.7\,GHz methanol maser detections in the Cygnus X region, followed by \citet{Nguyen2022} who reported methanol maser detections within $-2\degr < \ell < 60\degr$ and $|b| < 1\degr$.  The GLOSTAR VLA B-configuration data were used by \citet{dzib2023} to provide the first radio source catalog of compact sources in the pilot region and by \citet{yang2023} for the region within $2\degr < \ell < 28\degr$, $36\degr < \ell < 40\degr$, $56\degr < \ell < 60\degr$  and $|b| < 1\degr$. Moreover, GLOSTAR VLA and Effelsberg data were recently used by \citet{Dokara2023} to analyze nonthermal synchrotron emission from SNR candidates in the GLOSTAR pilot region, and by \citet{gong2023} for a deep analysis of formaldehyde absorption in the Cygnus X region, while \citet{khan2024} reports RRLs related to bright \hii regions of the full survey.\\

The focus of this paper is the characterization of continuum radio emission from GLOSTAR VLA D-configuration data (angular resolution ${\sim}18''$ and mean sensitivity ${\sim}128\,\mu{\rm Jy}\,{\rm beam}^{-1}$) to provide a reliable catalog of radio sources with angular sizes of $\sim18$\arcsec\ to a few arcminutes. The work presented in this paper extends the analysis already conducted on the radio continuum map of the pilot region \citep[][in which we reported 1\,575 radio sources]{Medina2019} to the rest of the inner Galactic disk covered by the survey. We follow a similar set of steps to create a complete catalog of the Galactic disk covered by the GLOSTAR survey ($2\degr < \ell < 60\degr$ and $|b| < 1\degr$). 

This paper is organized as follows. Section\,\ref{sec:obs} gives a description of the observations and the obtained data. Section\,\ref{sec:RCM} describes the resulting image of the radio continuum emission and the method used for source extraction. Section \ref{CatCons} discusses the catalog construction, and section \ref{CatSumm} summarizes and discusses the catalog properties. In Section \ref{Conc}, we give a summary of the work and outline our main findings.

\section{Observations and data reduction}\label{sec:obs}

The observations were performed with the VLA in the D-, \mbox{DnC-,} and C-configurations (see Table~\ref{tab:obs}). The hybrid DnC configuration was preferred for targets with low elevation (that is, $\ell<12\degr$) to recover a synthesized beam similar in shape to that obtained with D-configuration observations at higher elevations. However, the use of the hybrid configurations was discontinued in 2016. To compensate for this, the regions not observed in the DnC-configuration were observed in both D- and C-configurations, as was suggested by NRAO\footnote{\footnotesize\url{https://science.nrao.edu/facilities/vla/docs/manuals/propvla/array_configs}}. Though diverse array configurations were used, for the simplicity of convention, we refer to all these low-resolution VLA observations as the GLOSTAR VLA D-configuration observations. 
A detailed overview of the observing strategy, data reduction, and imaging of the observed data is given in \citet{brunthaler2021}, and more specific details on the calibration and imaging of the D-configuration data are described in \citet{Medina2019}. While we refer the reader to these works for details, a summary is provided below. 

\begin{figure*}
\begin{center}
\includegraphics[angle=90, width=0.81\textwidth, trim= 44 20 10 50, clip]{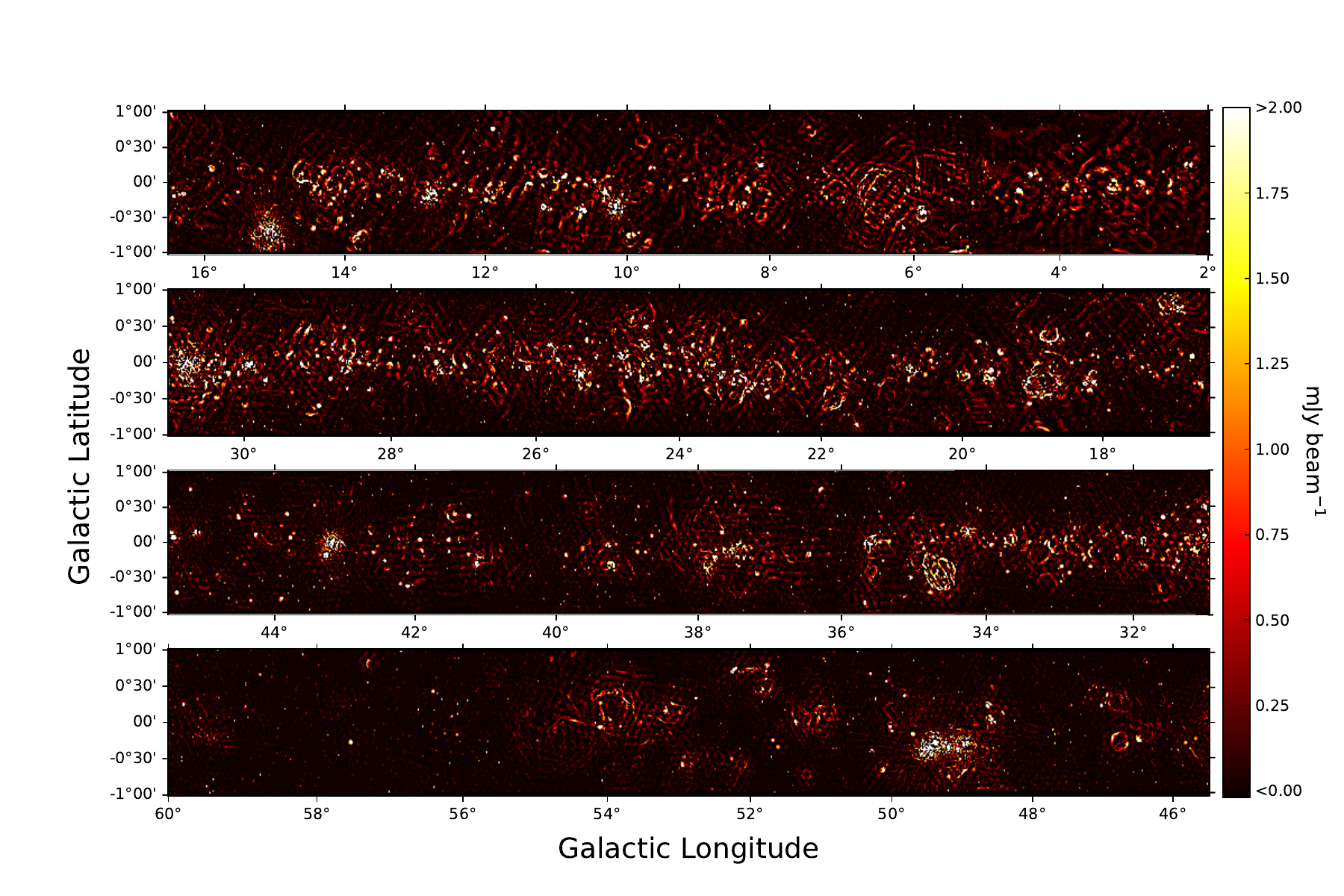}
\end{center}
\caption{ Full GLOSTAR radio continuum map at 5.8~GHz of 116 square degrees of the Galactic disk plane ($2\degr  < \ell < 60\degr$ and $|b| < 1\degr$).  Though the most significant peak emission has a brightness of $\sim10$~Jy~beam$^{-1}$, we have constrained the contrast by choosing an upper limit of 2~mJy~beam$^{-1}$, in order to make most of the weak extended and compact sources recognizable.
} 
\label{fig:FullRadioMap}

\end{figure*}

The observed band corresponds to the so-called C-band, covering a frequency range from 4.0 to 8.0~GHz. In particular, the correlator was tuned to record two 1.0~GHz wide sub-bands centered at 4.7 and 6.9\,GHz in a semi-continuum mode\footnote{The observations also comprised additional high spectral resolution correlation setups covering spectral lines  \citep[see][]{brunthaler2021}. In this paper, however, we focus on the continuum data.}. 
The portion of the Galactic disk for which results are reported in this paper has a size of 100 square degrees and is delimited by $|b| \leq 1\degr$ and  $2\degr\leq\ell\leq28\degr$ and $36\degr\leq\ell\leq60\degr$. 
The results for the areas corresponding to the Galactic center region ($-2\degr<\ell<+2\degr$) and the Cygnus X region are not included in this work and are left for future GLOSTAR catalog releases. 
A total of 52 observing sessions were performed to cover smaller areas using the hexagonal mosaicing scheme of pointed observations 
 \citep{Medina2019, brunthaler2021}\footnote{Note that the values given for the spacing and primary beam size in Sect. 2.1 of \citet{Medina2019} have incorrect units; they were given in arcseconds, whereas the correct units should be arcminutes. The hexagonal grid spacing is $\theta_{\rm hex}=3\rlap{.}'25$ and the FWHM primary beam size at the higher frequency band, 6.9\,GHz, is $\theta_{\rm B}=45/(\nu~{\rm [GHz]})=6\rlap{.}'5$. }.

The sessions observed while the VLA was in its  D- and DnC-configurations covered rectangular areas with sizes of two degrees in the Galactic latitude direction and one degree in the Galactic longitude direction, using $\sim$650 target pointings (spending $\sim22\,$s per target pointing). C-configuration observations, on the other hand, covered 1.5 degrees in the Galactic longitude direction and observed $\sim$950 target pointings (spending $\sim11\,$s per target pointing).

Quasars were observed for calibration purposes; in particular, 3C$~$286 and 
3C$~$48 were observed for amplitude and bandpass calibrations. A gain calibrator was observed for each session and chosen to be at an angular distance of ${\lesssim}10.0^{{\circ}}$ from the region covered in that session. The observations are part of the VLA projects with  IDs 11B-168, 13A-334, 15B-175, and 17A-197. The total telescope time used for these observations was 260 hours (that is, five hours per session). The observations occurred from December 2011 to June 2017. Individual observation dates, corresponding VLA project ID, VLA configuration, and used gain calibrators are listed in Table~\ref{tab:obs}.

\begin{table}[th]
\small
\begin{center}
\caption{ All observed epochs of the GLOSTAR-VLA survey presented in this work. }
\label{tab:obs}
\begin{tabular}{ccccc}
\hline
$\ell$ ($^\circ$) & VLA  & Date of Obs. & VLA &Phase calibrator \\
range & Project ID & dd/mm/yyyy& Conf.&\\
\hline
02	--	03	&  15B-175 	& 17/01/2016  & DnC &J1820--2528 \\ 
03	--	04	&  15B-175 	& 21/01/2016  & DnC &J1820--2528 \\ %
04	--	05	&  15B-175 	& 22/01/2016  & DnC &J1820--2528 \\
05	--	06	&  15B-175 	& 16/01/2016  & DnC &J1820--2528 \\
06	--	07	&  17A-197 	& 03/04/2017  &  D  &J1820--2528 \\ 
06	--	7.5	&  17A-197 	& 31/05/2017  &  C  &J1820--2528 \\
07	--	08	&  17A-197 	& 31/03/2017  &  D  &J1820--2528 \\
08	--	09	&  17A-197 	& 21/02/2017  &  D  &J1820--2528 \\
7.5	--	09	&  17A-197 	& 05/06/2017  &  C  &J1820--2528 \\
09	--	10	&  15B-175 	& 24/01/2016  & DnC &J1820--2528 \\
10	--	11	&  13A-334 	& 16/05/2013  & DnC &J1820--2528 \\
11	--	12	&  13A-334 	& 17/05/2013  & DnC &J1820--2528 \\ 
12	--	13	&  17A-197 	& 03/03/2017  &  D  &J1825--0737 \\ 
13	--	14	&  17A-197 	& 04/04/2017  &  D  &J1825--0737 \\
14	--	15	&  17A-197 	& 19/02/2017  &  D  &J1825--0737 \\
15	--	16	&  14A-420 	& 14/07/2014  &  D  &J1825--0737 \\
16	--	17	&  14A-420 	& 24/07/2014  &  D  &J1825--0737 \\
17	--	18	&  14A-420 	& 05/08/2014  &  D  &J1825--0737 \\
18	--	19	&  14A-420 	& 14/08/2014  &  D  &J1825--0737 \\
19	--	20	&  14A-420 	& 12/07/2014  &  D  &J1825--0737 \\
20	--	21	&  14A-420 	& 23/07/2014  &  D  &J1825--0737 \\
21	--	22	&  14A-420 	& 28/07/2014  &  D  &J1825--0737 \\
22	--	23	&  14A-420 	& 27/07/2014  &  D  &J1825--0737 \\
23	--	24	&  14A-420 	& 26/08/2014  &  D  &J1825--0737 \\
24	--	25	&  14A-420 	& 16/07/2014  &  D  &J1825--0737 \\
25	--	26	&  14A-420 	& 29/07/2014  &  D  &J1825--0737 \\
26	--	27	&  14A-420 	& 13/08/2014  &  D  &J1825--0737 \\
27	--	28	&  14A-420 	& 28/08/2014  &  D  &J1825--0737 \\ 
36	--	37	&  14A-420 	& 07/07/2014  &  D  &J1907+0127 \\
37	--	38	&  14A-420 	& 04/07/2014  &  D  &J1907+0127 \\
38	--	39	&  14A-420 	& 01/08/2014  &  D  &J1907+0127 \\
39	--	40	&  14A-420 	& 25/08/2014  &  D  &J1907+0127 \\
40	--	41	&  14A-420 	& 07/08/2014  &  D  &J1907+0127 \\
41	--	42	&  14A-420 	& 02/07/2014  &  D  &J1907+0127 \\
42	--	43	&  14A-420 	& 09/07/2014  &  D  &J1907+0127 \\
43	--	44	&  14A-420 	& 17/07/2014  &  D  &J1907+0127 \\
44	--	45	&  14A-420 	& 03/08/2014  &  D  &J1907+0127 \\
45	--	46	&  14A-420 	& 29/06/2014  &  D  &J1907+0127 \\
46	--	47	&  15B-175 	& 25/11/2015  &  D  &J1922+1530 \\
47	--	48	&  15B-175 	& 13/11/2015  &  D  &J1922+1530 \\
48	--	49	&  15B-175 	& 21/11/2015  &  D  &J1922+1530 \\
49	--	50	&  15B-175 	& 14/11/2015  &  D  &J1922+1530 \\
50	--	51	&  15B-175 	& 22/11/2015  &  D  &J1922+1530 \\
51	--	52	&  15B-175 	& 11/11/2015  &  D  &J1922+1530 \\
52	--	53	&  15B-175 	& 20/11/2015  &  D  &J1922+1530 \\
53	--	54	&  15B-175 	& 10/11/2015  &  D  &J1922+1530 \\
54	--	55	&  15B-175 	& 27/11/2015  &  D  &J1922+1530 \\
55	--	56	&  15B-175 	& 17/12/2015  &  D  &J1922+1530 \\
56	--	57	&  15B-175 	& 28/11/2015  &  D  &J1925+2106 \\
57	--	58	&  15B-175 	& 08/11/2015  &  D  &J1925+2106 \\
58	--	59	&  11B-168 	& 15/12/2011  &  D  &J1931+2243 \\
59	--	60	&  11B-168 	& 29/12/2011  &  D  &J1931+2243 \\
\hline
\end{tabular}
\end{center}
\end{table}

The data sets were edited, calibrated, and imaged using the \obit\ software package \citep{cotton2008}, designed 
for handling radio astronomical data\footnote{\obit\ applications can be accessed through a Python interface (ObitTalk). \obit\ also inter-operates with classic  AIPS  \citep[The Astronomical Image Processing System;][]{greisen2003} and has access to its tasks.}. Prior to calibration, the data sets were edited to remove the first 10 seconds of the scans on calibrators, which are affected by the antennas slewing. Then, a standard calibration is applied. This includes amplitude corrections based on the switched power signal, group delay, bandpass, amplitude and phase, and polarization calibration. The standard calibration was alternated with the editing of data affected by instrumental problems or radio frequency interference. Then, the data were reset and calibrated again, excluding the identified problematic data.

The images were produced using the \obit\ task MFImage. First, the data for every target pointing are imaged individually, covering the FWHM primary beam corresponding to the lowest continuum band, $4.7$\,GHz, which is $8\rlap{.}'4$.  
In cases for which the target pointing covered strong sources located outside the primary beam that are bright enough to affect the imaging, we add outlier fields at the location of these sources to account for them in the cleaning process. The observed bandwidth is divided into frequency bins that are narrow enough for the effects of variable spectral index and antenna pattern variations to be minor, enabling a joint spectral deconvolution. MFImage then generates an image for each individual frequency bin, and these images are then weighted averaged  
to produce the final map of the target field. The images are produced using a Briggs' weighting with robustness parameter 0.0. Finally, the frequency bins and the weighted averaged map are combined to produce the mosaiced images with a restoring beam of 18$''$ and a pixel size of $2\rlap{.}''5$ as described by \citet{Medina2019} and \citet{brunthaler2021}. The shortest baseline of the VLA in the D-configuration is 35\,m. Thus, sources with angular sizes larger than $\sim4'$ are filtered out from the images. In order to conserve computational resources, the total observed area was divided into six sub-mosaics that constitute the final radio continuum map. The full continuum image of the Galactic disk area covered by the GLOSTAR survey, $2\degr < \ell < 60\degr$ and $|b| < 1\degr$ (i.e., including the pilot region), is shown in Figure~\ref{fig:FullRadioMap}.

\section{ Analysis of the continuum map}\label{sec:RCM}

\subsection{Radio continuum map}

The observational data previously described were used to image 100 square degrees of the 
Galactic plane by producing six sub-mosaics (see Table\,\ref{tab:noise} 
for each longitude range, including the pilot region). Together with the 
sub-mosaic of the pilot region, previously reported by \citet{Medina2019}, 
we have a complete map of 116 square degrees of the Galactic
plane (that is, $2\degr  < \ell < 60\degr$ and $|b| < 1\degr$). 
{Figure~\ref{fig:FullRadioMap} presents the radio continuum map
of the complete region produced from the D, DnC, and C-configuration
data. This radio map has an effective central frequency of 5.8\,GHz 
and has been restored using a circular beam of $18''$.} 
Although the observed region extends slightly beyond $|b|=1.0\degr$,
the noise increases significantly in these outer regions because 
of the primary beam attenuation and because these areas 
do not overlap with other pointings. Hence, sources detected 
beyond this latitude range have been excluded 
from the analysis presented here, although they are listed in our final catalog.

\begin{figure}
\centering
\includegraphics[width=0.49\textwidth, trim= 0 0 0 0, angle=0] {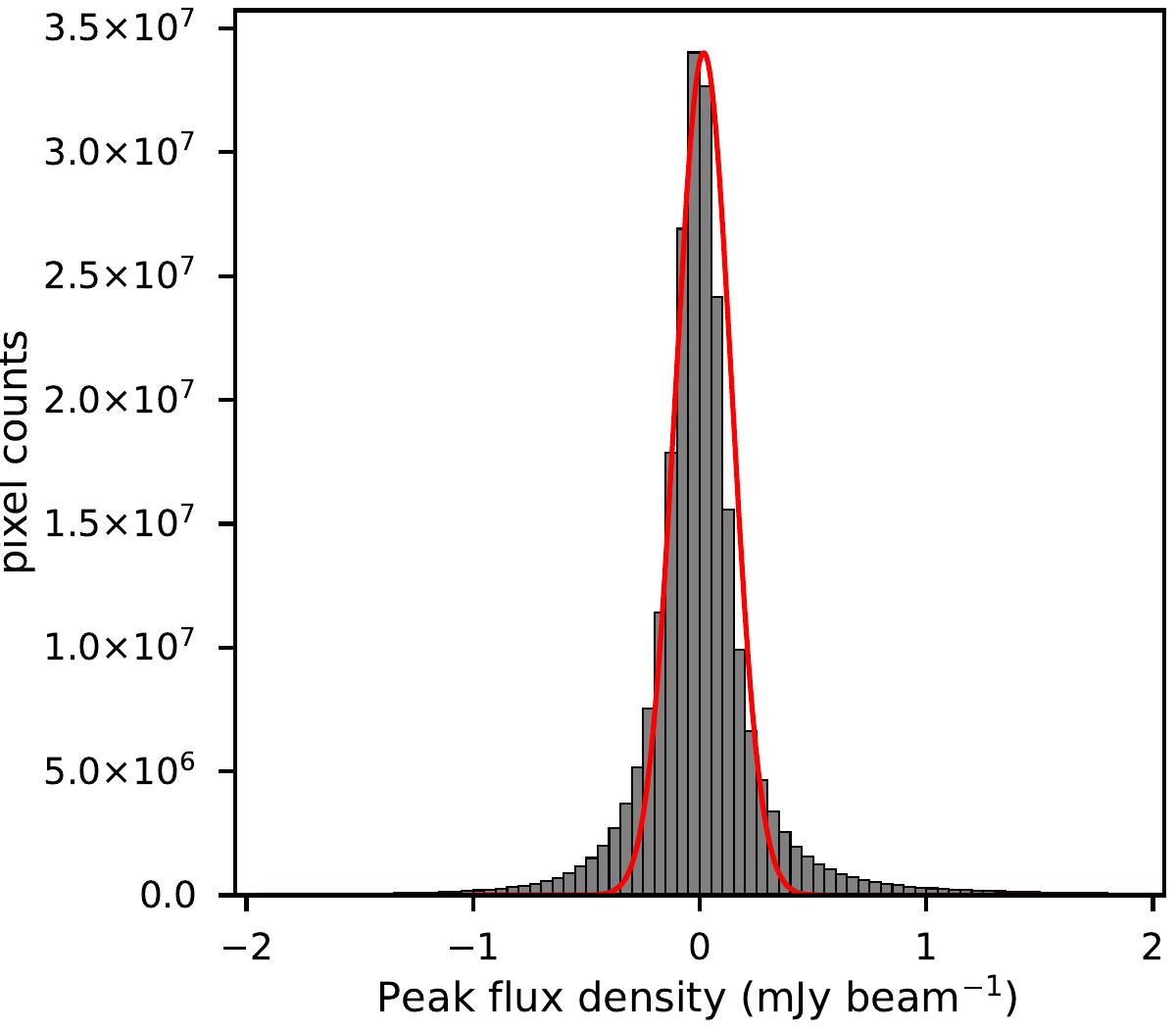} 
\caption{ Noise distribution of the full GLOSTAR VLA D-configuration mosaic. The range of pixel values ranges between $-$2 and 2\,mJy. The red line shows the results of a Gaussian fit to the distribution,  $\sigma_{rms}$ = 123 $\mu$Jy beam$^{-1}$. The bin size is 50\,$\mu$Jy beam$^{-1}$.}
\label{fig:noise_hist}
\end{figure}

Inspection of Fig.\,\ref{fig:FullRadioMap} reveals the presence of thousands of compact sources and many large-scale structures (LSSs, see Sect.~\ref{sect:Source_extraction} for details) associated, for example, with prominent star-forming complexes (e.g., W31, W33, W49, and W51) or SNRs (e.g., W28 and W44). The emission towards these more complex regions is not fully recovered due to the lack of short $(u,v)$ baselines, which also results in strong negative bowls that affect the noise level around them. Therefore, these LSSs cause significantly higher noise levels in the Galactic mid-plane and the inner part of the Galactic disk, where they are more densely concentrated, and affect the quality of the imaging around them. This is demonstrated in Table\,\ref{tab:noise} where the mean noise values determined from each of the mosaics can be seen to decrease with longitude from the Galactic center to the outer part of the disk where less star formation is occurring.  In Fig.\,\ref{fig:noise_hist} we present a histogram of the pixel values from the whole survey region. The significant noise variations across the GLOSTAR VLA D-configuration map produce the non-Gaussian profile of the pixel noise values.  We use a Gaussian fit to this pixel distribution to estimate the overall sensitivity of the final map. The standard deviation of the distribution of the pixel values,  $\sigma_{rms}$, is 123\,$\mu$Jy beam$^{-1}$; however, the noise values can be significantly higher towards prominent complexes concentrated towards the mid-plane. The noise level in regions free of extended emission is approximately 60\,$\mu$Jy\,beam$^{-1}$.

\begin{table}
\begin{center}
\caption{Average noise values estimated for the seven mosaic maps produced of the survey area across the full GLOSTAR radio map.}
\label{tab:noise}
\begin{tabular}{ |c |c |c |c |}
\hline\hline
Galactic longitude	&	$\sigma_{rms}$ & Radio  &  LSS \\
range& [$\mu$Jy beam$^{-1}$] & sources&\\
\hline
$\,\,\,2\degr$ --   $12\degr$  & 171  & 2026 & 31  \\   
       $12\degr$	-- $20\degr$  &	165  & 1655 & 19 \\
       $20\degr$	-- $28\degr$  &	157  & 1888 & 29\\
\,\,$\,28\degr$  --  $36\degr$\tablefootmark{*} & 150  & \,\,\,\,\,1770\tablefootmark{**} &27\\
$36\degr$	-- $44\degr$  &	119  & 1756 & 8\\
$44\degr$	-- $52\degr$  &	98   & 1766 & 13\\
$52\degr$	-- $60\degr$  &	84   & 2120 &  5\\
\hline
$2\degr$ -- $28\degr$; $36\degr$ -- $60\degr$\tablefootmark{***} & 123 & 11211 & 106 \\
$2\degr$ -- $60\degr$ & 123 & 12981 & 132 \\
\hline
\end{tabular}
\small
\tablefoottext{*}{Corresponding to the pilot region, previously reported by \citet{Medina2019}.}
\tablefoottext{**}{This number includes 195 sources related to the 27 LSS, as well as the 1575 individual sources. }
\tablefoottext{***}{This work.}
\end{center}
\end{table}

\begin{figure}[hp!]
\centering
\includegraphics[width=0.48\textwidth, trim= 5 50 5 50, angle=0] {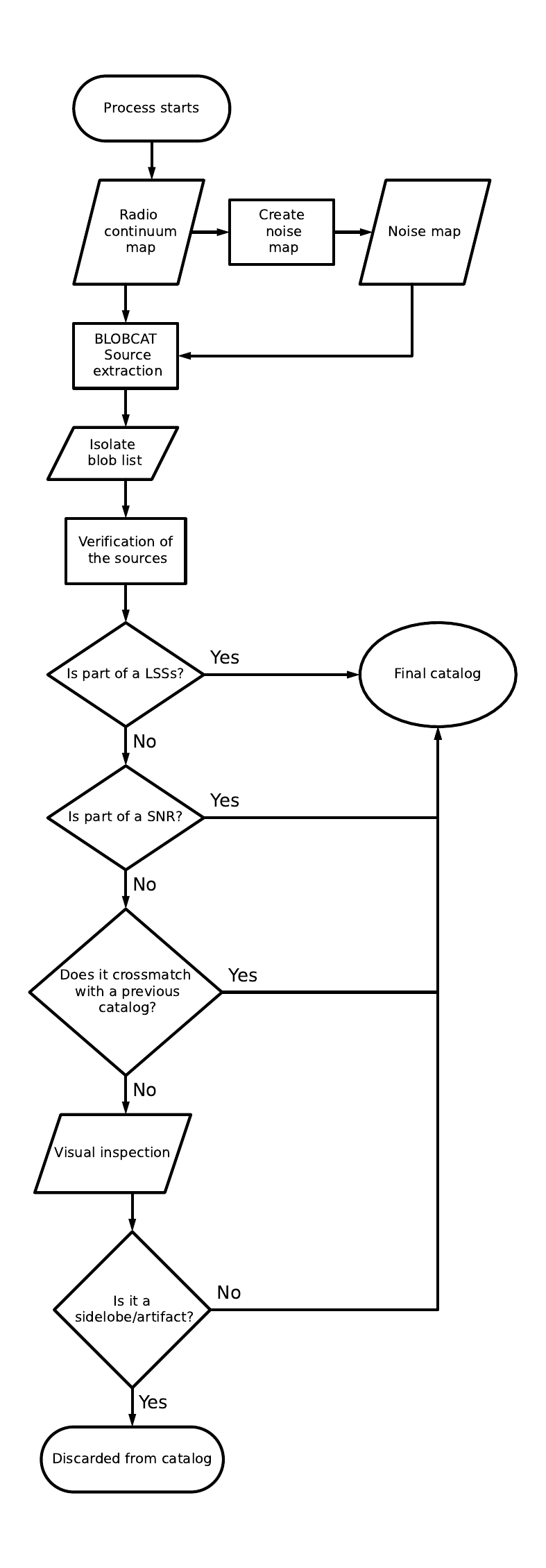} 
    \caption{Methodology of the source extraction and verification followed for the GLOSTAR-VLA survey data to create the final catalog. }
\label{fig:method}
\end{figure}

\subsection{Source extraction}\label{sect:Source_extraction}

The range of detected structures, level of complexity, variation, and survey coverage make source extraction and analysis difficult in the GLOSTAR low-resolution images presented here. In our previous analysis of the pilot region, we developed a number of steps to automatically identify sources and confirm them visually and by spatially correlating them with published source catalogs at different wavelengths \citep{Medina2019}.  In this paper, we followed the same process to create a catalog of radio sources for the rest of the covered Galactic plane, which were then combined with the sources detected in the pilot field to create a complete catalog of radio sources in the GLOSTAR VLA D-configuration radio map (excluding the Galactic center and the Cygnus X regions); that is $2\degr < \ell <  60\degr$ and $|b| <  1\degr$. We have again used the \blobcat\ software package to perform the source extraction; see \citet{hales2012} for a detailed description of this Python code. This code has also been previously used for the source extraction of THOR, a complementary 21\,cm radio continuum survey \citep{beuther2016,wang2018}), which will facilitate further comparisons. 
Below we provide a brief outline of the method used and refer the reader to \citet{Medina2019} for more details. The methodology steps are also visualized in Fig. \ref{fig:method}.

{\it Creation of the noise maps.} To perform the automatic source extraction, we first need to produce an independent noise map of the region due to the position-dependent nature of the noise in GLOSTAR VLA D-configuration radio continuum maps. We use the \textit{rms} estimation algorithm within the SExtractor package \citep{bertin1996, holwerda2005}. This algorithm defines the {\it rms} value for each pixel in an image by determining the distribution of pixel values within a local mesh until all values are around a chosen sigma value. Following our previous work \citet{Medina2019}, for the calculations we used a detection and analysis threshold of 5$\sigma$, a minimum size of 5 pixels, and a mesh size of $80 \times 80$ pixels$^2$. As a result, most real emission is removed from the noise image, and the determined noise map represents the correct noise level.

{\it Automatic source extraction with \blobcat.} We employ the \blobcat\ software package that uses an algorithm to detect and identify blobs, or islands of pixels representing sources, in two-dimensional astronomical radio-wavelength images \citep[see][ for details]{hales2012}. The algorithm makes a pixel-by-pixel comparison between the main image and the background noise map, locates pixels above a given threshold, and defines ``blobs'' around these pixels. Then, a first source parameter set is obtained by fitting a 2D elliptical Gaussian. Later, the peak and integrated flux densities are corrected by considering biases from the Gaussian fits, such as source morphology \citep[see][ for details]{hales2012}.  \blobcat\ returns several parameters from which we use the following (and their corresponding errors) for our catalog construction: the 2D source position (\verb|RA_p| and \verb|Dec_p|; notice these are software internal names and not the image coordinate system), peak flux density (\verb|S_p_CBBWS|), integrated flux density (\verb|S_int_CB|), and the number of pixels covered by the source (\verb|npix|).  

Following our previous work, we applied a detection threshold (dSNR) of 4$\sigma$ and a minimum source size of 5 pixels in diameter ($\sim$13\arcsec). This resulted in the extraction of \firstblobs\ blobs potentially associated with radio sources in the area delimited by  $2\degr < \ell < 28\degr$, $36\degr < \ell < 60\degr$ and $|b| <  1\degr$. All of them have flux above four times that in the local noise map and have a size comparable to or larger than the beam. However, this sample contains a significant number of artifacts that need to be removed, and also emission from LSSs; that is, sources fragmenting into multiple components that need to be grouped. Furthermore, at the edges of the map, the noise increases significantly; therefore, the sources located at $|b|>1\degr$ are excluded from further analysis and their identification names are 
marked with an $*$.  The following steps will address these issues.  Additionally, in our previous work, we also flagged (that is labeled) sources that appeared to consist of two or more distinct emission peaks; however, in this work, we have not attempted to flag them or to separate the emission because in our previous work, only around 3$\%$ of the sources are affected by this problem, and those are resolved in the  GLOSTAR VLA B-configuration images \citep[e.g.,][]{dzib2023}. Finally, as in our previous works \citep[e.g.,][]{Medina2019}, we considered the Y-factor (Y$_{\rm factor} = S_{\nu, {\rm Int}}/S_{\nu, {\rm Peak}}$)
 and roughly divided the sources into extended  (Y$_{\rm factor} >2.0$), compact 
($1.2<$Y$_{\rm factor} \leq2.0$) and unresolved  (Y$_{\rm factor}\leq$1.2).

\begin{figure*}[]
\centering
\includegraphics[width=0.49\textwidth, trim= 0 0 0 0 clip, angle=0]{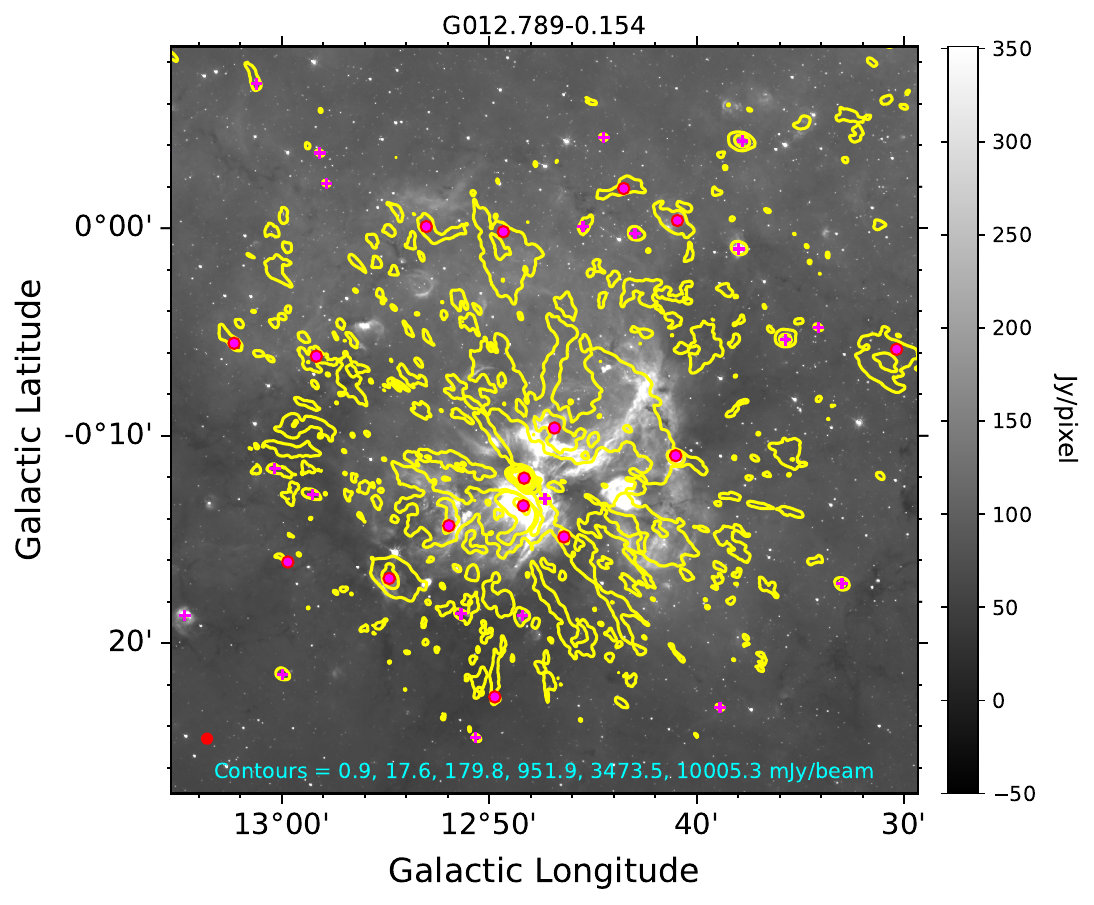} 
\includegraphics[width=0.49\textwidth, trim= 0 0 0 0 clip, angle=0]{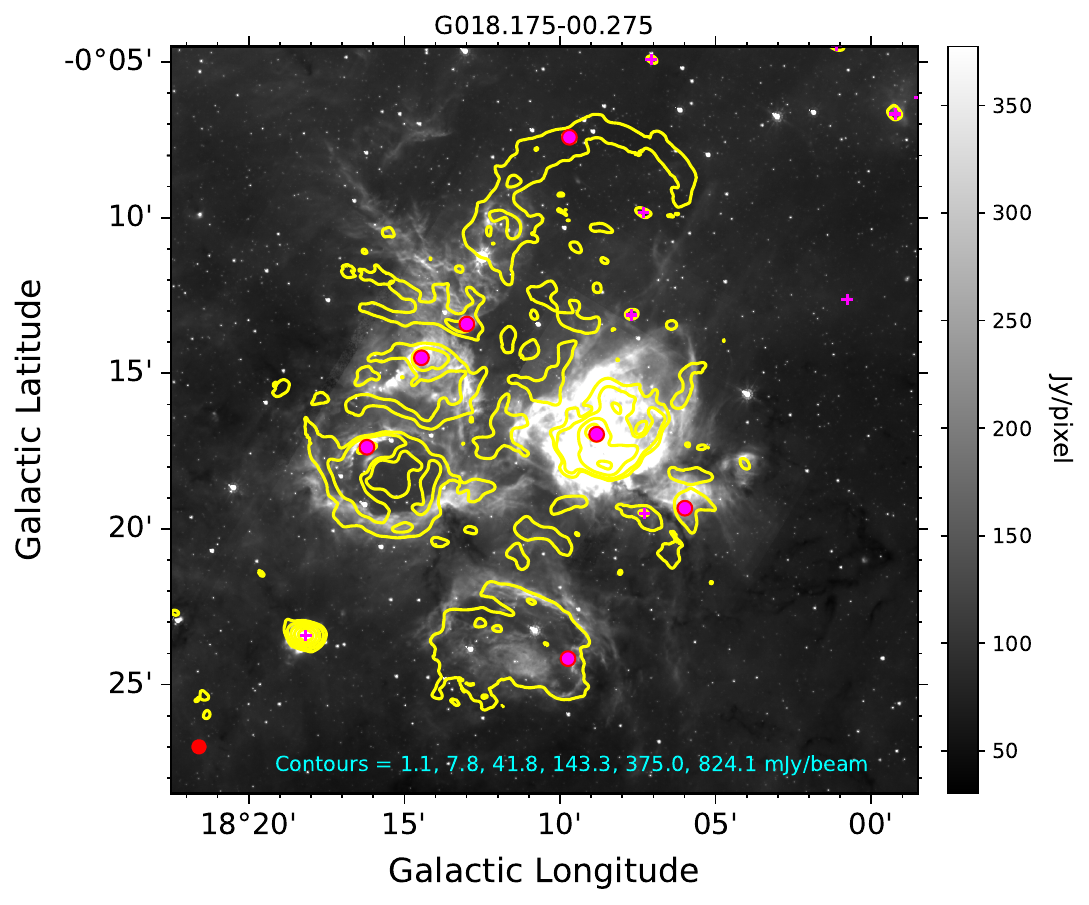}
\includegraphics[width=0.49\textwidth, trim= 0 0 0 0 clip, angle=0]{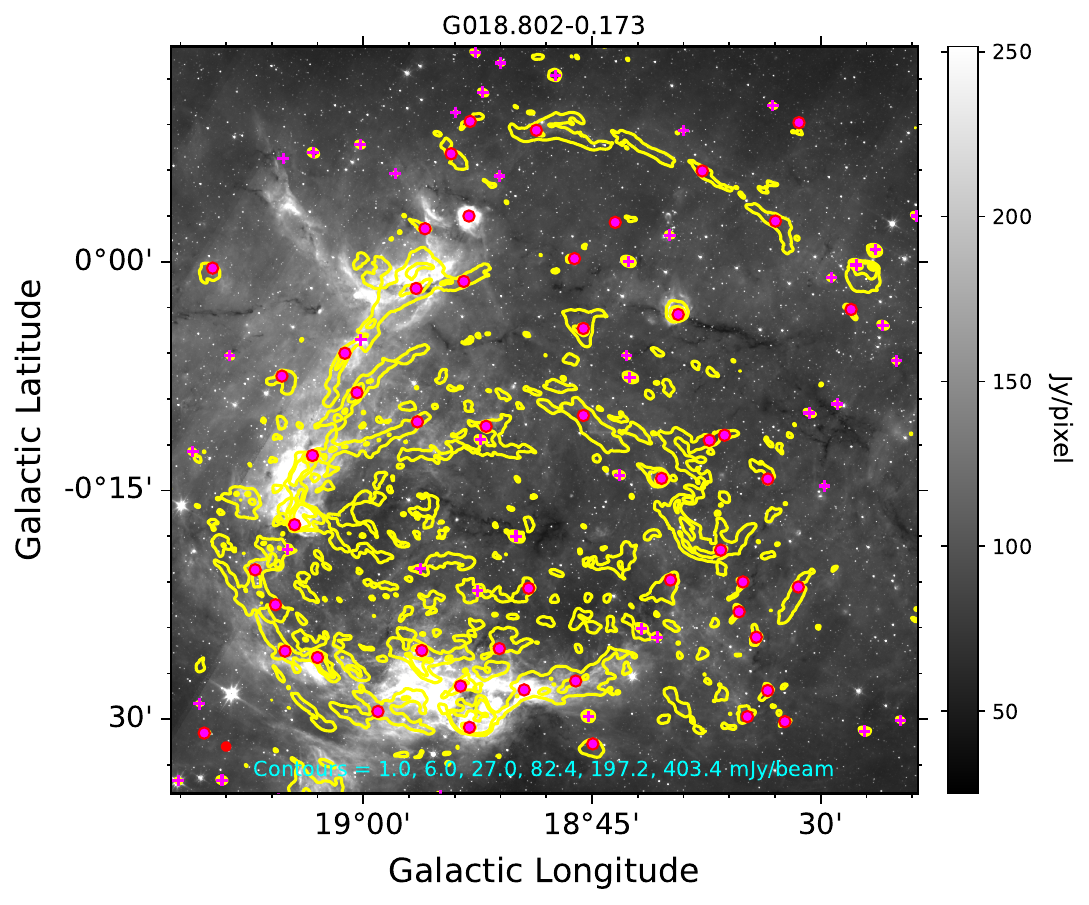}
\includegraphics[width=0.49\textwidth, trim= 0 0 0 0 clip, angle=0]{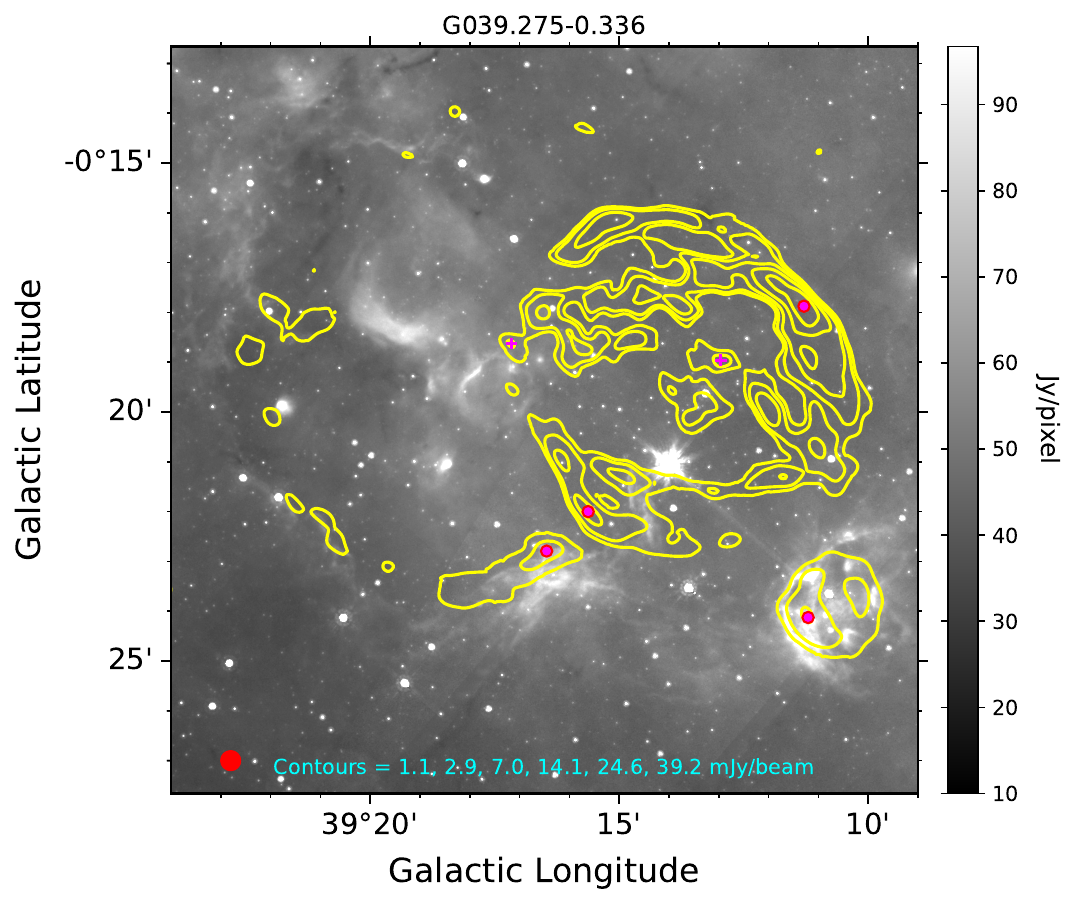}
\caption{Examples of radio emission {associated with} large-scale structures. The grayscale background represents GLIMPSE 8\,$\mu$m emission, while the yellow contours are the GLOSTAR 5.8\,GHz radio continuum emission. These regions often show coherent infrared structures that are morphologically correlated with {most of} the radio emission (see examples presented in the upper panels). Although this correlation is not always present, the different radio emission blobs are clearly related with each other (see examples presented in the lower panels). 
The contours are determined as described in the figures, and the red-filled circle shown in the lower left corners shows the GLOSTAR VLA D-configuration beam FWHM. The magenta-filled circles indicate the peak positions of radio sources associated with these regions, and magenta crosses indicate compact radio sources in the area that are treated as separate sources.
} 
\label{fig.zoom_maps}
\end{figure*}

{\it Identification of large-scale structures.} In our previous work, we defined LSSs as over-resolved structures whose emission is broken up into multiple independent fragments of radio sources. To identify LSSs, we performed a visual inspection and compared the radio emission distribution with their corresponding 8.0~$\mu$m maps extracted from the Galactic Legacy Infrared Mid-Plane Survey 
Extraordinaire (GLIMPSE) archive. 
Most GLOSTAR sources with corresponding 8.0~$\mu$m have thermal radio emission; that is they are \hii\ regions. SNRs are also generally faint or even not detected at mid-infrared (MIR) wavelengths \citep{Dokara2021}. 
In Fig.\,\ref{fig.zoom_maps} we show some examples of these LSSs. We delimited a rectangular area around the coherent infrared structures morphologically correlated with the radio emission (see the left panel of Fig. \ref{fig.zoom_maps}). Also, we have delimited a rectangular area around radio emission sources with shell-like morphology that are clearly part of one source even if there is no clear correlation with the  8.0~$\mu$m emission, which is the case for many SNRs (see the lower right panel of Fig. \ref{fig.zoom_maps}). Finally, we also classified as LSSs the complexes with two or more independent sources that clearly belong to the same star-forming region as the one seen in the 8.0~$\mu$m maps (see the top right panel of Fig. \ref{fig.zoom_maps}). All extended radio sources in these boxed regions are considered to be part of one LSS. Individual fragment components and their properties are given in the catalog; however, they have been excluded from the statistical analysis presented here.  
Compact and unresolved radio sources that are considered unlikely to be associated with LSSs have been retained in the final catalog as distinct sources, since these are possible individual sources and not part of the extended emission. They are included in our further analysis.

In this work, we have visually identified a total of \largestructure large-scale radio emission regions containing \includeLarge\ individual fragments within $2\degr < \ell < 28\degr$, $36\degr < \ell < 60\degr$ and $|b| <  1\degr$, which includes the SNRs like W28, W30, and W49 \citep[e.g., ][]{green2019} and star-forming region complexes such as W31 \citep{ghosh1989}, W33 \citep{messineo2011}, M16 \citep{oliveira2008}, M17 \citep{chini2008}, W39 \citep{kerton2013}, and  W51 \citep{gingsburg2017}.

In Table\,\ref{tbl:LargeStruc}, we give the list of LSSs, including the 27 from the pilot region. The name of each structure has been constructed from the central coordinates of the enclosing box. The table includes the extent of the boxed region in $\ell$ and $b$ and the integrated flux of the total emission in the box minus the emission from any compact sources that are considered as distinct sources. The total integrated flux density of each LSS, is calculated by adding the flux densities of all components together. The measured parameters of LSSs should be used with caution as the flux densities and sizes are unlikely to correspond to discrete sources, but rather to source complexes. We also point the readers to \cite{Dokara2021} and \cite{Dokara2023} for a discussion on the calculation of flux densities of sources identified as SNRs.

\setlength{\tabcolsep}{2pt}
\begin{table*}[htpb!]
\footnotesize
\renewcommand*{\arraystretch}{0.93}
  \begin{center}
  \caption{Large-scale structure sources.} 
  \label{tbl:LargeStruc}
 \begin{tabular}{lccccc|lccccc}\hline\hline
Name      & \multicolumn{1}{c}{$\ell_{\rm min}$} &\multicolumn{1}{c}{$\ell_{\rm max}$} &  \multicolumn{1}{c}{$b_{\rm min}$} & \multicolumn{1}{c}{$b_{\rm max}$} & \multicolumn{1}{c|}{Int flux}&Name      & \multicolumn{1}{c}{$\ell_{\rm min}$} &\multicolumn{1}{c}{$\ell_{\rm max}$} &  \multicolumn{1}{c}{$b_{\rm min}$} & \multicolumn{1}{c}{$b_{\rm max}$} & \multicolumn{1}{c}{Int flux}\\ 
      & \multicolumn{1}{c}{(\degr)} & \multicolumn{1}{c}{(\degr)} &  \multicolumn{1}{c}{(\degr)} &  \multicolumn{1}{c}{(\degr)} & \multicolumn{1}{c|}{(Jy)}&      & \multicolumn{1}{c}{(\degr)} & \multicolumn{1}{c}{(\degr)} &  \multicolumn{1}{c}{(\degr)} &  \multicolumn{1}{c}{(\degr)} & \multicolumn{1}{c}{(Jy)}\\ 
  \hline
G002.273$+$00.233&$2.2022$&$2.3444$&$0.1851$&$0.2809$&$ 0.876$          &G024.782$+$00.083&$24.6712$&$24.8938$&$-0.0231$&$0.1899$&$ 3.130$\\
G003.020$-$00.831&$2.6983$&$3.3427$&$-1.0227$&$-0.6386$&$ 0.413$        &G025.376$-$00.210 (W42)&$25.1914$&$25.5599$&$-0.4167$&$-0.0030$&$14.304$\\
G003.264$-$00.059&$3.1428$&$3.3843$&$-0.1798$&$0.0624$&$ 2.054$         &G025.571$+$00.208&$25.4747$&$25.6664$&$0.1571$&$0.2598$&$ 0.428$\\
G003.431$+$00.169&$3.3031$&$3.5599$&$0.1174$&$0.2205$&$ 0.285$          &G025.867$+$00.132&$25.7329$&$26.0002$&$-0.0196$&$0.2839$&$ 1.769$\\
G003.784$-$00.278$^\star$&$3.6846$&$3.8840$&$-0.4004$&$-0.1557$&$0.351$ &G026.123$+$00.021$^{\star\star}$&$25.9820$&$26.2630$&$-0.0301$&$0.0725$&$ 0.260$\\
G003.930$-$00.072&$3.8559$&$4.0048$&$-0.1254$&$-0.0180$&$ 0.416$        &G026.511$-$00.371&$26.3833$&$26.6394$&$-0.4873$&$-0.2551$&$ 0.305$\\
G004.382$+$00.125&$4.2975$&$4.4671$&$0.0245$&$0.2253$&$ 1.399$          &G027.240$-$00.108$^{\star\star}$&$27.1052$&$27.3742$&$-0.1894$&$-0.0275$&$ 0.975$\\
G005.151$-$00.321&$5.0827$&$5.2202$&$-0.4099$&$-0.2319$&$ 0.529$        &G027.301$+$00.265&$27.2457$&$27.3555$&$0.2408$&$0.2901$&$ 0.085$\\
G005.346$+$00.104&$5.2181$&$5.4741$&$-0.0436$&$0.2520$&$ 0.156$         &G027.395$-$00.008$^\star$&$27.3435$&$27.4462$&$-0.0489$&$0.0337$&$ 0.804$\\
G005.351$-$00.976&$5.1610$&$5.5414$&$-1.0161$&$-0.9357$&$ 0.645$        &G027.497$+$00.191&$27.4427$&$27.5518$&$0.1595$&$0.2235$&$ 0.781$\\
G005.886$-$00.453 (W28)&$5.7897$&$5.9815$&$-0.6064$&$-0.3000$&$ 8.493$  &G027.504$-$00.115&$27.4122$&$27.5952$&$-0.2009$&$-0.0291$&$ 0.168$\\
G006.430$-$00.218$^\star$&$6.0130$&$6.8468$&$-0.6657$&$0.2295$&$ 5.051$ &G036.705$-$00.175$^{\star\star}$&$36.5630$&$36.8467$&$-0.2978$&$-0.0518$&$ 0.184$\\
G007.113$-$00.138$^\star$&$6.9050$&$7.3214$&$-0.3177$&$0.0418$&$ 0.734$ &G036.886$+$00.475$^{\star\star}$&$36.7968$&$36.9760$&$0.4294$&$0.5202$&$ 0.124$\\
G007.420$+$00.710&$7.3358$&$7.5032$&$0.6228$&$0.7978$&$ 0.274$          &G037.855$+$00.319$^{\star\star}$&$37.6403$&$38.0700$&$0.0971$&$0.5416$&$ 0.667$\\
G008.134$+$00.239 (W30)&$7.9896$&$8.2776$&$0.1119$&$0.3667$&$ 5.097$    &G039.275$-$00.336$^\star$&$39.1547$&$39.3947$&$-0.4261$&$-0.2462$&$ 1.132$\\
G008.159$-$00.121 (W30)&$7.9677$&$8.3505$&$-0.1932$&$-0.0489$&$ 0.764$  &G041.087$-$00.213&$41.0224$&$41.1522$&$-0.2500$&$-0.1760$&$ 0.296$\\
G008.343$-$00.317 (W30)&$8.2847$&$8.4015$&$-0.3662$&$-0.2686$&$ 0.826$  &G041.479$+$00.408$^\star$&$41.3678$&$41.5892$&$0.2722$&$0.5442$&$ 0.666$\\
G008.557$-$00.291 (W30)&$8.4830$&$8.6311$&$-0.3580$&$-0.2242$&$ 0.223$  &G043.156$+$00.001 (W49)&$43.0275$&$43.2852$&$-0.1192$&$0.1211$&$32.464$\\
G008.838$-$00.310 (W30)&$8.6346$&$9.0410$&$-0.4385$&$-0.1816$&$ 2.070$  &G043.421$+$00.556&$43.3305$&$43.5112$&$0.4884$&$0.6243$&$ 0.153$\\
G009.603$+$00.203&$9.4802$&$9.7264$&$0.1173$&$0.2891$&$ 1.188$          &G044.129$+$00.051&$43.9331$&$44.3240$&$-0.0641$&$0.1653$&$ 0.914$\\
G009.679$-$00.078$^\star$&$9.6082$&$9.7503$&$-0.1704$&$0.0138$&$ 0.211$ &G044.797$-$00.496&$44.7431$&$44.8515$&$-0.5644$&$-0.4284$&$ 0.098$\\
G009.697$-$00.861&$9.6285$&$9.7648$&$-0.9388$&$-0.7840$&$ 0.271$        &G045.100$+$00.151&$45.0045$&$45.1957$&$0.0653$&$0.2370$&$ 6.241$\\
G009.782$+$00.571$^\star$&$9.6491$&$9.9141$&$0.4650$&$0.6773$&$ 0.385$  &G045.453$+$00.083&$45.2807$&$45.6244$&$-0.0524$&$0.2177$&$ 7.326$\\
G009.949$-$00.809$^\star$&$9.8684$&$10.0300$&$-0.9148$&$-0.7034$&$1.101$&G045.644$-$00.439$^\star$&$45.4986$&$45.7898$&$-0.5602$&$-0.3175$&$ 0.169$\\
G010.094$-$00.306&$9.7930$&$10.3944$&$-0.5601$&$-0.0509$&$28.705$       &G046.205$+$00.017$^{\star\star}$&$46.1323$&$46.2778$&$-0.0698$&$0.1028$&$ 0.039$\\
G010.642$-$00.393 (W31)&$10.5689$&$10.7153$&$-0.4431$&$-0.3432$&$5.594$ &G046.388$+$00.880&$46.3046$&$46.4717$&$0.8094$&$0.9497$&$ 0.083$\\
G011.181$-$00.348$^\star$&$11.1308$&$11.2306$&$-0.3936$&$-0.3018$&$1.957$&G046.763$-$00.272$^\star$&$46.6174$&$46.9084$&$-0.4204$&$-0.1237$&$ 0.719$\\
G011.184$+$00.128&$11.0846$&$11.2825$&$0.0497$&$0.2053$&$ 0.777$        &G046.771$+$00.257&$46.6066$&$46.9354$&$0.0866$&$0.4265$&$ 0.324$\\
G011.194$-$00.687&$11.1459$&$11.2415$&$-0.7555$&$-0.6180$&$ 0.095$      &G048.597$+$00.036&$48.5073$&$48.6868$&$-0.0476$&$0.1189$&$ 2.242$\\
G011.386$-$00.060$^\star$&$11.3002$&$11.4722$&$-0.1233$&$0.0038$&$0.771$&G048.630$+$00.243&$48.5466$&$48.7136$&$0.1891$&$0.2970$&$ 0.632$\\
G011.902$-$00.141&$11.7201$&$12.0838$&$-0.3116$&$0.0300$&$ 3.402$       &G049.341$-$00.462 (W51)&$48.8824$&$49.7988$&$-0.8138$&$-0.1105$&$72.024$\\
G012.789$-$00.154 (W33)&$12.4841$&$13.0933$&$-0.4020$&$0.0936$&$27.244$ &G051.084$+$00.016&$50.7336$&$51.4349$&$-0.2626$&$0.2946$&$ 1.315$\\
G013.445$+$00.146$^\star$&$13.3955$&$13.4942$&$0.0943$&$0.1976$&$ 0.437$&G051.203$-$00.753&$51.1109$&$51.2943$&$-0.8542$&$-0.6516$&$ 0.150$\\
G013.644$+$00.252&$13.5986$&$13.6898$&$0.2122$&$0.2919$&$ 0.068$        &G051.926$+$00.567&$51.6540$&$52.1973$&$0.3706$&$0.7628$&$ 1.025$\\
G013.732$-$00.026&$13.6419$&$13.8220$&$-0.1202$&$0.0688$&$ 0.169$       &G052.860$-$00.527&$52.7256$&$52.9947$&$-0.6574$&$-0.3974$&$ 0.377$\\
G013.787$-$00.843&$13.5905$&$13.9839$&$-1.0315$&$-0.6550$&$ 1.322$      &G053.938$+$00.236&$53.6859$&$54.1899$&$0.0142$&$0.4575$&$ 1.442$\\
G013.987$-$00.133&$13.9315$&$14.0431$&$-0.1734$&$-0.0928$&$ 0.708$      &G054.564$-$00.074&$54.2844$&$54.8440$&$-0.2404$&$0.0922$&$ 0.241$\\
G014.509$+$00.071&$14.2662$&$14.7516$&$-0.1981$&$0.3409$&$ 3.568$       &G057.239$+$00.839$^\star$&$57.1496$&$57.3291$&$0.7303$&$0.9475$&$ 0.185$\\
G015.098$-$00.668 (M17)&$14.7537$&$15.4420$&$-1.0629$&$-0.2741$&$18.100$&G059.600$-$00.180&$59.5392$&$59.6608$&$-0.2312$&$-0.1282$&$ 0.072$\\ \cline{7-12}
G015.665$-$00.212&$15.5952$&$15.7341$&$-0.2672$&$-0.1567$&$ 0.094$&\multicolumn{6}{l}{Pilot Region \citep{Medina2019}:}\\ \cline{7-12}
G016.664$-$00.338&$16.5305$&$16.7973$&$-0.4606$&$-0.2155$&$ 0.447$&G028.026$-$00.045&27.970&28.082&-0.100&0.010&0.32\\
G016.920$+$00.832 (M16)&$16.1891$&$17.6513$&$0.5999$&$1.0636$&$ 2.881$&G028.520+00.132&28.470&28.570&0.049&0.215&0.30\\
G017.478$-$00.114$^\star$&$17.3969$&$17.5591$&$-0.1994$&$-0.0285$&$ 0.183$&G028.6$-$00.1 (SNR)$^{\star}$&28.544&28.694&-0.188&-0.038&0.95\\
G017.796$-$00.029$^{\star\star}$&$17.7099$&$17.8825$&$-0.1043$&$0.0458$&$ 0.079$&G029.087$-$00.682&28.976&29.199&-0.756&-0.608&0.22\\
G018.175$-$00.275&$18.0359$&$18.3139$&$-0.4474$&$-0.1033$&$ 5.957$&G029.219+00.415&29.095&29.343&0.344&0.486&0.50\\
G018.772$+$00.417$^\star$&$18.6292$&$18.9155$&$0.2781$&$0.5559$&$ 2.154$&G029.6+00.1 (SNR)$^{\star}$&29.512&29.612&0.071&0.165&0.13\\
G018.802$-$00.173 (W39)&$18.4183$&$19.1855$&$-0.5303$&$0.1836$&$ 7.595$&W43$-$south  center&29.805&30.131&-0.195&0.071&9.08\\
G018.948$-$00.934$^\star$&$18.6930$&$19.2028$&$-1.0442$&$-0.8241$&$ 0.631$&G029.986$-$00.582&29.913&30.060&-0.650&-0.513&0.11\\
G019.621$-$00.224&$19.4219$&$19.8206$&$-0.3935$&$-0.0536$&$ 6.335$&G030.462+00.449&30.405&30.519&0.393&0.506&0.16\\
G019.988$-$00.180$^\star$&$19.8842$&$20.0916$&$-0.2779$&$-0.0823$&$ 1.733$&W43 center&30.525&30.908&-0.238&0.080&13.43\\
G020.485$+$00.181&$20.4336$&$20.5371$&$0.1370$&$0.2245$&$ 0.332$&G031.053+00.482&30.992&31.115&0.435&0.529&0.59\\
G020.727$-$00.154&$20.6068$&$20.8465$&$-0.2996$&$-0.0075$&$ 1.837$&G031.166$-$00.106&31.117&31.214&-0.050&-0.162&0.42\\
G021.023$-$00.469$^\star$&$20.9295$&$21.1160$&$-0.5487$&$-0.3893$&$ 0.100$&G031.411$-$00.234&31.355&31.467&-0.286&-0.183&0.60\\
G021.076$-$00.281&$20.9225$&$21.2286$&$-0.3555$&$-0.2057$&$ 0.324$&G031.5$-$00.6 (SNR)$^{\star}$&31.400&31.700&-0.790&-0.490&0.14\\
G021.553$-$00.098&$21.5013$&$21.6043$&$-0.1442$&$-0.0510$&$ 0.132$&G031.9+00.0 (SNR)$^{\star}$&31.821&31.920&-0.053&0.071&1.01\\
G021.633$-$00.237$^{\star\star}$&$21.5136$&$21.7515$&$-0.3378$&$-0.1368$&$ 0.238$&G032.4+00.1 (SNR)$^{\star}$&32.322&32.488&0.027&0.172&0.10\\
G021.810$-$00.515$^\star$&$21.6236$&$21.9964$&$-0.7031$&$-0.3263$&$ 1.896$&G032.586$-$00.172&32.534&32.637&-0.284&-0.059&0.26\\
G022.885$-$00.328&$22.7114$&$23.0583$&$-0.4449$&$-0.2110$&$ 1.468$&G032.8$-$00.1 (SNR)$^{\star}$&32.649&32.950&-0.269&0.115&0.67\\
G023.100$+$00.553&$23.0305$&$23.1686$&$0.4874$&$0.6179$&$ 0.311$&G033.179$-$00.010&33.118&33.240&-0.075&0.055&0.66\\
G023.279$-$00.343$^\star$&$23.0626$&$23.4947$&$-0.5853$&$-0.1008$&$ 4.065$&G033.2$-$00.6 (SNR)$^{\star}$&33.033&33.327&-0.708&-0.395&0.09\\
G023.575$-$00.046&$23.4805$&$23.6688$&$-0.1478$&$0.0556$&$ 0.881$&G033.6+00.1 (SNR)$^{\star}$&33.567&33.781&-0.067&0.132&0.93\\
G023.846$-$00.185$^{\star\star}$&$23.8014$&$23.8903$&$-0.2344$&$-0.1350$&$ 0.072$&G034.260+00.125&34.177&34.343&0.043&0.207&10.17\\
G024.186$+$00.233&$24.1238$&$24.2477$&$0.1552$&$0.3108$&$ 0.443$&G034.6$-$00.5 (SNR)$^{\star}$&34.390&34.930&-0.760&-0.093&2.96\\
G024.385$+$00.688&$24.3285$&$24.4413$&$0.5469$&$0.8300$&$ 0.159$&G035.032$-$00.283&34.994&35.070&-0.363&-0.204&0.04\\
G024.461$+$00.271&$24.3315$&$24.5912$&$0.1915$&$0.3500$&$ 1.139$&G035.506$-$00.026&35.316&35.696&-0.157&0.106&2.36\\
G024.725$-$00.083&$24.6746$&$24.7749$&$-0.1392$&$-0.0266$&$ 0.567$&G035.6$-$00.4 (SNR)$^{\star}$&35.502&35.692&-0.593&-0.258&2.36\\
G024.739$-$00.658$^\star$&$24.6763$&$24.8022$&$-0.8330$&$-0.4831$&$ 0.687$&G035.680$-$00.868&35.610&35.750&-0.943&-0.792&0.19\\
\hline
\end{tabular}
\end{center}
Notes: The LSS name is constructed from the central position of the box used to encapsulate the regions. When the LSS is known to be (part of) a well-known region, this is mentioned within brackets next to the GLOSTAR name. 
The sources with $^\star$ and $^{\star\star}$  are associated with SNRs from the catalog of \cite{green2019} and SNR candidates from the catalog of \citet{anderson2017}, respectively.
The values in the Int flux column are obtained by adding the flux densities of individual components as obtained from the flux extraction procedure described in Section~\ref{sect:Source_extraction}. 
\end{table*}

{\it Crossmatch with recently identified SNR candidates.} SNRs are sources of nonthermal radio emission \citep{anderson2017} that can also be detected with the GLOSTAR observations \citep[see][]{Medina2019, Dokara2021,Dokara2023}. The radio emission from most of these radio sources is extended and, in many cases, has a weak surface brightness. Because of these properties, automatic source extraction is not optimal to fully recover these sources. Their identification and total flux determination require a more detailed visual inspection that goes beyond the scope of this work. SNRs and SNR candidates in the observed area of the GLOSTAR survey are the subject of the studies presented by  \citet{Dokara2021} and \cite{Dokara2023}. However, our procedure recovered sources located in the regions containing well-known SNRs and previously identified SNR candidates. Previously, well-known SNRs with large angular scales were identified as LSSs (see also Table~\ref{tbl:LargeStruc}). Additionally, we also identified \includeSNR\ extended sources ($Y_{\rm factor}>2.0$) 
located in areas of other SNR candidates, including the 80 new SNR candidates from the GLOSTAR survey \citep{Dokara2021}. These sources are labeled in the final catalog.

{\it Cross match with previous catalogs.} In order to verify the remaining sources, the next step was to assume that any radio source that has a counterpart in a published radio or MIR catalog is real. 
We used several complementary multiwavelength surveys to cross match with our detected radio sources to identify possible counterparts. The selected surveys  are better described in the following sections. They either have comparable angular resolution: GLIMPSE \citep{fazio2004}, WISE \citep[Wide-Field Infrared Survey Explorer;][]{Wright2010}, HiGAL \citep[the Herschel infrared Galactic Plane Survey;][]{molinari2010,Molinari2016}, ATLASGAL \citep[the APEX Telescope Large Area Survey of the Galaxy;][]{schuller2009}, GLOSTAR VLA B-configuration \citep{yang2023}, CORNISH \citep{hoare2012}, RMS \citep[the Red MSX Source survey;][]{urquhart2009}, and THOR \citep{beuther2016,wang2020} all have angular resolutions between 6.0 and 20\arcsec. The CORNISH and RMS VLA surveys, both at 5.0\, GHz, comparable to the GLOSTAR survey's wavelength, have resolutions of 1.5\arcsec\ and 2\arcsec, respectively.  We used an initial cross match radius of 10\arcsec, the same as chosen in our previous paper \citep{Medina2019}.

\begin{figure}[hbt!]
    \centering
    \includegraphics[width=0.47\textwidth]{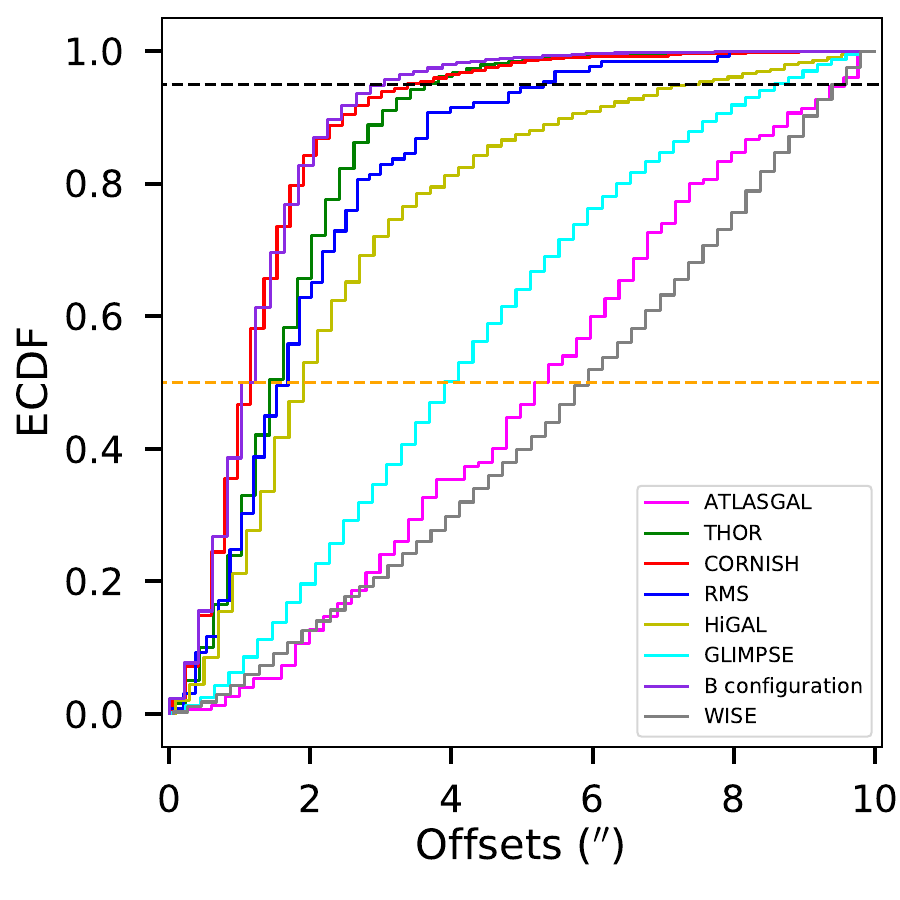} 
    \caption{Empirical cumulative density function of the position offsets among GLOSTAR compact sources ($Y$-factor $\leq$ 2.0) with the other survey counterparts. The dashed lines indicate the 50\% (orange) and 95\% (black) of matches.}
    \label{fig:cumulative_offsets}
\end{figure}

The offset distribution between GLOSTAR VLA D-configuration sources and the above catalogs was analyzed based on the empirical cumulative distribution function (ECDF), see Figure \ref{fig:cumulative_offsets}. The offsets between GLOSTAR and other radio centimeter wavelength surveys (THOR, CORNISH, and RMS) in Fig. \ref{fig:cumulative_offsets} show 
a good position match. This is because they are tracing radio emission with similar morphology. For ATLASGAL, GLIMPSE, and WISE, the situation is different. ATLASGAL $870\,\mu$m submillimeter continuum sources trace compact high column density emission of cold dust (and associated molecular gas) in star-forming regions and are quite rare.
Thus, a radio source the direction of an ATLASGAL source may be related to it even if the angular separation between the ATLASGAL and GLOSTAR positions is larger than 10\arcsec. 
The cases of the WISE and GLIMPSE catalogs are more complex: due to the high density of foreground field stars detected in these survey there is a significant probability of false line of sight matches with the GLOSTAR detections.  As a precaution, a smaller matching radius of $6''$ was used for WISE and GLIMPSE in our final counterpart search. 

We have found \WITHtenarcsec\ radio sources with at least one counterpart at a 10\arcsec\ cross match radius. In Table \ref{tab:surveys_matches}, we summarize the counterparts of the GLOSTAR-VLA sources with the complementary multiwavelength surveys.

{\it Visual inspection.} After our cross match analysis, we find \visualSources\ blobs to be unrelated to extended sources and to have no counterparts in any other survey. To understand the nature of these sources, we first compare the statistics with the expected number of false detections. As discussed in our previous works \citep[e.g.,][]{Medina2019,yang2023} the number of false detections can be calculated using the complementary cumulative 
distribution function  $\Phi(x)=1- \phi(x),$ where $\phi(x)$ is the  cumulative distribution 
function. Assuming that the noise in our radio maps 
follows a Gaussian distribution,
\begin{equation} 
\phi(x)=\frac{1}{2} \bigg [  1 + erf \bigg( \frac{x}{\sqrt{2}}  \bigg) \bigg ]
\end{equation}
where $erf$ is the error function given by 
\begin{equation} 
erf(x)= \frac{1}{\sqrt{\pi}} \int_{-x}^{x} e^{-t^2} dt.
\end{equation}
$\Phi(x)$ is the probability that the value of a  
random variable $x$ with a standard normal distribution will exceed the value $x$. Considering the 100 square degrees of the analyzed radio map, the expected false detection is calculated like $\theta^2\times\phi(x)$, where $\theta$ is the synthesized beam size. Therefore, considering sources above 4$\sigma$, we expect only 127 false detections, which is much smaller than the \visualSources\ sources without counterparts.

Next, we carried out a visual inspection to identify and remove {side lobes} (which appear as elongated blobs in the direction of a very bright source) and {artifacts} (weak sources, between 4--5$\sigma$, and have a size much smaller than the beam size). This led to the removal of \rejectedUnclear\ blobs. 
We also identified \NumUnknown\ sources as {unclear}; these sources have a signal-to-noise ratio (S/N) between 4.0 and 5.0; they have no counterpart and do not comply with the characteristics of a side-lobe or an artifact; however, they are located in a noisy region or close to a very bright source.  Finally, we identified \realsVI\ as {real sources}, and included all detections with a S/N ratio $\,\geq$\,5 that are not elongated (and thus a not part of sidelobes). Table \ref{tab:id_breakdown} summarizes the numbers of detected sources ascribed to the different categories.

\setlength{\tabcolsep}{6pt}
\begin{table*}[!th]
\footnotesize
  \begin{center}
  \caption{Statistics of the matches between GLOSTAR VLA D-configuration and other published surveys. A match radius of 10\arcsec\ and 6\arcsec\ has been used, centered on the peak of the GLOSTAR emission. }   
 \begin{tabular}{lrrrrrrr}\hline\hline
Survey & Ref. &Wavelength & Resolution &\emph{rms} level       & Num. of sources & Num. of  & Num. of \\ 
       &      &      & (\arcsec)  &mJy beam$^{-1}$ & in GLOSTAR region        & matches (10\arcsec ) &matches (6\arcsec )	\\ 

\hline
THOR &(1) &20\,cm            & 25 &0.3-1 &  \textcolor{black}{6\,337}  & \Gthor & \sGthor \\
RMS& (2) &6\,cm              & 1.0-2.0 & 0.22 & \textcolor{black}{737}& \Grms & \sGrms \\
CORNISH & (3)   &6\,cm       & 1.5 & 0.33 & \textcolor{black}{1\,754}& \Gcornish &\sGcornish \\
GLOSTAR B configuration & (4) &6\,cm & 1.0 & 0.08 & \textcolor{black}{5\,437}& \GBconf &\sGBconf \\
ATLASGAL & (5)&870\,$\mu$m & 19.2  &50-70 & \textcolor{black}{3\,096} & \Gatlasgal & \sGatlasgal \\ 
HiGAL &(6) &70-500\,$\mu$m & 5.8-35 & $\cdots$ & \textcolor{black}{35\,344}  & \Ghigal &\sGhigal \\
WISE & (7)&3.4-22\,$\mu$m & 6-12 & $\cdots$ &\textcolor{black}{2\,426\,781} & \Gwise &\sGwise \\
GLIMPSE & (8,9)&3.6-8.0\,$\mu$m & 1.4-2 & $\cdots$ &\textcolor{black}{18\,100\,849}  & \Gglimpse& \sGglimpse\\
\hline
\label{tab:surveys_matches}
\end{tabular}
\end{center}
{References are:  (1) \citet{wang2018}, (2) \citet{urquhart2009}, (3) \citet{hoare2012}, 
(4) \citet{yang2023}, (5) \citet{schuller2009}, (6) \citet{Molinari2016},  (7) \citet{Wright2010},
(8) \citet{churchwell2009} and (9) \citep{fazio2004}.}
\end{table*}

\begin{table}[!t]
\footnotesize
  \caption{Summary of detection categories. \label{tab:id_breakdown}}
 \begin{tabular}{lr}\hline\hline
Description   	   & Number \\ 
			       &  of sources\\ 
 \hline
Total number of  extracted sources& \firstblobs\\
Number of sources assoc. with large structures & \includeLarge\\
Number of sources assoc. with SNR & \includeSNR\\
Number of sources with a counterpart  & \WITHtenarcsec\\
Number of sources with $Y_{\rm factor}>3$ & \YfMthree \\
Number of recovered real sources & \realsVI\\
Number of unclear sources & \NumUnknown \\
Number of artifacts or side lobes\tablefootmark{a}  & \rejectedUnclear  \\
\hline
Number of sources in final catalog  & \FinCatNum \\
\hline\hline
\tablefoottext{a}{Excluded from final catalog.}
\end{tabular}
\end{table}

\section{Final catalog}\label{CatCons}

After removing all artifacts extracted by \blobcat, the final catalog 
consists of \FinCatNum\ entries. From these, 9\,254 represent discrete 
sources; the remaining are part of LSS and SNR candidates. 
In this section, we use the discrete sources and compare them with 
previous surveys in order to quantify the reliability of their properties based on our source extraction process (see Sect. \ref{sect:Source_extraction}).
Additionally, we use the information from the counterparts of previous 
radio and infrared (IR) surveys to classify the sources.

\subsection{Astrometry}

To check the quality of the GLOSTAR source positions, we have compared 
the positions of unresolved (Y$_{\rm factor}<\,1.2$) sources that have counterparts in the radio fundamental catalog of compact radio sources\footnote{This  
catalog is provided via the project webpage at 
\url{https://astrogeo.smce.nasa.gov}. Responsible NASA (National Aeronautics and Space Administration) official: L. Petrov.}.
These radio sources are quasars that were observed with Very Long 
Baseline Interferometry (VLBI) and their celestial positions 
are known with accuracies better than a few milli-arcseconds.  

\begin{figure}
    \centering
    \includegraphics[width=0.48\textwidth, trim= 20 50 0 24,clip, angle=0] {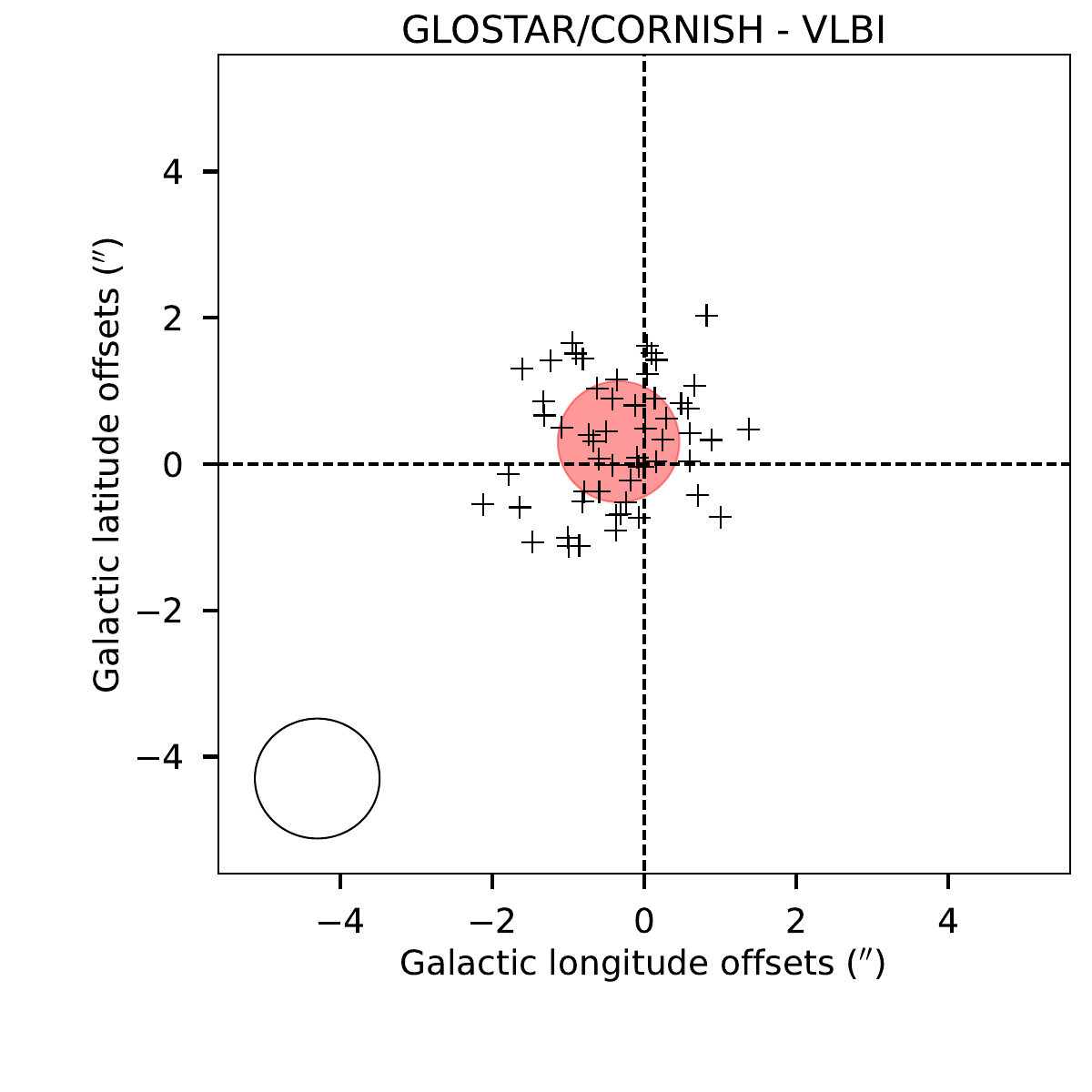}\\ 
    \includegraphics[width=0.48\textwidth, trim= 20 50 0 24,clip, angle=0] {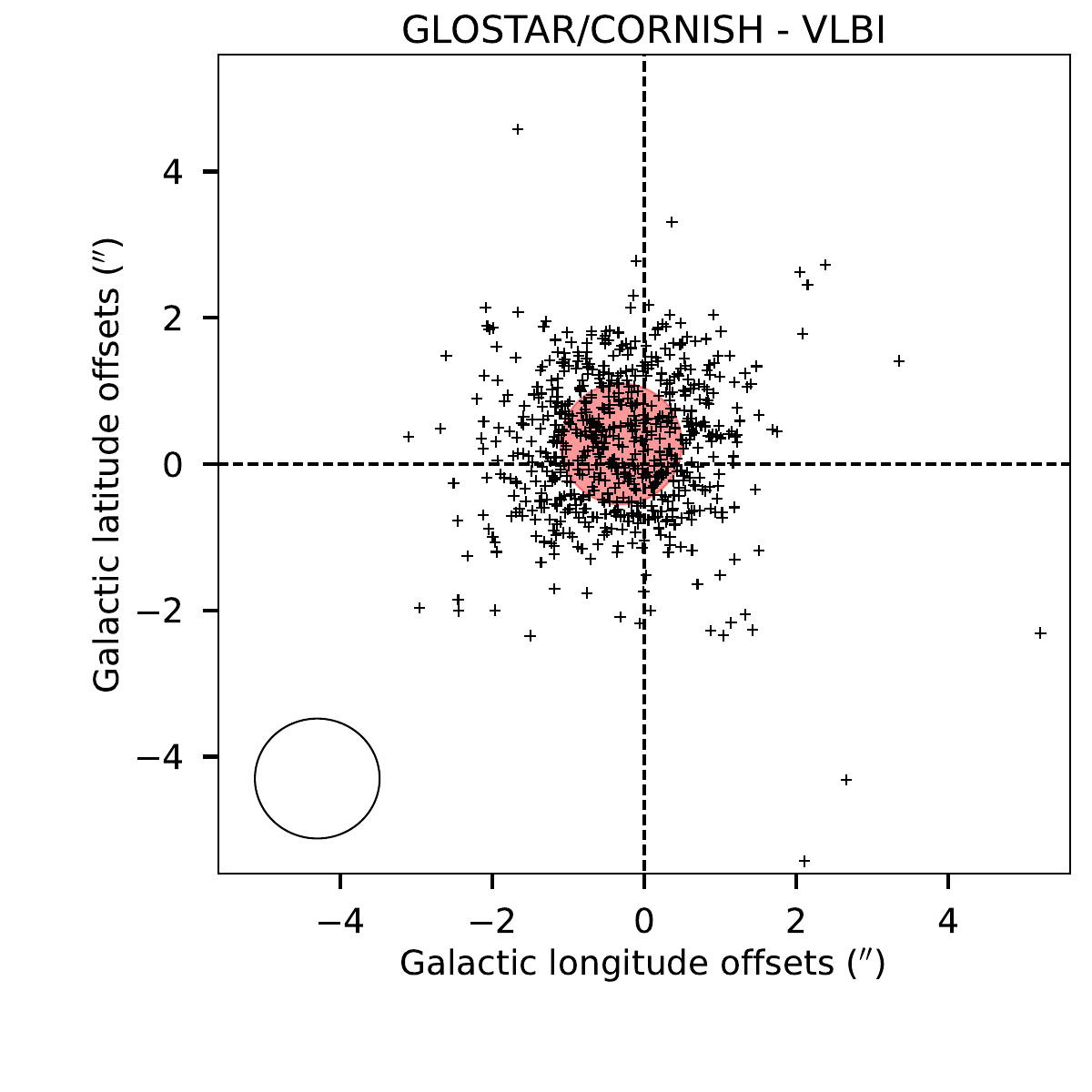}
    \caption{Position offsets between GLOSTAR unresolved sources (Y factor $\leq1.2$) and VLBI extragalactic sources (upper panel), and CORNISH unresolved sources (lower panel). Red circles are centered in the mean values of the offsets, and the radii sizes are equal to the standard deviations. The circles in the bottom left corners of both panels represent the mean position error of GLOSTAR-VLA D-configuration radio sources reported in this work, which is $1\rlap{.}''6$. }
    \label{fig:GaVaC}
\end{figure}

Using a position matching of $6''$ between both catalogs, we found 
55 sources in common. In the upper panel of  Fig.~\ref{fig:GaVaC}, 
we show the $\ell$ and $b$ position offsets between GLOSTAR D-configuration 
and VLBI sources. 
The mean values of the position offsets are $-0\rlap{.}''34\pm0\rlap{.}''10$\footnote{The errors reported on the mean values, here and through the manuscript, are estimated using the standard error of means (${\rm SEM}={\sigma}/{\sqrt{N}},$ where $\sigma$ is the standard deviation and $N$ is the number of elements in the sample).} 
and $+0\rlap{.}''31\pm0\rlap{.}''11$ in Galactic longitude and latitude, respectively. 
The standard deviations of the offsets are $0\rlap{.}''8$ in both directions. 
Thus, the positions of the GLOSTAR catalog presented in this paper are 
accurate to within~$1''$.

A second test can be done by comparing the GLOSTAR sources with the
CORNISH catalog \citep{purcell2013}.  CORNISH used the VLA prior 
to its upgrade to observe at 5.0\,GHz an area similar to that covered 
by us. Most of the sources 
in this catalog are expected to be background extragalactic objects, which is 
also the case observed for GLOSTAR. No detectable proper motions are expected from 
these sources. We have also used a matching radius of $6''$, and we have
restricted our analysis to sources that are unresolved in both catalogs.
In the lower panel of Fig.~\ref{fig:GaVaC}, the measured position offsets 
of 668 sources meeting these criteria are shown. The mean values of the position 
offsets are $-0\rlap{.}''30\pm0\rlap{.}''04$ and $+0\rlap{.}''28\pm0\rlap{.}''04$ in  $\ell$ and $b$, 
respectively. The standard deviation of the offsets 
is $0\rlap{.}''80$ in both Galactic coordinates. 
This result is in agreement with that found for the VLBI sources.

\subsection{Flux density levels}\label{Sec:Snu}

In this section, we check the accuracy of flux density measurements. By using the 668 CORNISH unresolved sources 
detected in the GLOSTAR survey, we can check the quality of 
the measured flux densities in GLOSTAR \citep{Medina2019,dzib2023,yang2023}. This can be done since most of these sources 
are background extragalactic objects whose radio 
emission is expected to have a low degree of variability. 

\begin{figure}
    \centering
    \includegraphics[width=0.48\textwidth, trim= 0 0 0 0, angle=0] {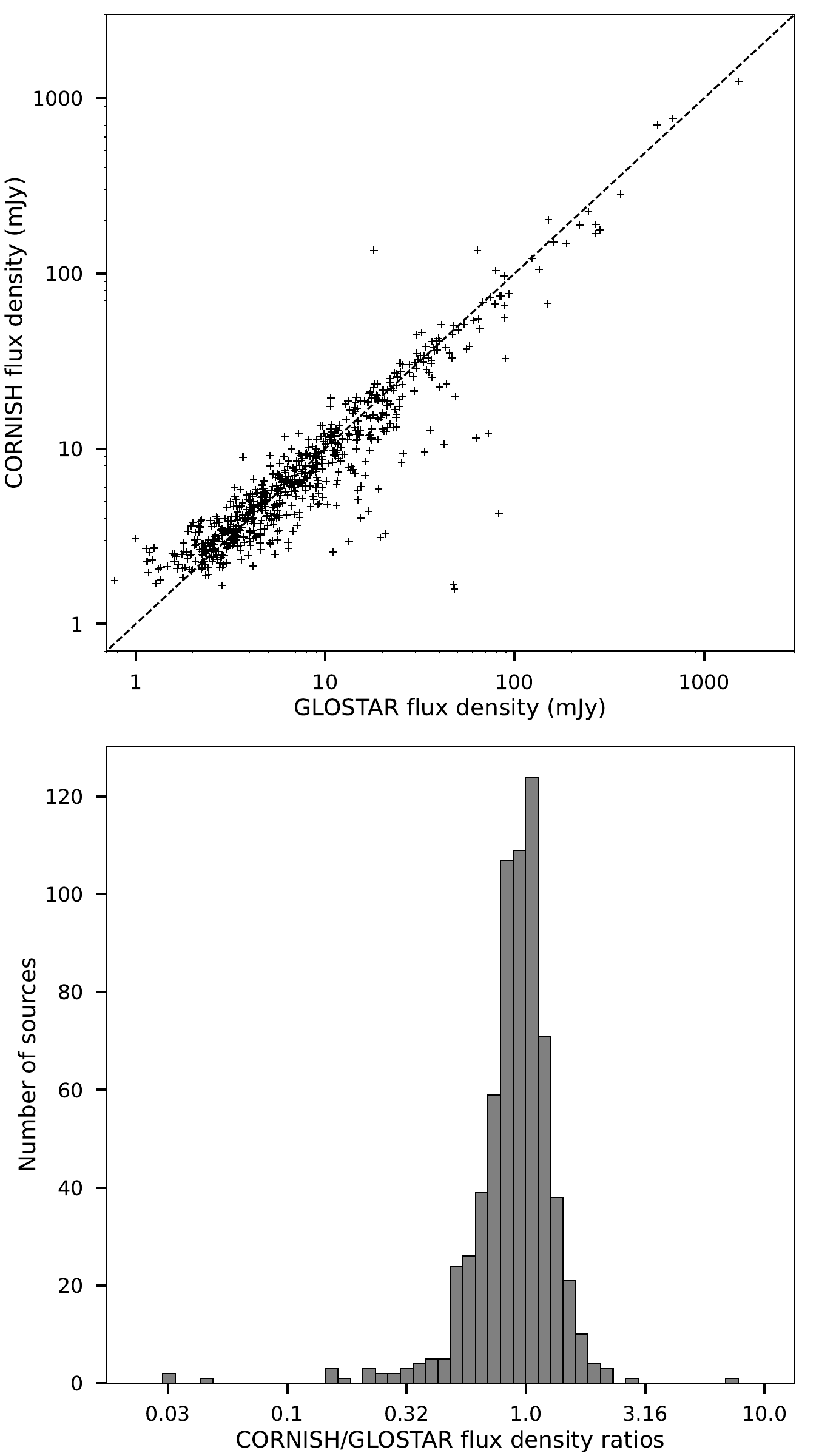}
    \caption{{\it Upper panel:} Comparison of the flux densities of CORNISH unresolved sources that are also detected as unresolved sources in the GLOSTAR survey (this work). The dashed black line is the line of equality. {\it Lower panel:} Flux density ratio distribution of sources plotted in the upper panel. The mean value of the flux density ratios is $1.01\pm0.02$ and the standard deviation is 0.42.}
    \label{fig:FGaC}
\end{figure}

In the upper panel of Fig.~\ref{fig:FGaC}, the flux densities of the CORNISH 
sources are plotted against the GLOSTAR flux densities. For a visual guide, 
the equality line is also shown. This panel shows that the radio
emission from both catalogs is consistent. In the lower panel of
Fig.~\ref{fig:FGaC}, the distribution of the ratios of the flux densities of sources from  
both catalogs are shown, with mean and standard deviation values of 
$1.01\pm0.02$ and 0.42, respectively. Taking a 3$\sigma$ value, it is then 
concluded that the fluxes of the GLOSTAR survey have an accuracy better than 
6\%.

\subsection{Source effective radius}

The BLOBCAT software returns the number of pixels within a source (output 
parameter {\verb|npix|}). This number does not contain any information on the structure,
elongation, or position angle of the source. While source size cannot be recovered from 
\blobcat, following the strategy by \citet{Medina2019} we can determine the effective radius 
of a source. 

We aim to estimate the radius of a circular source that covers the same 
number of pixels as the source extracted using \blobcat. The area of a single pixel  in the 
GLOSTAR D-configuration images is $2\rlap{.}''5\times2\rlap{.}''5$. The area, $A$,
covered by a source is {\verb|npix|} times the pixel size squared, and the effective source radius 
is given by

$$R_{\rm source}=\sqrt{\frac{A}{\pi}}.$$

\noindent The effective radius, as obtained from this procedure, is listed for all sources 
in Table~\ref{tab:glostar_cat}. 

\subsection{Spectral indices}

The flux density of radio sources as a function of the observed frequency can expressed in a power law form: 

\begin{ceqn}
\begin{align}\label{Eq:SI}
    S_\nu\propto\nu^\alpha,
\end{align}
\end{ceqn}

\noindent where $S_\nu$ is the flux density at the observed frequency, $\nu$,
and $\alpha$ is the spectral index. 
The sensitive and wideband GLOSTAR-VLA observations allow an in-band 
spectral index ($\alpha_{\rm GLOSTAR}$) to be determined. We followed the successful strategy used
in the other GLOSTAR-VLA catalogs \citep{Medina2019,dzib2023,yang2023}.
First, to avoid possible spectral index biases, we constrain our spectral index 
determinations to sources with S/N $\geq10$ and $Y_{\rm factor}\leq2.0$; that is, compact 
sources. A lower limit for S/N was chosen to ensure that the source could be detected in most of the seven frequency bin images. For instance, as the noise level per frequency bin image is expected to be $\sqrt{7}\times$ larger than that of the combined image, for sources with S/N=10 we expect detection of the order of $10/\sqrt{7}=3.8\sigma_{\rm bin}$ in each frequency bin image. This rough value ensures detections in most frequency bin images. For extended sources, as the $(u,\,v)$-coverage is different across the band,  the fraction of 
total flux density recovered at the lower sideband can greatly exceed that at 
the upper sideband. In these cases, the spectral index derived will have apparently 
steeper (more negative) spectra to values that can be physically impossible. 
Second, source flux densities are measured on the imaged frequency bins (see Sect. 2). 
We use the logarithmic form of Equation~\ref{Eq:SI}, which is a linear 
equation, and the measured flux densities in the individual frequency bin images
to perform a least-squares fitting and determine the spectral index.
Following these steps, we have estimated the spectral indices of 3\,968
 sources. 

In \citet{dzib2023} and \cite{yang2023}, we have compared the spectral indices 
determined from GLOSTAR data with those determined by the THOR survey \citep{wang2018} for compact common sources.
We have found a good consistency between the spectral index determination by both surveys. 
The THOR survey covered the full L-band (1.0--2.0 GHz) with the VLA, and they also  
 split their observed bandwidth into smaller frequency chunks to 
estimate the in-band spectral index \citep[see][for details]{wang2020}. Though 
THOR observed at lower frequencies
frequency than GLOSTAR, the resulting angular 
resolution of their images is similar to that of the  GLOSTAR-VLA D-configuration images ($18''$ for GLOSTAR 
and $25''$ for THOR).  
Thus, following \citet{Medina2019}, we now proceed to determine the source spectral indices
from the peak flux densities measured by GLOSTAR and THOR ($\alpha_{\rm GLOSTAR-THOR}$). We consider mid-frequencies 
of 1.5 GHz and 5.8 GHz for THOR and GLOSTAR, respectively. As only two flux density
values are considered, the spectral index can be determined as

\[
\alpha_{\rm GLOSTAR-THOR}=\frac{{\rm ln}\left(S_{\rm GLOSTAR}/S_{\rm THOR} \right)}{{\rm ln}{\rm \left(5.8/1.5 \right)}},
\]

\noindent where $S_{\rm THOR}$ and $S_{\rm GLOSTAR}$ are the peak flux densities 
from the respective survey. The spectral index uncertainty is calculated using:

\[
\sigma_{\alpha_{\rm GLOSTAR-THOR}} = \frac{\sqrt{\left(\sigma_{S_{\rm THOR}}/S_{\rm THOR}\right)^2 + \left(\sigma_{S_{\rm GLOSTAR}}/S_{\rm GLOSTAR}\right)^2}}{{{\rm ln}\left(5.8/1.5 \right)}},
\]

\noindent which is obtained following standard error propagation theory. 
$\sigma_{S_{\rm GLOSTAR}}$ and $\sigma_{S_{\rm THOR}}$ are the source flux density 
uncertainties for GLOSTAR and THOR.  In the case of the THOR survey, their catalog 
does not contain direct values for flux density uncertainties.
However, since their S/N ratio is given, the THOR flux density uncertainty can be  estimated as 

\[
\sigma_{S_{\rm THOR}}= \frac{S_{\rm THOR}}{\rm S/N}.
\]

\noindent With this procedure, we estimate $\alpha_{\rm GLOSTAR-THOR}$ for 4\,127 radio sources, of which 1\,308 do not have an estimated value of $\alpha_{\rm GLOSTAR}$. 

The total number of radio sources with an estimated radio spectral index, either the GLOSTAR-inband or GLOSTAR-THOR, is 5\,276. Results of both $\alpha_{\rm GLOSTAR}$ and $\alpha_{\rm GLOSTAR-THOR}$ on individual sources are listed in Table~\ref{tab:glostar_cat} and they will be compared in Sect. 5.3.

\subsection{Counterparts in other surveys}

During the last few decades, Galactic Plane surveys have addressed many aspects of high-mass star ($>8\,$M$_\odot$) formation.
As these surveys image large areas of the Galaxy, they provide unbiased samples
of star-forming regions with different properties \citep[see, for example][]{urquhart2018_csc,urquhart2022,elia2017}.

As we have seen in previous sections, information from other radio surveys,
such as CORNISH and THOR, can corroborate and complement the source parameters
from our catalog. Additionally, information at shorter wavelengths
can be used to give insights into the nature of the observed radio sources. 
In the context of GLOSTAR, particularly interesting radio sources are those related to star formation. 
CORNISH and THOR were briefly described in previous sections and in the following
we describe the other Galactic plane surveys used to characterize the
GLOSTAR radio sources. 

Recently, we have presented the GLOSTAR VLA B-configuration catalog of the Galactic plane 
using higher angular resolution (1\arcsec) images \citep{yang2023}. The imaged field covers the area 
$|b| \leq 1.0^\circ$, and the Galactic longitude ranges $2^\circ< \ell < 40^\circ$
and $56^\circ< \ell < 60^\circ$; that is, 32 square degrees less than the GLOSTAR VLA D-configuration.
About 5\,500 radio sources were reported and classified. Using a crossmatch radius of 10\arcsec, 
 we found a match for 2\,497 compact radio sources. 

The RMS survey was a multiwavelength project aiming to identify massive 
young stellar objects (MYSOs) in the Galactic plane \citep{hoare2005, lumsden2013}. Using a multiwavelength
classification scheme, RMS identifies the nature of radio sources; particularly within the 
northern hemisphere they identified 79 PNe and 391 \hii\ regions with radio emission
\citep{urquhart2009}. We retrieved 150 of the RMS sources within a search 
radius of 10\arcsec.

Emission at sub-millimeter wavelengths is dominated by dense cool dust and gas, 
which is intimately related to star formation. At these wavelengths, ATLASGAL \citep{schuller2009} 
is the first high-resolution ($\approx 20''$ FWHM) ground-based submillimeter
(870\,$\mu$m) survey of the thermal dust emission in the entire inner Galactic plane. 
The ATLASGAL survey has presented $>10\,000$ dense clumps in the Galactic plane 
\citep{contreras2013,csengeri2014,urquhart2014c}. Correlating radio sources (such
as those detected in GLOSTAR) with ATLASGAL clumps is an excellent way to identify 
embedded or dust-enshrouded objects such as \uchii\ regions \citep[see, for example,][]{Medina2019,irabor2018,urquhart_radio_south,urquhart2009,urquhart2013,purcell2013}.
Correlating our radio source catalog with the ATLASGAL compact source
catalog \citep{urquhart2014c}, we find cross match 281 sources within $10''$. 

HiGAL is a photometric survey in five far-infrared (FIR) 
bands between 70 and 500\,$\mu$m \citep{Molinari2016}. Its observations covered the whole Galactic 
plane with a varying latitude range \citep{elia2021}. The FWHM beam sizes range 
from $6''$ to $35''$, and the mean position uncertainty is 1\rlap{.}$''$2 \citep{Molinari2016}. Using a 
cross matching radius of $10''$ we found 1527 sources in both our GLOSTAR and the HiGAL catalogs.

WISE was a NASA IR-wavelength 
astronomical space telescope mission \citep{Wright2010} that mapped the entire 
sky in four MIR bands W1, W2, W3, and W4  centered  at  3.4,  4.6,  12,  
and  22\,$\mu$m, respectively, using a 40\,cm telescope feeding an array with a total of 4 million 
pixels; these wavelengths correspond to angular resolutions of 6\rlap{.}$''$1, 6\rlap{.}$''$4, 6\rlap{.}$''$5, and 12\arcsec. The WISE All-sky release source catalog contains the source properties of $\sim563$ million sources. The WISE All-sky release source catalog has been filtered out for sources with S/N$<5$, spurious detections, and image artifacts\footnote{Detailed information on WISE data processing and source catalogs (including rejected detections) can be found in \url{https://wise2.ipac.caltech.edu/docs/release/allsky/}.}. 
Using a radius of $10''$ for cross matching GLOSTAR radio sources and sources from the WISE All-sky release source catalog, we found 7\,298 common sources. However, 
as was mentioned earlier, because of the high density of WISE detected sources, mostly foreground 
field stars, there is a significant probability of false matches and we have reduced 
the cross matching criteria to 6\arcsec\ and found 3\,034 matching sources. 

Finally, we have also used the GLIMPSE point source catalog\footnote{The GLIMPSE point Source Catalog (GLMC) is the most reliable of GLIMPSE catalogs with a reliability $\geq95\%$ \citep{churchwell2009}.  The GLMC can be found at the following sites:
\url{http://ssc.spitzer.caltech.edu/legacy/glimpsehistory.html} and \url{http://irsa.ipac.caltech.edu/data/SPITZER/GLIMPSE}.} \citep{churchwell2009}. This legacy project was conducted 
with the Spitzer space telescope and observed the shorter wavelength part of the MIR range; that is, at 
3.6, 4.5, 5.8, 8.0 $\mu$m. GLIMPSE is composed of several surveys, and particularly interesting 
for GLOSTAR are GLIMPSEI and GLIMPSEII which, combined, effectively observed 
the Galactic plane from $-70^{\circ} < \ell < 65^{\circ}$ and  $|b|<1^{\circ}$.
Using a cross matching criterion of 10\arcsec, we found 7\,326 GLIMPSE counterparts to radio sources.
However, similar to the WISE catalog, a large number of these are expected to be foreground objects. Thus, we reduced the cross match 
radius to 6\arcsec\ for the final counterpart search and found 5\,525 counterparts.

\subsection{Source classification}\label{Sec:SClass}

The present catalog is composed of \FinCatNum\ radio sources, and thus one-by-one source classification, as performed in our previous work
\citep[i.e.,][]{Medina2019},
is a time-consuming task. Furthermore, in our previous catalog work, we have
shown that the classification of radio sources detected within the 
GLOSTAR-VLA project is largely ($>90\%$) compatible with the classifications 
made by  other radio surveys \citep{Medina2019,dzib2023,yang2023}. 
Notably, in the GLOSTAR-VLA pilot region, the source classifications 
in the high-resolution and low-resolution maps are also highly consistent 
\citep{dzib2023}. Thus, our source classification approach for the present 
catalog is slightly different and is presented below. We focus the source 
classification analysis on the \FinDiscNum\ sources that are not related to LSSs (\includeLarge\ sources) 
or SNRs (\includeSNR\ sources) or Unclear (\NumUnknown). 

\subsubsection{Source classification from other surveys}

First, we identify sources with counterparts in other catalogs and use their 
classification. We prioritize the catalogs in the following 
order: GLOSTAR-VLA B-configuration \citep[][]{yang2023}, 
CORNISH \citep[][]{purcell2013}, RMS \citep[][]{urquhart2009},
WISE \hii\ regions \citep[][]{anderson2014}, and THOR \citep{wang2020}. 
We also check the GLOSTAR-VLA 6.7\,GHz methanol maser catalog \citep[][]{Nguyen2022}, as this maser line is only detected toward MYSOs.

The above catalogs use different names for different classes of sources. To homogenize them,
we have used the classifications chosen in previous GLOSTAR-VLA catalogs.
These classifications are {\it HII} for \hii\ regions and \hii\ region candidates,
{\it PN} for planetary nebulae and planetary nebula candidates, {\it EgC} for background 
extragalactic candidate sources, {\it Psr} for pulsars, {\it Radio-star} for stars with radio emission,
and {\it Unclassified} for sources that could not be classified within our classification scheme. 
The CORNISH survey used the terms UCHII and HII-Region, which, in our final classification, 
are renamed as HII. Sources CORNISH classified as Radio Galaxy (Central Source), 
Radio Galaxy (lobe), Galaxy, and IR-Quiet are 
renamed as EgC. Sources classified as \hii\ region and HII/YSO in the RMS survey are categorized as HII, while YSO and Evolved star classifications are renamed as Radio-star in our catalog.
Regardless of the \hii\ region sub-classification from \citep{anderson2017},
we rename these sources as HII. From the THOR survey, we discarded the classification X-ray (which only accounts
for a counterpart at another wavelength) and renamed their {\it jet} classification 
to EgC \citep[see also][]{wang2018}. Sources associated with methanol masers are labeled 
as HII, as they represent objects in the very early phases of high-mass star formation. While most class II methanol masers are not associated with compact radio emission, some of them are: in the GLOSTAR D array data, \citet{Nguyen2022} find that toward 12\%\ of the 554 detected methanol masers also radio continuum is detected, while 97\% are associated with dust emission. 

The number of classified radio sources from other surveys is 3\,302. The numbers of adopted classifications are 
2\,286 from GLOSTAR-VLA B-configuration, 537 from CORNISH, 57 from RMS,
167 from WISE \hii\ regions, 249 from THOR, and six sources from the 6.7\,GHz methanol 
maser catalog. Classifications and references to the classification 
origins are given in Table~\ref{tab:glostar_cat}. 
We note that a large 
fraction of GLOSTAR radio sources could not be classified by only using the information from other surveys.

\subsubsection{Source classification from information at shorter wavelengths}

After the cross match classification, there are still 5\,925 unclassified 
sources. To classify them, we use the information obtained in surveys conducted  at
other than radio wavelengths. We follow similar criteria as used in the previous 
GLOSTAR-VLA catalogs. 

First, compact (Y$_{\rm factor}\leq2.0$) radio sources that do not have counterparts 
at submm,  FIR, and MIR wavelengths are likely to be background extragalactic sources 
\citep{hoare2012,lucas2008,marleau2008,Medina2019}, and thus they are classified as EgC.
By matching these criteria,  we classified 1\,569 radio sources as EgC.

The {\it HII} region candidate classification is assigned to radio sources that have 
a counterpart at submm wavelengths or that have a counterpart only 
at FIR wavelengths \citep{hoare2012,anderson2012, urquhart2013, yang2021}.  There are
32 radio sources with counterparts in the ATLASGAL submm survey,
and 82 sources with a counterpart only in the HiGAL FIR survey. All of these 114
sources are classified as \hii\ region candidates.

If radio sources have counterparts only at FIR and MIR wavelengths, additional criteria 
are needed to classify them. Extended sources can, for example, represent photo dissociated regions (PDRs), 
parts of SNRs, or parts of extended \hii\ regions and complicate the classification 
procedure. While these may include interesting cases, for the purpose of this paper we concentrate on 
compact radio sources with the signatures of early phases of star formation. Thus, in the following, we focus on the classification of compact
radio sources (Y$_{\rm factor}\leq2.0$). The 1\,463 sources with Y$_{\rm factor}>2.0$ that have not been previously classified, are classified as 
{\textit{Unclassified}} in our final catalog.

\begin{figure}
\centering
\includegraphics[width=0.46\textwidth, trim= 0 0 0 0, angle=0]{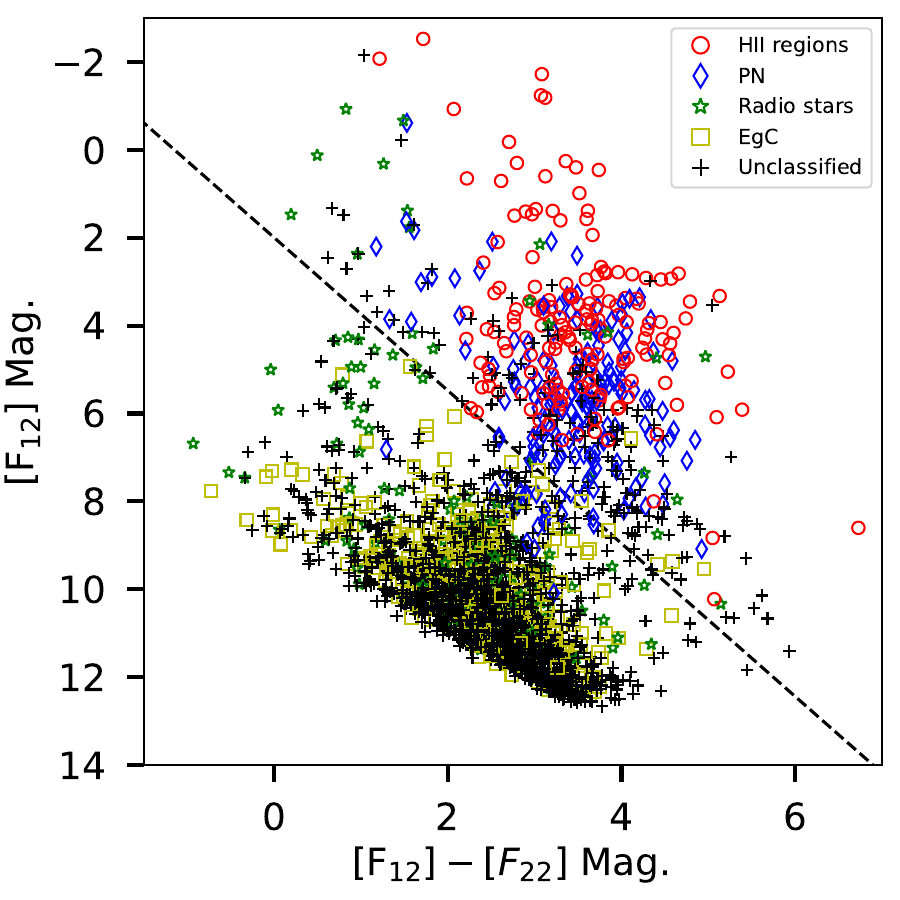} 
\caption{Mid-infrared color-magnitude diagram of GLOSTAR radio sources with a WISE counterpart.
The dashed line indicates the constraint used by \citet{Medina2019} to distinguish between different types of objects.}
\label{fig:MCC}
\end{figure} 

The 2\,779 remaining unclassified compact radio sources have counterparts at FIR or MIR wavelengths.
\citet{Medina2019}  constructed [12] vs. [12]--[22]  color-magnitude diagram using mid IR magnitudes from the WISE catalog 
and showed that in the resulting diagram the \hii\ regions and PNe occupy a
different area than radio stars and other unclassified objects 
\citep[which have more likely an extragalactic origin, see discussion by][]{Medina2019}. 
As is discussed by \citet{Medina2019}, \hii\ regions and PNe fulfill the constraint (hereafter 
the color-magnitude constraint):

$$[12] {\rm Mag.} < 1.74\times ([12]-[22])+2,$$

\noindent while Radio-stars and EgC do not. 

To illustrate the situation for the GLOSTAR radio sources discussed in this work, we use the flux 
information from WISE. A total of 2\,401 GLOSTAR radio sources have counterparts detected in the 12 and 22 $\mu$m WISE bands.
These include 
178 \hii\ regions, 175 PNe, 452 EgC, 159 radio-stars, and 1\,437  
unclassified sources. In Fig.~\ref{fig:MCC}, 
we plot the color-magnitude diagram [12] vs. [12]--[22] of these sources, where the source 
types are distinguished with different markers and colors. The line that follows 
the color-magnitude constraint is also plotted as a dashed line. Fig.~\ref{fig:MCC} clearly 
shows that \hii\ regions and PNe occupy different areas in this diagram than 
the radio stars and EgC objects. 

From the already classified sources, we notice that 360 lie above the color-magnitude constraint line,
and these are divided into 178 (50\%) \hii\ regions, 152 (42\%) PNe, 7 (2\%) EgC,
and 23 (6\%) radio stars. Thus, we can assume that sources that fall above the color-magnitude constraint
are either \hii\ regions or PNe. There are 153 unclassified sources fulfilling these criteria, and
in this work we classify them as \hii\ region candidates, labeled as HII in the final catalog;  we notice that up to 50\% will likely be PNe; however, given the current analysis, it 
is not possible to definitively characterize them.

On the other hand, 604 classified sources 
do not comply with the color-magnitude constraint
and these are divided into 23 (4\%) PNe, 445 (74\%) EgC, and 136 (22\%) radio stars.
We notice that EgC sources constitute the majority in this region of the color-magnitude diagram. Following these counts, we classify the 1\,284 previously unclassified sources in this area as EgC, noting that
about 25\%  of them may likely be galactic sources; however, it is not possible to differentiate them with the current data. 

The IR information of the remaining sources is scarcer than for the previous cases, and 
MIR colors cannot be obtained as fluxes are only reported in a single band. We noticed that 86 of these sources have at least 
a counterpart in one of the FIR bands. As FIR emission may trace cold dust, radio sources associated 
with FIR emission might be related to \hii\ regions, and accordingly we classify them as \hii\ region candidates (labeled as HII in the final catalog).
However, further studies are required to confirm the nature of these sources.
The 1\,256 remaining sources have counterparts only in the  MIR range. These can be either EgC or 
radio stars; however, our previous work \citep{Medina2019,dzib2023,yang2023} has shown that
most sources with only MIR and near-infrared counterparts are background extragalactic sources. 
Accordingly, these 1256 sources are classified as EgC. A summary of the classified sources
per method is given in Table~\ref{tab:Class}. Class and classification methods are given in the
catalog and Table~\ref{tab:glostar_cat}. 

\begin{table}[!h]
\footnotesize
  \caption{Summary of classification categories. \label{tab:Class}}
 \begin{tabular}{llr}\hline\hline
        &	         & Number \\ 
Class	&	Method   &  of sources\\ 
 \hline
        & Other catalog & 410\\
        & Methanol maser & 6 \\ 
        & SMM counterpart& 32\\
HII        & Only FIR counterpart &82\\
        & Color-magnitude constrain &153\\
        & FIR counterpart, no color with MIR & 86\\
\cline{2-3}
        &   Total    & 769 \\
        \hline
PNe        & Previously known source & 240 \\
    \hline
Radio-star & Previously known source & 425 \\  
    \hline
PSR & Previously known source & 6 \\
    \hline
PDR & Previously known source & 1 \\    
\hline
& Previously known source & 2203 \\
    & No SMM or IR counterpart & 1569 \\
EgC    & Color-magnitude constrain & 1284 \\
    & Only MIR counterparts  & 1256 \\ \cline{2-3}
     & Total & 6312 \\ 
\hline
         & Previously known source & 19 \\
Unclassified    & Y$_{\rm factor}>2$ \& no previous class & 1463 \\\cline{2-3}
         & Total & 1482 \\
        \hline
\hline
\end{tabular}
\end{table}

\newpage
\setlength{\tabcolsep}{3.5pt}
\begin{sidewaystable*}[p]
\centering
\scriptsize
\renewcommand{\arraystretch}{1.1}
\caption{GLOSTAR source catalog for the observed region. }
\label{tab:glostar_cat}
\begin{minipage}{\linewidth}
\begin{tabular}{lccccccccccccccccccccccccc}

\hline \hline

\multicolumn{1}{c}{GLOSTAR name}&  \multicolumn{1}{c}{$\ell$}&\multicolumn{1}{c}{$b$}&	\multicolumn{1}{c}{S/N}  &\multicolumn{1}{c}{$S_{\rm peak}$}  &  \multicolumn{1}{c}{$\sigma_{S_{\rm peak}}$}  &\multicolumn{1}{c}{$S_{\rm int}$} &\multicolumn{1}{c}{$\sigma_{S_{\rm int}}$} & \multicolumn{1}{c}{Y$_{\rm factor}$} & \multicolumn{1}{c}{Radius} & \multicolumn{4}{c}{Spectral index}  & \multicolumn{5}{c}{Radio} & submm &\multicolumn{3}{c}{IR} \\ \cline{15-19} \cline{21-23}

\multicolumn{1}{c}{}&  \multicolumn{1}{c}{(\degr)}&\multicolumn{1}{c}{(\degr)}&	\multicolumn{1}{c}{}& \multicolumn{2}{c}{(mJy\,beam$^{-1}$)}  & \multicolumn{2}{c}{(mJy)}  & \multicolumn{1}{c}{} & \multicolumn{1}{c}{(\arcsec)} &  \multicolumn{1}{c}{$\alpha$} & \multicolumn{1}{c}{$\Delta \alpha$} &  \multicolumn{1}{c}{$\alpha$*} & \multicolumn{1}{c}{$\Delta \alpha$*} & GB& C & R &T &MM & A&H& W &G &Class& \multicolumn{1}{c}{Ref.}& Method\\

\multicolumn{1}{c}{(1)}&  \multicolumn{1}{c}{(2)}&\multicolumn{1}{c}{(3)}&	\multicolumn{1}{c}{(4)}& \multicolumn{1}{c}{(5)} & \multicolumn{1}{c}{(6)} & \multicolumn{1}{c}{(7)}  & \multicolumn{1}{c}{(8)}  & \multicolumn{1}{c}{(9)} & \multicolumn{1}{c}{(10)} &  \multicolumn{1}{c}{(11)} & \multicolumn{1}{c}{(12)} & \multicolumn{1}{c}{(13)} & \multicolumn{1}{c}{(14)} & \multicolumn{1}{c}{(15)} & \multicolumn{1}{c}{(16)} & (17)&(18)&(19)&(20)&(21)&(22)&(23)&(24)&(25)&(26)\\
\hline
G002.005+00.348&02.0047&+0.3479&09.2&1.29&0.16&1.34&0.16& 1.0&13.10&...&...&...&...& & & & & & & & & &EgC&1&No SMM or IR counterpart\\
G002.007--00.681&02.0072&-0.6806&170.3&26.07&1.41&121.27&6.07& 4.7&52.00&...&...&...&...& & & & & & &Y& & &HII&1&Only FIR counterpart\\
G002.011+00.108&02.0111&+0.1076&05.4&0.75&0.14&1.10&0.15& 1.5&12.50&...&...&...&...& & & & & & &Y& & &EgC&1&No SMM or IR counterpart\\
G002.012+00.744&02.0120&+0.7438&178.3&15.20&0.82&13.67&0.69& 0.9&21.10&...&...&-0.16&0.05&Y& & & & & &Y&Y& &PN&2&Previously known source\\
G002.013+00.896&02.0125&+0.8959&05.3&0.46&0.09&0.57&0.09& 1.2&11.10&...&...&...&...& & & & & & & &Y& &EgC&1&Color-magnitude constrain\\
G002.018+00.669&02.0177&+0.6688&32.2&3.12&0.19&3.52&0.20& 1.1&20.20&...&...&-0.33&0.25&Y& & & & & & &Y& &Other&2&Previously known source\\
G002.021--00.824&02.0210&-0.8243&19.0&1.92&0.14&1.65&0.13& 0.9&14.40&...&...&0.07&0.18& & & & & & &Y&Y& &HII&1&Color-magnitude constrain\\
.&.&.&.&.&.&.&.&.&.&.&.&.&.&.&.&.&.&.&.&.&.&.&.&.&.\\
.&.&.&.&.&.&.&.&.&.&.&.&.&.&.&.&.&.&.&.&.&.&.&.&.&.\\
.&.&.&.&.&.&.&.&.&.&.&.&.&.&.&.&.&.&.&.&.&.&.&.&.&.\\
G021.103--00.976&21.1032&-0.9757&15.3&2.89&0.24&2.67&0.23& 0.9&14.20&0.26&0.15&0.04&0.24&Y& & &Y& & & & &Y&Radio-star&2&Previously known source\\
G021.103+00.664&21.1034&+0.6639&07.4&0.60&0.09&0.51&0.09& 0.8&10.60&...&...&...&...&Y& & & & & & &Y&Y&EgC&2&Previously known source\\
G021.114+00.872&21.1137&+0.8723&05.1&0.43&0.09&0.44&0.09& 1.0&9.70&-0.96&0.20&...&...&Y& & &Y& & & &Y&Y&Radio-star&2&Previously known source\\
G021.121--00.956&21.1206&-0.9563&05.6&0.91&0.17&1.29&0.17& 1.4&12.10&...&...&...&...&Y& & & & & & & &Y&EgC&2&Previously known source\\
G021.126--00.294&21.1258&-0.2944&17.2&5.42&0.43&204.65&10.24&37.7&106.60&0.11&0.12&...&...& & & &Y& & & & & &LSS&1&Visual Inspection\\
G021.130--00.247&21.1300&-0.2465&05.4&1.63&0.31&17.96&0.95&11.0&31.80&...&...&...&...& & & & & & & & & &LSS&1&Visual Inspection\\
G021.145+00.585&21.1445&+0.5854&05.9&0.49&0.09&0.68&0.09& 1.4&12.60&...&...&...&...& & & & & & & & & &EgC&1&No SMM or IR counterpart\\
.&.&.&.&.&.&.&.&.&.&.&.&.&.&.&.&.&.&.&.&.&.&.&.&.&.\\
.&.&.&.&.&.&.&.&.&.&.&.&.&.&.&.&.&.&.&.&.&.&.&.&.&.\\
.&.&.&.&.&.&.&.&.&.&.&.&.&.&.&.&.&.&.&.&.&.&.&.&.&.\\
G047.145--00.166&47.1451&-0.1660&04.1&0.25&0.06&0.23&0.06& 0.9&7.70&...&...&...&...& & & & & & & & &Y&EgC&1&Only MIR counterparts\\
G047.147+00.730&47.1471&+0.7299&20.1&1.36&0.10&11.02&0.56& 8.1&46.90&...&...&...&...& & & & & & & & & &SNR&7&Previously known SNR\\
G047.147--00.113&47.1472&-0.1132&05.1&0.31&0.06&0.28&0.06& 0.9&9.00&...&...&...&...& & & & & & & & &Y&EgC&1&Only MIR counterparts\\
G047.154+00.890&47.1540&+0.8903&24.6&1.52&0.10&1.72&0.11& 1.1&18.90&-0.22&0.07&-0.64&0.30& & & &Y& & & & &Y&EgC&1&Only MIR counterparts\\
G047.162--00.513&47.1617&-0.5132&04.5&0.28&0.06&0.21&0.06& 0.8&7.50&-0.55&0.22&...&...& & & &Y& & & & & &EgC&1&No SMM or IR counterpart\\
G047.171--00.621&47.1708&-0.6208&04.0&0.24&0.06&0.16&0.06& 0.6&6.30&...&...&...&...& & & & & & & &Y&Y&EgC&1&Color-magnitude constrain\\
G047.172--00.481&47.1722&-0.4806&08.5&0.52&0.07&0.39&0.06& 0.8&10.60&-0.52&0.13&...&...& & & &Y& & & & &Y&EgC&1&Only MIR counterparts\\
.&.&.&.&.&.&.&.&.&.&.&.&.&.&.&.&.&.&.&.&.&.&.&.&.&.\\
.&.&.&.&.&.&.&.&.&.&.&.&.&.&.&.&.&.&.&.&.&.&.&.&.&.\\
.&.&.&.&.&.&.&.&.&.&.&.&.&.&.&.&.&.&.&.&.&.&.&.&.&.\\
G059.932+00.837&59.9320&+0.8368&55.7&5.88&0.33&7.02&0.37& 1.2&21.60&-0.47&0.04&-0.66&0.11& & & &Y& & & & & &EgC&1&No SMM or IR counterpart\\
G059.937--00.353&59.9365&-0.3528&05.2&0.88&0.18&0.93&0.18& 1.1&10.00&-0.43&0.16&...&...& & & &Y& & & &Y&Y&SNR&7&Previously known SNR\\
G059.940--00.157&59.9399&-0.1569&39.4&6.92&0.41&7.06&0.39& 1.0&18.60&-0.03&0.05&0.36&0.12& &Y& &Y& & & & & &EgC&3&Previously known source\\
G059.941--00.042&59.9413&-0.0417&28.6&4.86&0.31&5.58&0.33& 1.1&18.80&-0.52&0.05&-0.88&0.23& & & &Y& & & & & &EgC&1&No SMM or IR counterpart\\
G059.948+00.555&59.9478&+0.5549&14.6&1.83&0.16&1.67&0.15& 0.9&14.00&-0.13&0.09&-0.28&0.28& & & &Y& & & &Y&Y&EgC&1&Color-magnitude constrain\\
G059.963+00.036&59.9629&+0.0361&04.6&0.94&0.21&1.26&0.21& 1.3&10.20&...&...&...&...& & & & & & & &Y&Y&EgC&1&Color-magnitude constrain\\
G059.968+00.660&59.9680&+0.6597&07.1&1.17&0.18&0.89&0.17& 0.8&11.00&-0.04&0.14&...&...& & & &Y& & & &Y&Y&EgC&1&Color-magnitude constrain\\
\hline
\end{tabular}\\
Notes: The columns are, from left to right, the source name, Galactic coordinates, the S/N, the peak and integrated flux densities and their errors, the Y-factor, the effective radius, the estimated spectral indices determined from the GLOSTAR and THOR flux densities and from only the in-band GLOSTAR sub-images (*). Columns 15 to 23 list if the source has a counterpart in other surveys. Radio: GB for GLOSTAR B-configuration catalog \citep{yang2023}, C for CORNISH \citep{hoare2012,purcell2013}, R for RMS \citep{urquhart2009}, T for THOR \citep{beuther2016,wang2020}, and MM for methanol masers surveys \citep{Nguyen2022,green2009}. 
Submillimeter/FIR: A for ATLASGAL \citep{urquhart2018_csc}. Infrared: H for HiGAL \citep{Molinari2016}, W for WISE \citep{Wright2010}, and G for GLIMPSE \citep{churchwell2009}. Column 24 lists the final classification of the radio sources. 
References in column 25 are as follows: (1) This work, (2) \citet{yang2023}, (3) 
\citet{purcell2013}, (4) \citet{urquhart2009}, (5) \citet{anderson2014}, 
(6) \citet{wang2020}, (7) SNR remnants catalogs by \citet{Dokara2021} or \citet{green2019}, 
and (8) Methanol maser catalogs by \citet{Nguyen2022} or \citet{green2009}. Finally, column 27 lists the method used for the source classification.

Only a small portion of the data is provided here, and the full table is available in electronic form at the CDS via anonymous ftp to cdsarc.u-strasbg.fr (130.79.125.5) or via \url{http://cdsweb.u-strasbg.fr/cgi-bin/qcat?J/A\&A/} and through the dedicated GLOSTAR webpage at \url{https://glostar.mpifr-bonn.mpg.de/glostar/}.  \\ 
\end{minipage}
\end{sidewaystable*}
\setlength{\tabcolsep}{6pt}

\section{Summary of full catalog}\label{CatSumm}

The final catalog consists of \FinCatNum\ entries, \includeLSS\ of which are related to
extended structures (for example, SNRs or extended \hii\ regions) and \NumUnknown\ are sources of uncertain nature
that could not be identified as real or spurious sources. Source classification 
was possible for most of the remaining sources, but 1468 extended sources 
(Y$_{\rm factor}>2$) were classified as {\textit{Unclassified}} as their extended structures require a deeper analysis to classify them, which goes beyond the scope
of the present work.  In the following,  we discuss the remaining classified 
sources. This discussion will focus on the spatial distribution,
flux densities, sizes, and spectral indices of the sources. Also, we 
discuss the identified \hii\ region candidates.

\subsection{Spatial distribution}

The Galactic disk hosts most of its star-forming regions in its inner parts. Basic source properties, namely intensity and angular size, depend on their distance, which is determined by their location in the Galaxy. 

\begin{figure}[!h]
    \centering
    \includegraphics[width=0.5\textwidth, trim= 0 0 0 0, angle=0] {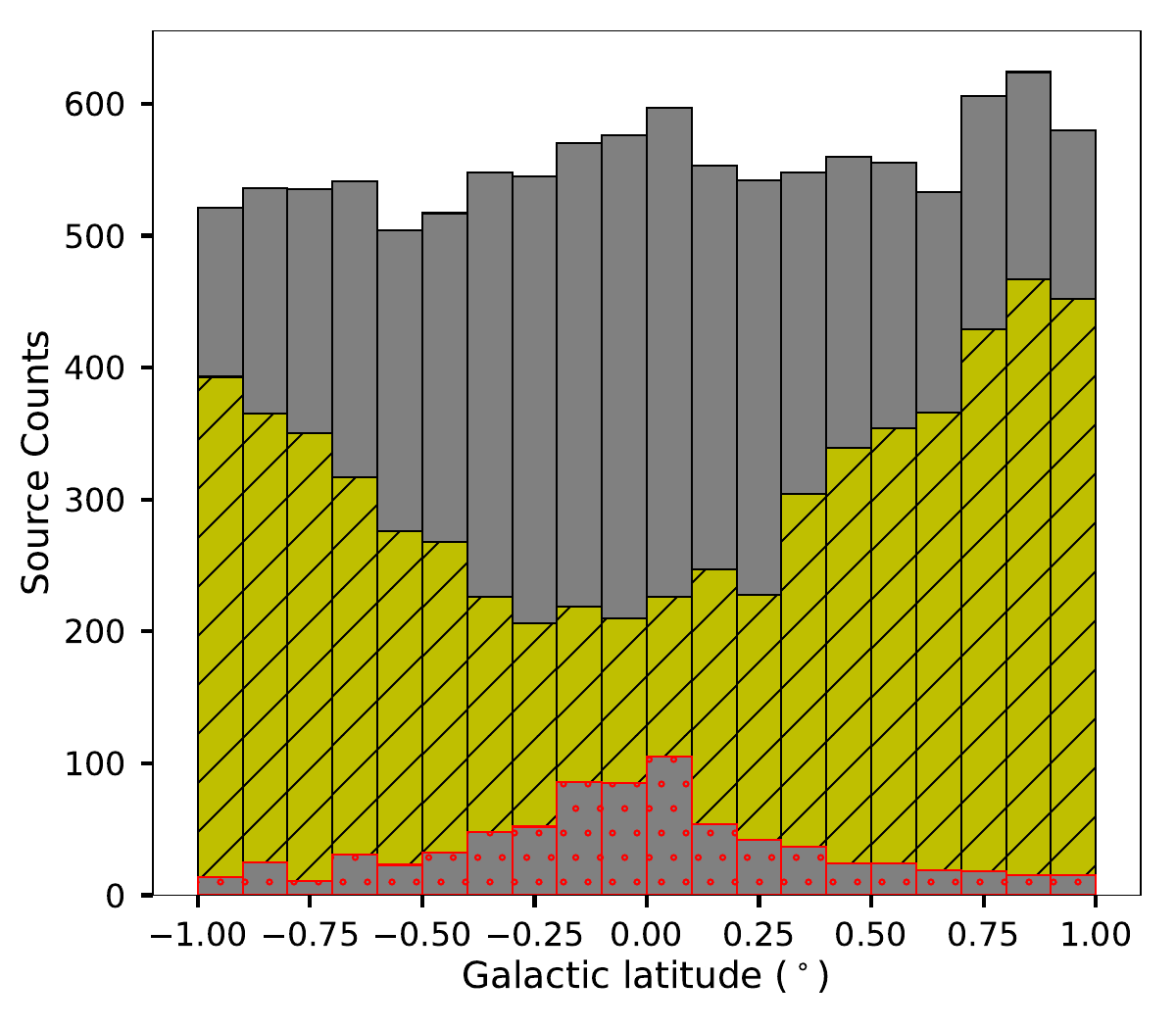} 
    \caption{Source distribution of GLOSTAR D-configuration sources in the Galactic latitude. 
    Color bars are: full sample in gray, yellow for EgC sources and red for 
    \hii\ regions. The bin size is $0.1\degr$.}
    \label{fig:SDlat}
\end{figure}

\begin{figure*}
    \centering
    \includegraphics[width=1.0\textwidth, trim= 0 0 0 0, angle=0] {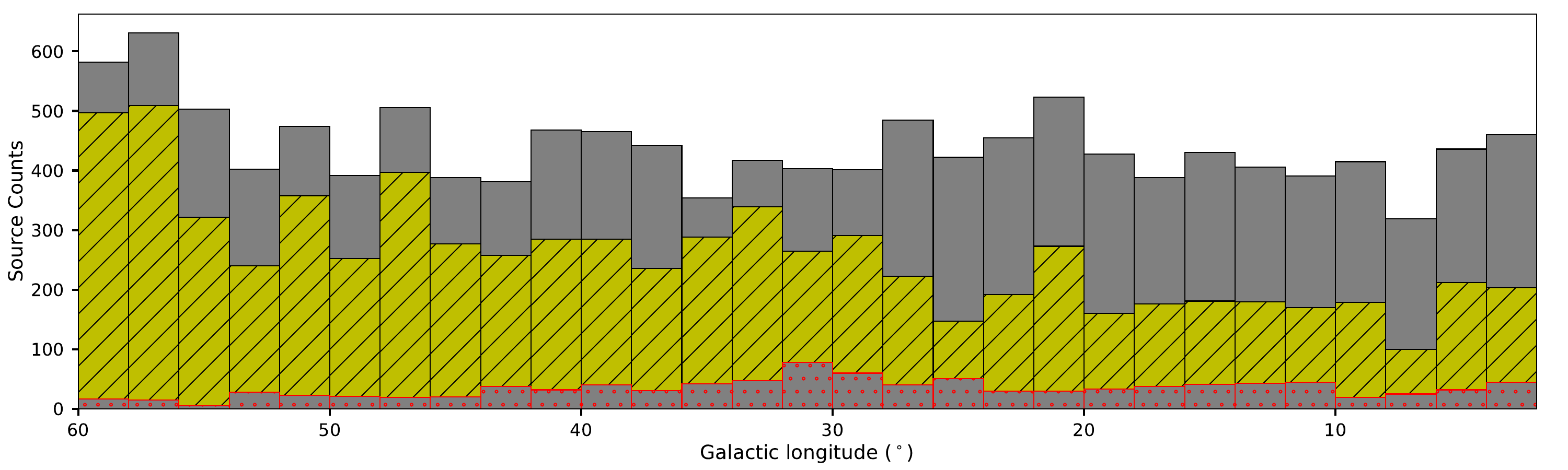} 
    \caption{Source distribution of GLOSTAR D-configuration sources in Galactic longitude. Sources in the range $28\degr<\ell<36\degr$ are taken from \citet{Medina2019}.
    The color of the bars is the same as for Fig.~\ref{fig:SDlat}. The bin size is $2\degr$.}
    \label{fig:SDlon}
\end{figure*}

Figures~\ref{fig:SDlat} and \ref{fig:SDlon} show the distribution of all 
detected GLOSTAR sources. 
The overall distribution in the observed Galactic latitude range
is robustly uniform (Fig.~\ref{fig:SDlat}). When we only consider the EgC sources (yellow bars), it can be noticed 
that their number diminishes towards the Galactic mid-plane ($b=0\degr$).
Moreover, as can be seen from Fig.~\ref{fig:FullRadioMap}, most of the extended sources are found around this 
latitude line, and it is also where most of the star formation is expected. The extended 
structures have two effects on source identification: they confuse background sources and, in interferometric images
with no zero-spacing information, they also increase the noise levels. Thus, background extragalactic 
sources that are intrinsically uniformly distributed in the sky show a decreasing number 
towards the mid-plane.
On the contrary, \hii\ regions (red bars) are more numerous towards low Galactic latitudes; that is, where most of the star formation in the Galaxy is occurring. 

The source distribution in Galactic longitude, Fig.~\ref{fig:SDlon}, also looks uniform for $\ell<56^\circ$, with the source counts increasing at larger longitudes. The area with $\ell>56^\circ$ contains a small 
number of LSSs, resulting in an improvement of the noise level (see also Table~\ref{tab:noise}) and 
less compromised sources. As a consequence, a higher number of sources are detected at these longitudes.
However, the number of \hii\ regions in this area is also lower, indicating a lower number of star-forming sites. This is in concordance with the dearth of 
MYSOs that \citet{zhang2019} noted in the segment of the Perseus arm stretching
between $\ell = \sim50^\circ$ and $90^\circ$.

\begin{figure*}
    \centering
    \includegraphics[scale=1.35,trim= 0 0 0 5,clip]{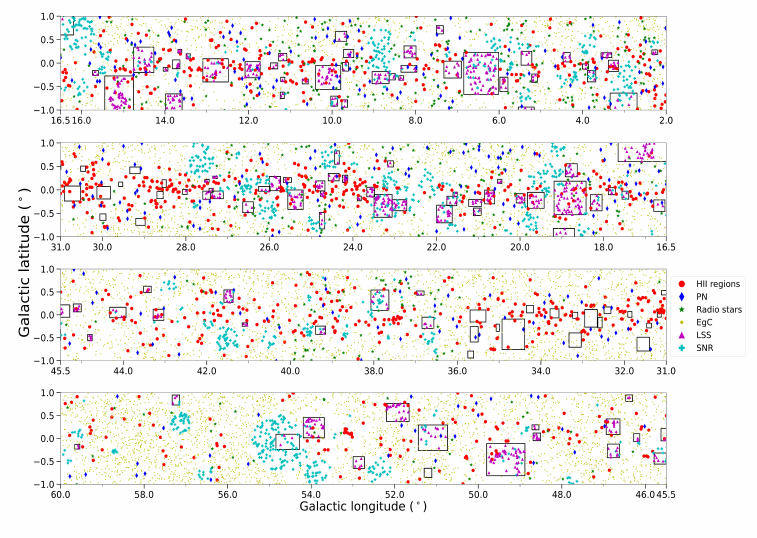} 
    \caption{Spatial distribution of GLOSTAR radio sources reported in this work. Different markers are used for the different resulting source classifications and are labeled on the right side of the plot. Squares delimit the area of the LSSs also identified in this work (see Sec. \ref{sect:Source_extraction}). Sources in the range $28\degr <\ell <36\degr$ are taken from \citet{Medina2019}.  }
    \label{fig:2DSD}
\end{figure*}

Finally, in Fig.~\ref{fig:2DSD}, we show the complete spatial distribution of radio sources detected in the
course of
this work. We have included the sources with LSS and SNRs. This figure shows that the number of sources is lower in the presence of extended sources because of the effects mentioned earlier.

\subsection{Fluxes and angular sizes}

\begin{figure}
    \centering
    \includegraphics[width=0.5\textwidth, trim= 0 0 0 0, angle=0] {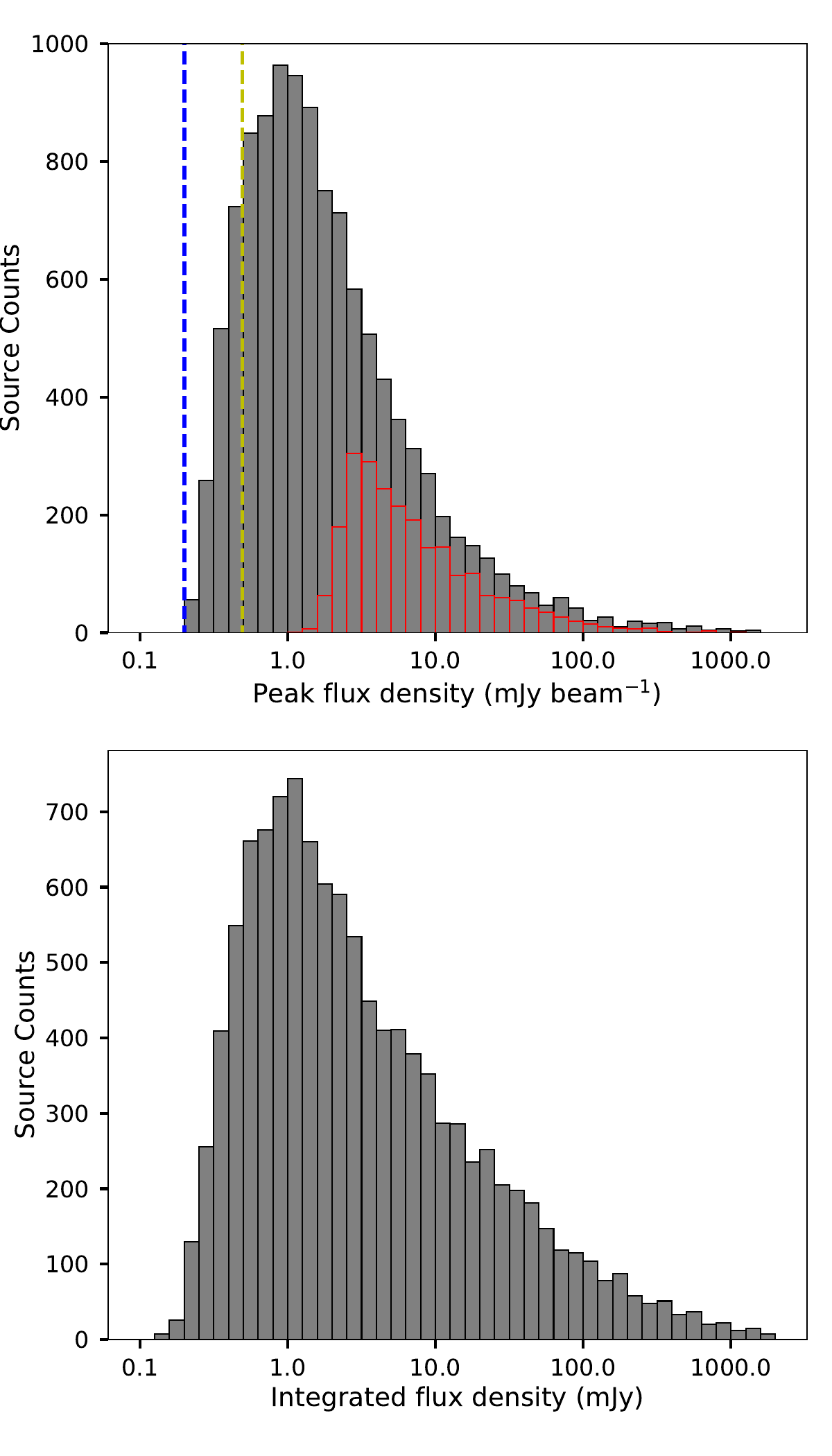} 
    \caption{Peak (top) and integrated (bottom) flux distribution of GLOSTAR sources. The dashed blue and yellow lines in the upper panel delimit the 4$\sigma_{\rm noise}$ detection limit in areas free of extended emission ($\sigma_{\rm noise}\simeq50 \mu$Jy beam$^{-1}$) and for the nominal noise level of the D-configuration images ($\sigma_{\rm noise}\simeq123 \mu$Jy beam$^{-1}$).
    The red histogram in the upper panel shows the flux distribution of sources previously identified by CORNISH. The bin width is 0.1~dex.}
    \label{fig:histoPF}
\end{figure}

The source flux density is one of the most critical parameters in the GLOSTAR source catalog. 
By comparing our results with the CORNISH survey, we have shown that the flux density
parameter from the GLOSTAR survey is accurate within 6\% (see Section~\ref{Sec:Snu}). 
The CORNISH survey, however, had a nominal noise level of $\sim$0.4 mJy beam$^{-1}$, significantly 
higher than the GLOSTAR D-configuration images. In the top panel of Fig.~\ref{fig:histoPF}, 
we show the source peak flux density
distribution for GLOSTAR and CORNISH for the Galactic plane area covered by this work. 
The better sensitivity of the GLOSTAR survey resulted in a larger number of detected sources, indicating 
that most sources with a brightness lower than a few mJy beam$^{-1}$ are new detections.

\begin{figure}
    \centering
    \includegraphics[width=0.5\textwidth, trim= 0 0 0 0, angle=0] {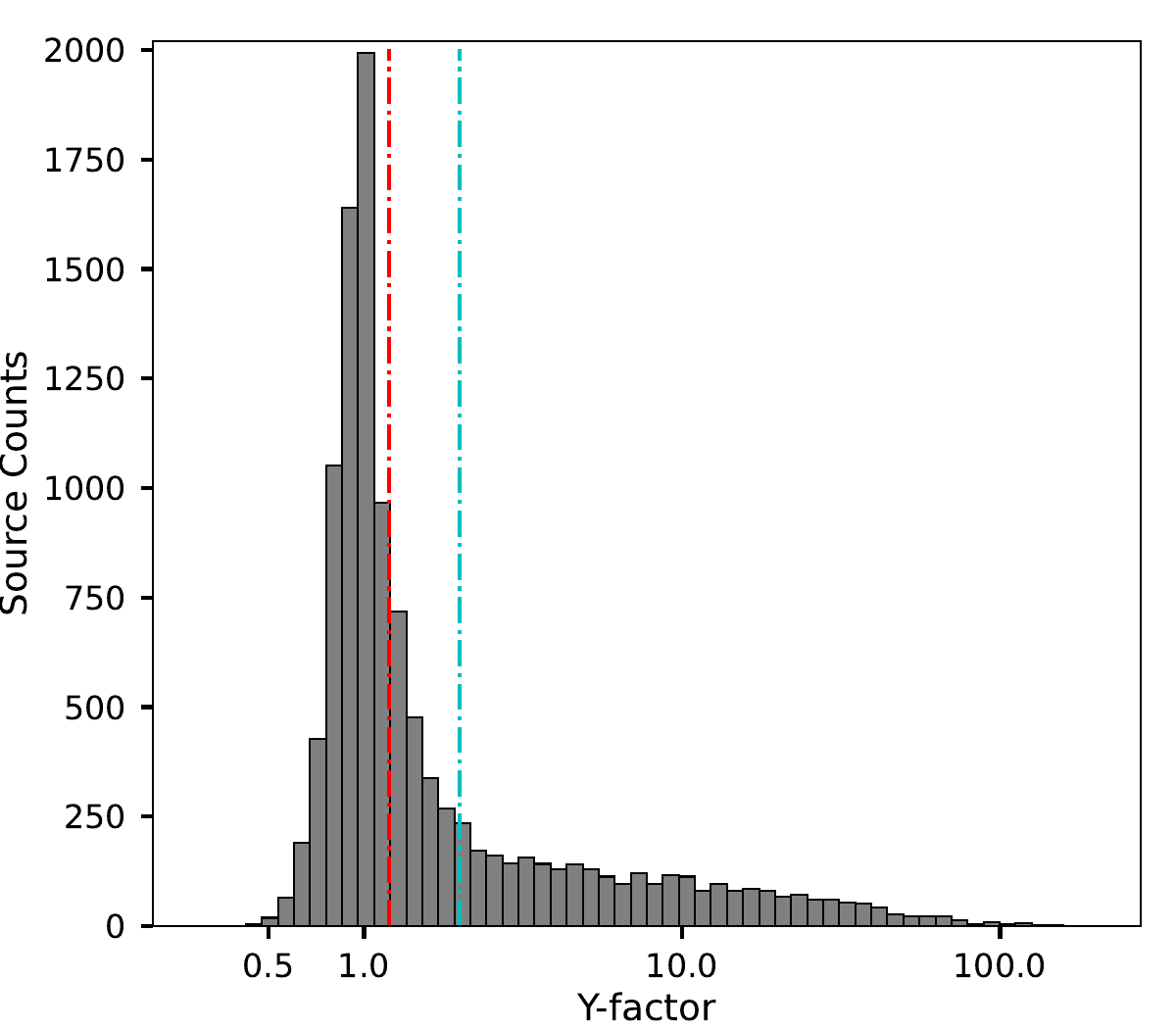} 
    \caption{Distribution of Y-factor of GLOSTAR sources. The dash-dotted red (Y$_{\rm factor} = 1.2$) and cyan (Y$_{\rm factor} = 2.0$) lines indicate the criterion used to distinguish between unresolved, compact, and extended sources, respectively. The bin width is 0.05~dex.}
    \label{fig:histoYf}
\end{figure}

\begin{figure}
    \centering
    \includegraphics[width=0.5\textwidth, trim= 0 0 0 0, angle=0] {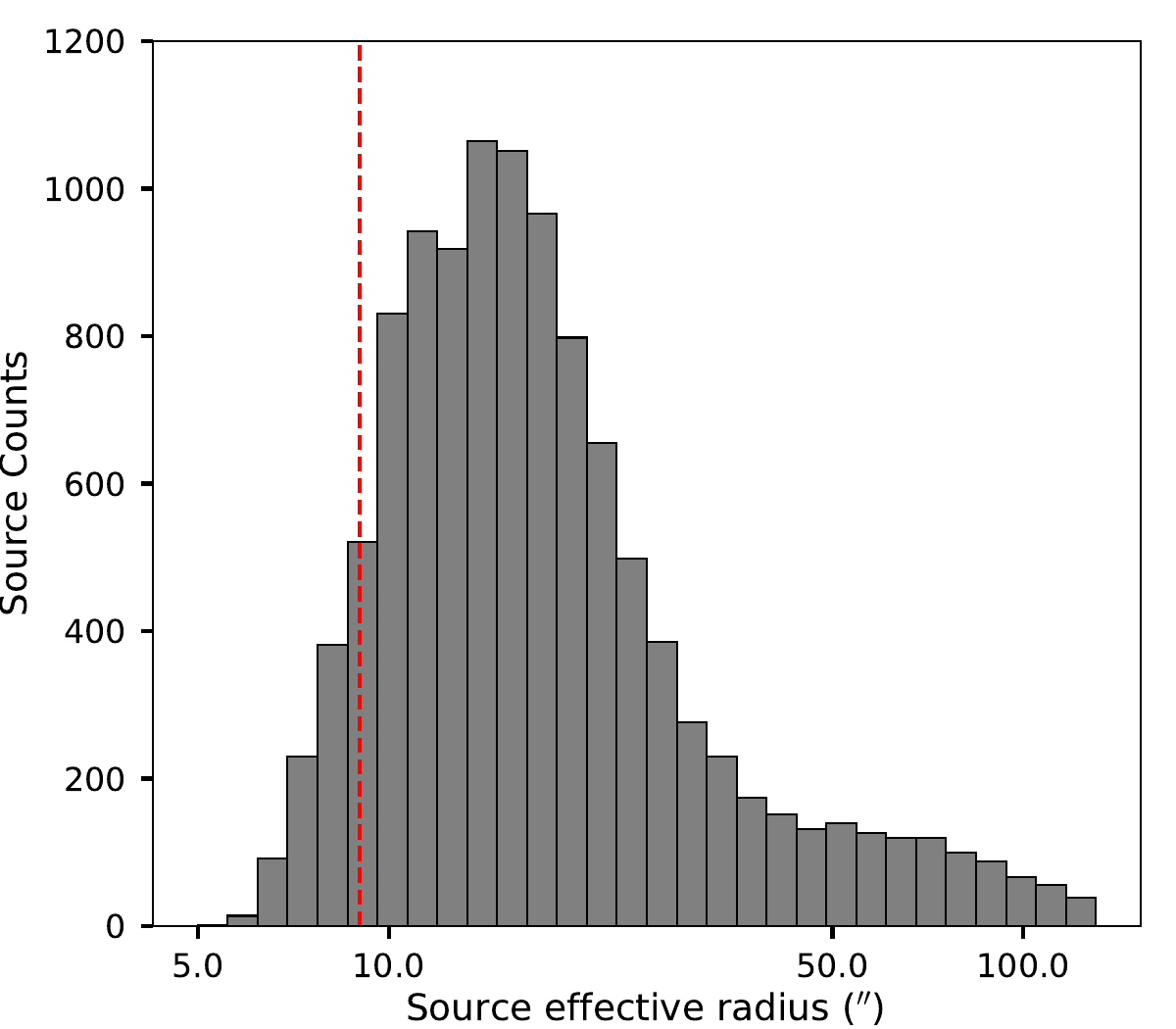} 
    \caption{Distribution of the source radii. The dashed red line indicates the resolution of the observation (radius of the synthesized beam). The bin size used is 0.05~dex.}
    \label{fig:histoR}
\end{figure}

The ratio $S_{\rm int}/S_{\rm peak}$ is also known as the Y-factor, a parameter that can
be used to infer the size of a source. In the GLOSTAR survey, we have used it to define an 
unresolved source when the Y$_{\rm factor}\leq1.2$, a compact source when $1.2<$Y$_{\rm factor}\leq2.0$, 
and an extended source when Y$_{\rm factor}>2.0$. In Fig.~\ref{fig:histoYf}, we plot the Y-factor 
distribution of the GLOSTAR sources. Similarly, Fig.~\ref{fig:histoR} shows the distribution of source effective radius. 
Both results show that our catalog mainly comprises
compact and unresolved sources.

\subsection{Spectral indices}

Sources that exhibit radio emission produced from ionized thermal gas with temperature $> 10^4$~K (such as  \hii\ regions 
and PNe) have spectral indices ranging from --0.1 to 2.0, representing optical thin and thick bremsstrahlung, respectively. On the other hand, 
non-thermal radio emission is produced in high-energy processes. The most common, resulting in synchrotron radiation, is produced by relativistic electrons spiralling around magnetic field lines and reaches brightness temperatures above 10$^6$\,K. Active galactic nuclei, for example, are common nonthermal radio emitters.
Other compact  Galactic nonthermal radio sources are pulsars (PSRs) and micro-quasars. 
Interestingly, also in the star formation context nonthermal radio emission may arise from compact sources, such as strong wind collision regions in massive multiple stellar systems \citep[][]{dzib2013,yanza2024}, 
strong shocks of jets of very young massive stars \citep{Reid1995, carrasco2010}, and magnetically active young low-mass stars \citep{forbrich2021,dzib2021}. Typical spectral index values for non-thermal radio sources related to massive stars are about $-0.5$ \citep[e.g., ][]{Reid1995, doug2003, mohan2022}.  However, the spectral index 
of radio emission from magnetically active stars ranges from $-2.0$ to $+2.0$ \citep{dulk1985,dzib2015}. 
\begin{figure}[!h]
    \centering
    \includegraphics[width=0.41\textwidth, trim= 0 0 0 0, angle=0] {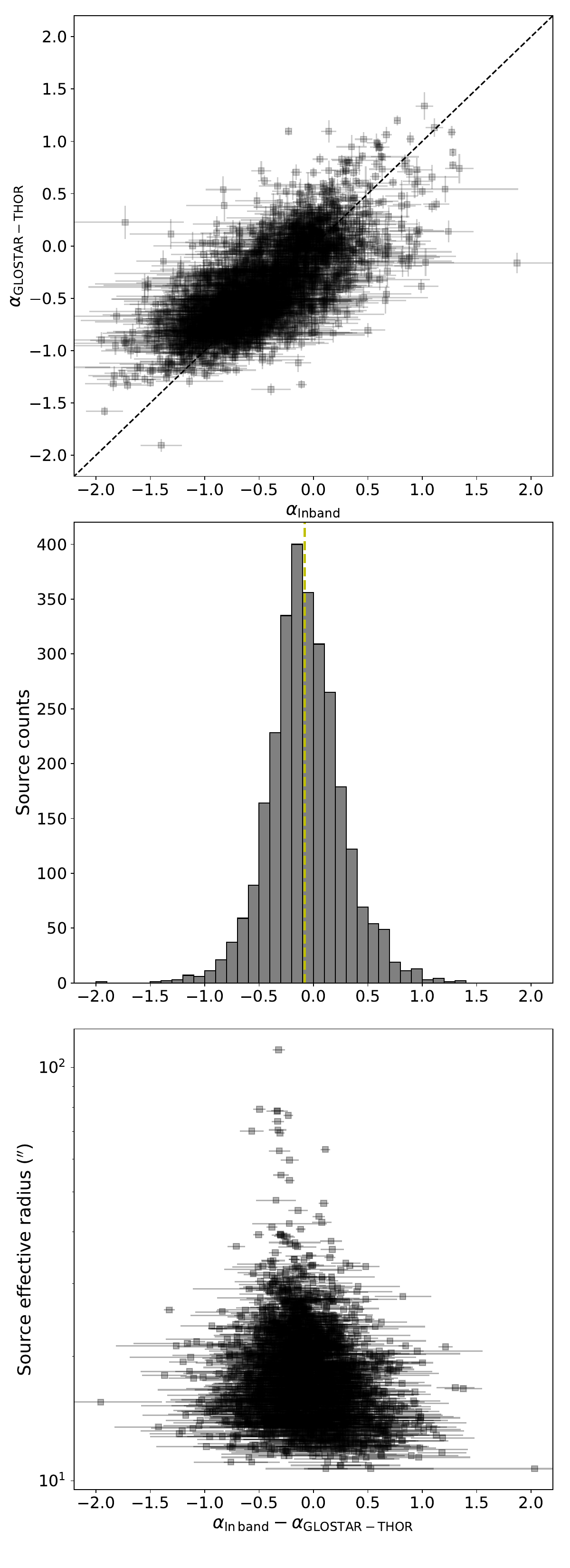} 
    \caption{Comparison of GLOSTAR in-band and GLOSTAR-THOR spectral indices. \textit{Top:} Scatter plot of $\alpha_{\rm In band}$ vs. $\alpha_{\rm GLOSTAR-THOR}$. The dashed black line is the equality line. \textit{ Middle:} Distribution of the differences of the spectral indices.
    The dashed yellow line indicates the mean value of $-0.08\pm0.01$. The standard deviation is 0.35. The histogram used a bin size of 0.1.  \textit{ Bottom:} Source effective radius vs. $\mathbf{\alpha_{\rm In band}-\alpha_{\rm GLOSTAR-THOR}}$. }
    \label{fig:SIcomp}
\end{figure}

In this work, we have measured the spectral index of over 5\,000 sources. First, by
using the upgraded capabilities of the VLA interferometer that offer a total bandwidth of 4\,GHz in C-band,
we have produced images in smaller frequency bins and determined {in-band} spectral
indices of compact sources detected with S/N ratio $\,>\,$10.0. Second, we have used the flux densities from the THOR
survey and computed GLOSTAR-THOR spectral indices of compact sources detected in both surveys.
Using the flux densities reported by THOR has two advantages. First, it increases the frequency
baseline of the spectral index determination. Second, the total flux densities of both surveys 
can be used without splitting the data into smaller and less sensitive frequency bins. By using the full-band images of both surveys, we optimize the $(u, v)$-coverage of both surveys. This results in smaller errors in the spectral index determination. Quantitatively, this is reflected by the fact that 
we obtain mean error values for the in-band and the GLOSTAR-THOR spectral indices of
$0.2$ and $0.07$, respectively. We also notice the recent findings by \citet{rashid2024}, whose results indicate that broadband spectral indices are more reliable than in-band spectral indices. For the 2\,819 sources in which both spectral indices could be determined, we plot their values in the top panel of Fig.~\ref{fig:SIcomp}. In the middle panel of Fig.~\ref{fig:SIcomp} we show the distribution
of the differences between these two spectral index determinations, and in the low panel we plot this difference against the GLOSTAR source effective radius. We found a mean difference of $-0.08\pm0.01$ and a standard deviation 
of 0.35. To check if the small negative difference is correlated with size, we ran a Pearson correlation test between the source effective radius and the difference of spectral indices. We obtain a correlation coefficient $r=-0.14$, with a significance value of $p=6\times10^{-14}$. This indicates a weak negative correlation (larger sources have more negative values) and could explain the small negative difference between the two spectral index determinations. While the in-band spectral indices are slightly more negative than the GLOSTAR-THOR
spectral index, the independent spectral index determinations are consistent considering the mean values of the errors obtained for both methods. Because the error bars are smaller, however, a preference is given
to the GLOSTAR-THOR spectral indices.

\begin{figure}
    \centering
    \includegraphics[width=0.5\textwidth, trim= 0 0 0 0, angle=0] {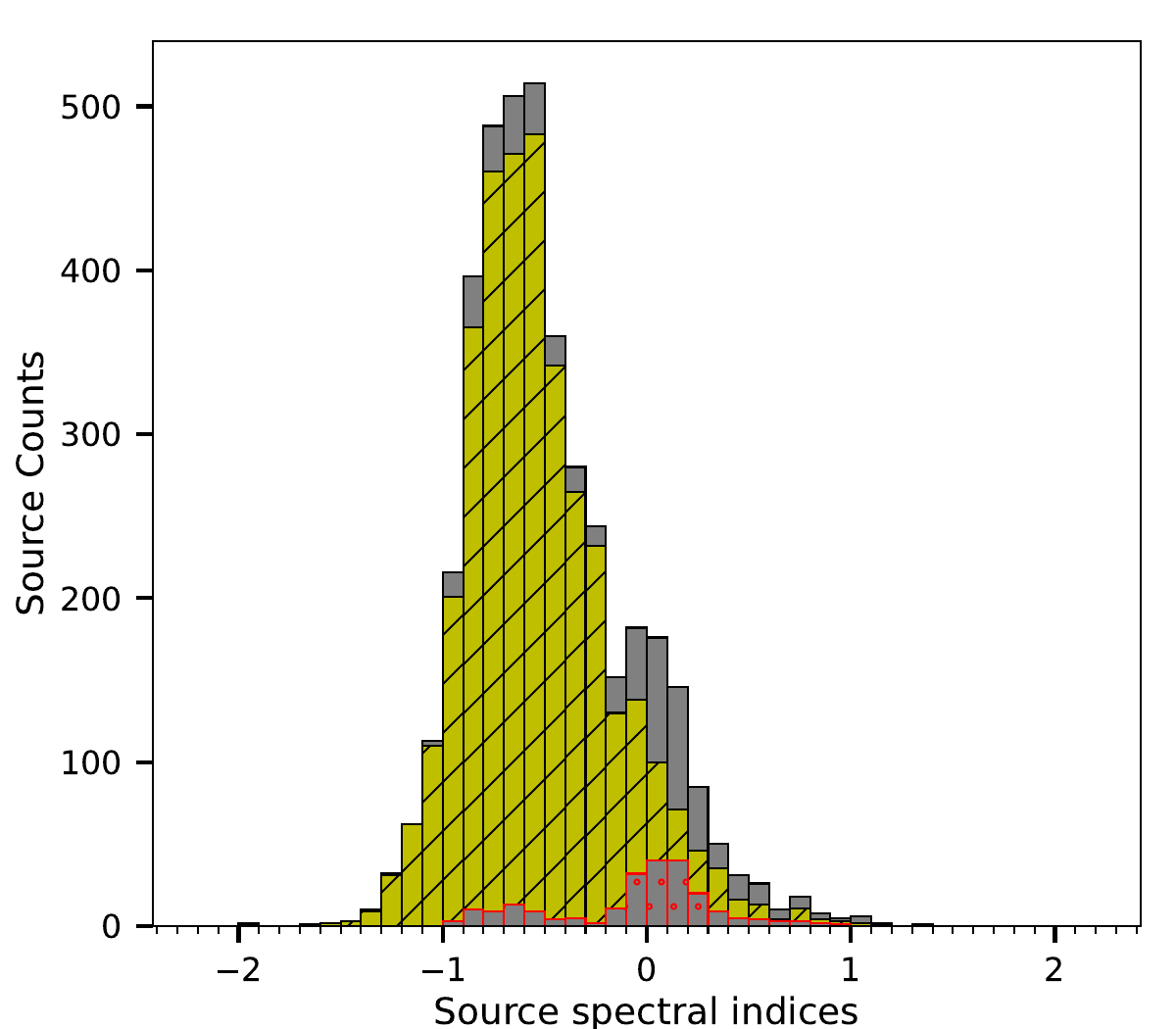} 
    \caption{GLOSTAR-THOR spectral index distribution of radio sources reported in this work. In gray is the full sample, yellow is for sources related to EgC and red for  \hii\ region candidates. The histogram used a bin size of 0.1. }
    \label{fig:SI}
\end{figure}

Figure~\ref{fig:SI} shows the GLOSTAR-THOR spectral index distribution of GLOSTAR sources. 
The distribution shows two peaks, the first at a value $\sim-0.6$ and the second at a value
$\sim0.0$, pointing to two populations of radio sources. The distribution of EgC sources (yellow histogram) shows only one peak 
and the mean is $\sim-0.6$, similar to the canonical value of $-0.7$ for extragalactic nonthermal radio sources 
\citep{condon1984}.
For the 225 \hii\ regions with measured spectral index (red histogram), the distribution also shows a single peak at $\alpha=0.14\pm0.02$, consistent with
the values for (almost) optically thin free-free emission (thermal radio emission). 
We discuss \hii\ regions in more detail in the following section.

\subsection{\hii regions}

\begin{figure}
    \centering
    \includegraphics[width=0.5\textwidth, trim= 0 0 0 0, angle=0] {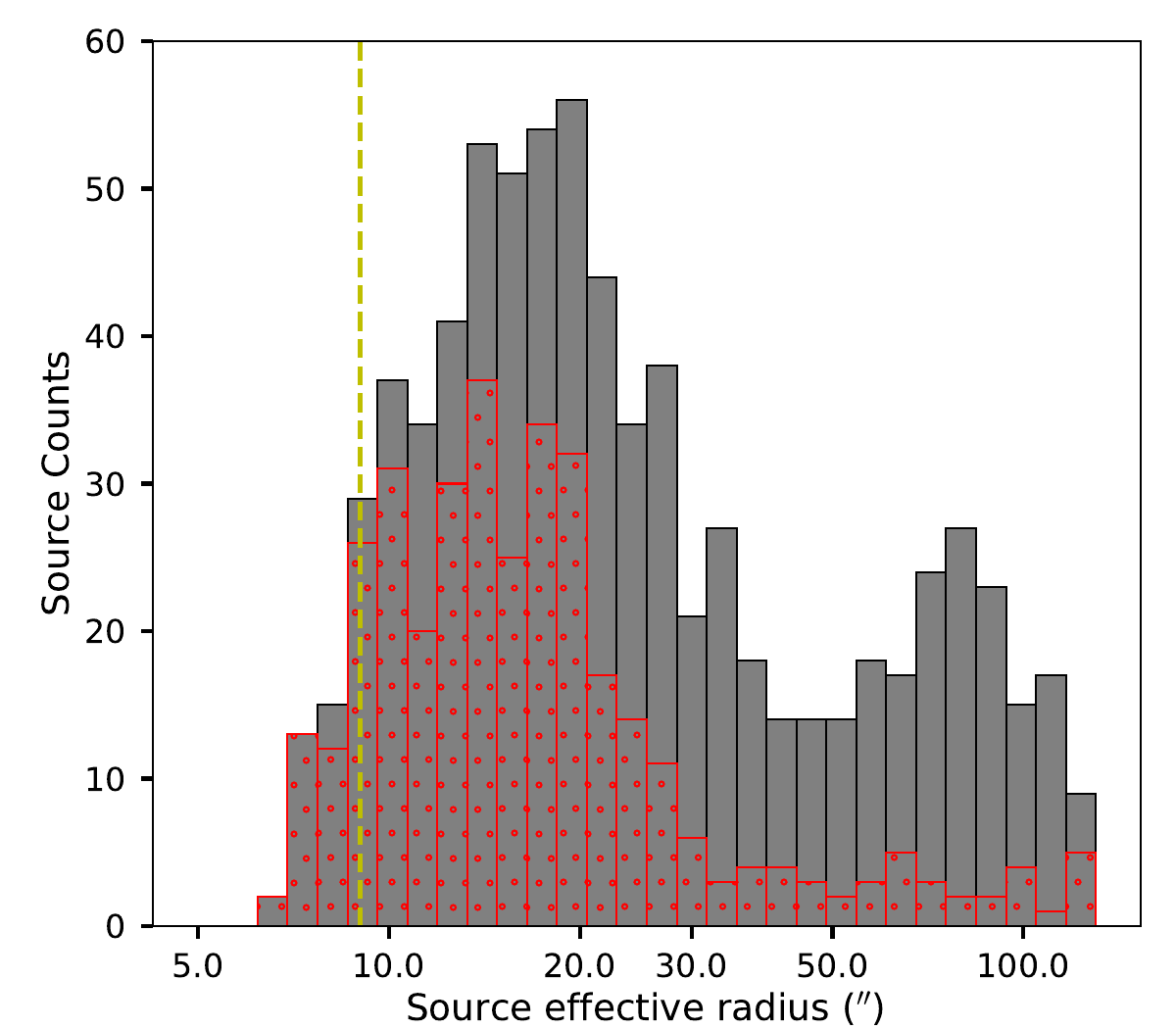} 
    \caption{Distribution of \hii\ region effective radius. The gray bars show the distribution for all identified \hii regions. Red bars are the new \hii\ regions candidates identified in this work.  The dashed yellow line indicates the resolution of the observation (radius of the synthesized beam). The bin size used is 0.05~dex. }
    \label{fig:RHii}
\end{figure}

Following the classification criteria discussed in Section~\ref{Sec:SClass}, we have identified 
769 \hii\ regions, of which 359 are new \hii\ region candidates. Most of them are located close to the mid-plane of the Galactic disk ($b\simeq0.0$, see Figure~\ref{fig:SDlat}).  
Most of the newly identified \hii\ region candidates are compact, with an effective 
radius smaller than $20''$ (see Fig.~\ref{fig:RHii}). 

We have estimated the spectral index for 225 \hii\ region candidates, and their mean 
spectral index value is $+0.14\pm0.02$ (Fig.~\ref{fig:SI}).
Surprisingly, 55 (24\%) of the \hii\ region candidates have a spectral index  
$<-0.2$, which is smaller than $-0.1$, the minimum spectral index value 
expected for optically thin free-free radio emission. 
In the GLOSTAR higher angular resolution images, we have also found such sources
\citep{dzib2023,yang2023}. In those cases, we have speculated that in \hii\ regions associated
with radio sources with negative spectral indices, the radio sources do not trace radio 
emission from the ionized gas of the \hii\ region, which will be resolved out, but rather  
trace stellar processes producing non-thermal radio emission or a mixture of emission produced by thermal and non-thermal processes. Sensitive radio observations have indeed revealed compact non-thermal radio sources in the vicinity of the \hii\ regions, up to several tens in some cases  \citep[e.g.,][]{wilner1999,medina2018,yanza2022}. \citet{wilner1999} argue that the non-thermal compact radio sources they find around the archetypal ultracompact \hii\ region W3(OH), several of which are time variable, represent low-mass stars in the stellar cluster that surrounds the MYSO that excites the \hii\ region.

In our VLA D-configuration 
images, we do not expect that \hii\ regions related to compact radio sources ($Y_{\rm factor}\leq2.0$) have resolved out emission. This excludes the possibility of imaging artifacts.

\begin{figure}[h!]
    \centering
    \includegraphics[width=0.5\textwidth, trim= 0 0 0 0, angle=0]{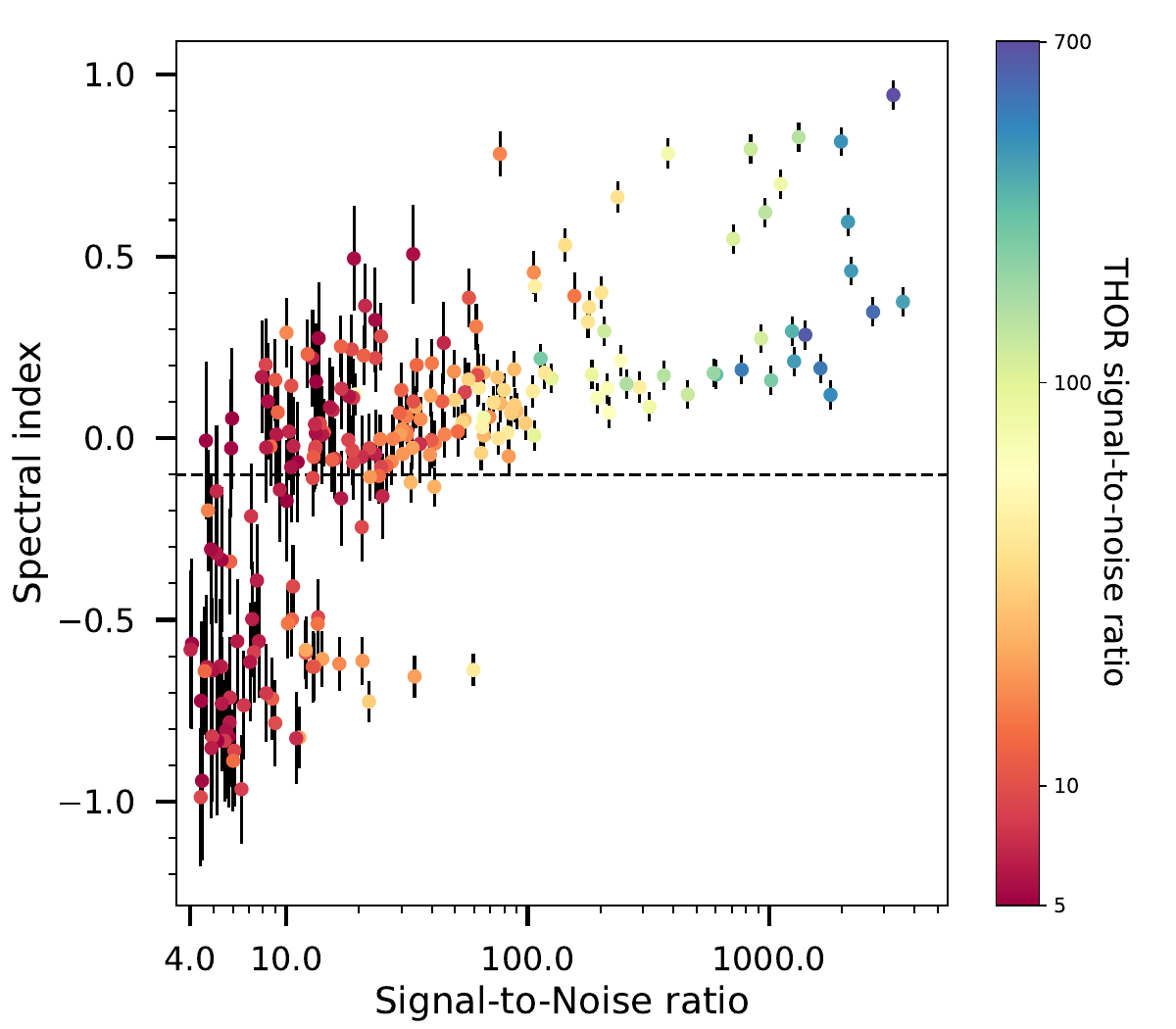} 
    \caption{Spectral indices of \hii\ regions as a function of the S/N. The circles are color-coded with their S/N in the THOR survey. The dashed line shows the spectral index value of --0.1, the minimum expected for optically thin free-free radio emission.}
    \label{fig:HII_a}
\end{figure}

In  Fig.~\ref{fig:HII_a} we plot the 225 spectral indices measured for \hii\ region 
candidates as a function of their S/N from the GLOSTAR maps.
From this plot, it can be seen that most of the sources with negative spectral index
also have a low S/N ratio. Most of them also have a low S/N ratio in the THOR survey. 
The low brightness of these sources could have biased the spectral index determination. 
However, there are still some sources with a S/N ratio $>$10 in both GLOSTAR and THOR
that clearly have a negative spectral index.
These radio sources require further studies to confirm or refute their classification as \hii\ regions. 

\subsection{Extragalactic background sources}

Along with Galactic sources, our observations are expected to detect a large number of extragalactic radio sources. In fact, it has been shown that most of the unclassified
sources in our previous catalog can indeed be related to galaxies
with radio emission \citep{Medina2019,Chakraborty2020}. While selections effects could play a role
on their distribution, the early studies by \citet{fomalont1991}
show that, at 5\,GHz, the rough number of expected background sources 
per square arcminute, with flux levels above 
$S[\mu{\rm Jy}]$, is described by:
$$N(S[\mu{\rm Jy}])=(0.42\pm0.05)(S[\mu{\rm Jy}]/30)^{-1.18\pm0.19}.$$

Considering a mean noise level of $\sigma=$123~$\mu$Jy, and a 4$\sigma$ threshold
for our detected sources, the number of expected extragalactic radio sources
in our images is $6\, 400\pm3\, 750$. 
This number is in excellent agreement with the  
6 312 sources that we have classified as EgC. 

\section{Summary and conclusions}\label{Conc}

The GLOSTAR-VLA Galactic plane survey \citep{Medina2019,brunthaler2021} is currently 
the most sensitive mid-radio wavelength survey covering a large fraction of the Galactic 
plane observed from the northern hemisphere.  Its main objective is unveiling 
signatures of recent massive star formation.
However, as many of these sources are expected to be extragalactic background sources,
it is necessary to classify the detected sources, a task that has 
many challenges. 

In this work, we have presented radio images of a 100 square degree area of the Galactic plane, with an angular resolution of $18''$. 
The mapped region covers the area 
delimited by the coordinates $2\degr < \ell < 28\degr$, $36\degr < \ell < 60\degr$  
and $|b| < 1\degr$. We have used a combination of the \blobcat\ software and visual 
inspection procedures to identify \FinCatNum\ radio sources in the GLOSTAR map.
We have identified \includeLSS\ that are part of very extended sources (\hii\ 
region complexes or SNRs). The radio source catalog presents the results of
the source extraction performed with \blobcat, such as positions, S/N, 
flux densities, Y-factor, and effective radius. We have also obtained the spectral indices of 5 276 radio sources, which are also listed in the catalog. 
We have cross-matched the GLOSTAR radio sources with the radio sources 
reported from other radio surveys; for example, THOR \citep{beuther2016}, CORNISH \citep{hoare2012}, and RMS \citep{urquhart2009}. These radio surveys
were used to, first, verify the measured source parameters and, second,  
classify sources. 
Source classification was also performed with the information on
(sub)millimeter and IR wavelength counterparts. Source classes are also listed in the catalog, in which
a large number (6\,312) are extragalactic background sources. 

With the performed multiwavelength analysis, we identified 769\,\hii\ region candidates. 
Previous works have reported 410 of these as \hii\ regions or \hii\ region candidates, and the remaining 359 we identify as \hii\ candidates for the first time. The spatial distribution of these sources 
is concentrated around the Galactic mid-plane 
and their numbers decrease in the outer parts of the Galactic disk ($\ell>56^\circ$),
indicating zones with higher and lower star formation, respectively. 
Using additional flux density measurements at 1.4~GHz from the THOR survey, we have  
also determined spectral indices of 225 \hii region candidates. Their mean spectral index is $\sim0.1$, consistent with thermal free-free radio emission. 
Interestingly we found several \hii region candidates with negative spectral index, although this could be partly due to their low S/N ratio in GLOSTAR and THOR. However, there is an interesting sample 
of \hii\ region candidates with S/N$>$10 and negative spectral indices that need 
to be studied further to establish their true nature.

Combining the information from large surveys is
an excellent way to obtain an unbiased look for tracers of early star formation. The derived properties and classification of several thousands of
new and known radio sources are invaluable information and truly show the legacy nature of the GLOSTAR radio survey.

\begin{acknowledgements}
This research was partially funded by the ERC Advanced Investigator Grant GLOSTAR (247078). 
S.A.D. acknowledge the M2FINDERS project from the European Research
Council (ERC) under the European Union's Horizon 2020 research and innovation programme
(grant No 101018682). AYY acknowledges support from the NSFC grants No. 11988101 and No. NSFC 11973013. M.R.R. is a Jansky Fellow of the National Radio Astronomy Observatory.
This work uses information from the GLOSTAR databases at 
\url{http://glostar.mpifr-bonn.mpg.de} supported by the MPIfR (Max-Planck-Institut für Radioastronomie), Bonn, which is based on observations with the Karl G. Jansky Very Large Array (VLA) of NRAO (The National Radio Astronomy Observatory is a facility of the National Science Foundation operated under cooperative agreement by Associated Universities, Inc.) and 100-m telescope of the MPIfR at Effelsberg. 
 It also made use of information from the ATLASGAL
database at 
\url{http://atlasgal.mpifr-bonn.mpg.de/cgi-bin/ATLASGAL_DATABASE.cgi}
supported by the MPIfR, Bonn, as well as information from the CORNISH
database at \url{http://cornish.leeds.ac.uk/public/index.php} which
was constructed with support from the Science and Technology 
Facilities Council of the UK.
This work has used data from GLIMPSE survey of the \emph{Spitzer} Space Telescope, which is operated by the Jet Propulsion Laboratory, California Institute of Technology under a contract with NASA.
This publication also makes use of data products from the Wide-field Infrared Survey Explorer, which is a joint project of the University of California, Los Angeles, and the 
Jet Propulsion Laboratory/California Institute of Technology, funded 
by the National Aeronautics and Space Administration. 
This paper used the data products from the Hi-GAL survey of the \emph{Herschel} telescope which is an ESA space observatory with science instruments provided by European-led Principal Investigator consortia and with important participation from NASA. 
This document was prepared using the collaborative tool Overleaf available at: 
\url{https://www.overleaf.com/}. 
\end{acknowledgements}

\bibliographystyle{aa_url}
\bibliography{references}

\begin{thebibliography}{70}
\expandafter\ifx\csname natexlab\endcsname\relax\def\natexlab#1{#1}\fi

\bibitem[{{Anderson} {et~al.}(2014){Anderson}, {Bania}, {Balser}, {Cunningham}, {Wenger}, {Johnstone}, \& {Armentrout}}]{anderson2014}
{Anderson}, L.~D., {Bania}, T.~M., {Balser}, D.~S., {et~al.} 2014, \href{http://dx.doi.org/10.1088/0067-0049/212/1/1}{\color{magenta}\apjs}, \href{http://adsabs.harvard.edu/abs/2014ApJS..212....1A}{212, 1}

\bibitem[{{Anderson} {et~al.}(2017){Anderson}, {Wang}, {Bihr}, {Rugel}, {Beuther}, {Bigiel}, {Churchwell}, {Glover}, {Goodman}, {Henning}, {Heyer}, {Klessen}, {Linz}, {Longmore}, {Menten}, {Ott}, {Roy}, {Soler}, {Stil}, \& {Urquhart}}]{anderson2017}
{Anderson}, L.~D., {Wang}, Y., {Bihr}, S., {et~al.} 2017, \href{http://dx.doi.org/10.1051/0004-6361/201731019}{\color{magenta}\aap}, \href{https://ui.adsabs.harvard.edu/abs/2017A&A...605A..58A}{605, A58}

\bibitem[{{Anderson} {et~al.}(2012){Anderson}, {Zavagno}, {Barlow}, {Garc{\'\i}a-Lario}, \& {Noriega-Crespo}}]{anderson2012}
{Anderson}, L.~D., {Zavagno}, A., {Barlow}, M.~J., {Garc{\'\i}a-Lario}, P., \& {Noriega-Crespo}, A. 2012, \href{http://dx.doi.org/10.1051/0004-6361/201117640}{\color{magenta}\aap}, \href{https://ui.adsabs.harvard.edu/abs/2012A&A...537A...1A}{537, A1}

\bibitem[{{Bertin} \& {Arnouts}(1996)}]{bertin1996}
{Bertin}, E. \& {Arnouts}, S. 1996, \href{http://dx.doi.org/10.1051/aas:1996164}{\color{magenta}\aaps}, \href{https://ui.adsabs.harvard.edu/abs/1996A&AS..117..393B}{117, 393}

\bibitem[{{Beuther} {et~al.}(2016){Beuther}, {Bihr}, {Rugel}, {Johnston}, {Wang}, {Walter}, {Brunthaler}, {Walsh}, {Ott}, {Stil}, {Henning}, {Schierhuber}, {Kainulainen}, {Heyer}, {Goldsmith}, {Anderson}, {Longmore}, {Klessen}, {Glover}, {Urquhart}, {Plume}, {Ragan}, {Schneider}, {McClure-Griffiths}, {Menten}, {Smith}, {Roy}, {Shanahan}, {Nguyen-Luong}, \& {Bigiel}}]{beuther2016}
{Beuther}, H., {Bihr}, S., {Rugel}, M., {et~al.} 2016, \href{http://dx.doi.org/10.1051/0004-6361/201629143}{\color{magenta}\aap}, \href{http://adsabs.harvard.edu/abs/2016A%26A...595A..32B}{595, A32}

\bibitem[{{Brunthaler} {et~al.}(2021){Brunthaler}, {Menten}, {Dzib}, {Cotton}, {Wyrowski}, {Dokara}, {Gong}, {Medina}, {M{\"u}ller}, {Nguyen}, {Ortiz-Le{\'o}n}, {Reich}, {Rugel}, {Urquhart}, {Winkel}, {Yang}, {Beuther}, {Billington}, {Carrasco-Gonzalez}, {Csengeri}, {Murugeshan}, {Pandian}, \& {Roy}}]{brunthaler2021}
{Brunthaler}, A., {Menten}, K.~M., {Dzib}, S.~A., {et~al.} 2021, \href{http://dx.doi.org/10.1051/0004-6361/202039856}{\color{magenta}\aap}, \href{https://ui.adsabs.harvard.edu/abs/2021A&A...651A..85B}{651, A85}

\bibitem[{{Carrasco-Gonz{\'a}lez} {et~al.}(2010){Carrasco-Gonz{\'a}lez}, {Rodr{\'\i}guez}, {Anglada}, {Mart{\'\i}}, {Torrelles}, \& {Osorio}}]{carrasco2010}
{Carrasco-Gonz{\'a}lez}, C., {Rodr{\'\i}guez}, L.~F., {Anglada}, G., {et~al.} 2010, \href{http://dx.doi.org/10.1126/science.1195589}{\color{magenta}Science}, \href{https://ui.adsabs.harvard.edu/abs/2010Sci...330.1209C}{330, 1209}

\bibitem[{{Chakraborty} {et~al.}(2020){Chakraborty}, {Roy}, {Wang}, {Datta}, {Beuther}, {Medina}, {Menten}, {Urquhart}, {Brunthaler}, \& {Dzib}}]{Chakraborty2020}
{Chakraborty}, A., {Roy}, N., {Wang}, Y., {et~al.} 2020, \href{http://dx.doi.org/10.1093/mnras/stz3621}{\color{magenta}\mnras}, \href{https://ui.adsabs.harvard.edu/abs/2020MNRAS.492.2236C}{492, 2236}

\bibitem[{{Chini} \& {Hoffmeister}(2008)}]{chini2008}
{Chini}, R. \& {Hoffmeister}, V. 2008, in Handbook of Star Forming Regions, Volume II, ed. B.~{Reipurth}, Vol.~5, 625

\bibitem[{{Churchwell} {et~al.}(2009){Churchwell}, {Babler}, {Meade}, {Whitney}, {Benjamin}, {Indebetouw}, {Cyganowski}, {Robitaille}, {Povich}, {Watson}, \& {Bracker}}]{churchwell2009}
{Churchwell}, E., {Babler}, B.~L., {Meade}, M.~R., {et~al.} 2009, \href{http://dx.doi.org/10.1086/597811}{\color{magenta}\pasp}, \href{http://adsabs.harvard.edu/abs/2009PASP..121..213C}{121, 213}

\bibitem[{{Condon}(1984)}]{condon1984}
{Condon}, J.~J. 1984, \href{http://dx.doi.org/10.1086/162705}{\color{magenta}\apj}, \href{https://ui.adsabs.harvard.edu/abs/1984ApJ...287..461C}{287, 461}

\bibitem[{{Contreras} {et~al.}(2013){Contreras}, {Schuller}, {Urquhart}, {Csengeri}, {Wyrowski}, {Beuther}, {Bontemps}, {Bronfman}, {Henning}, {Menten}, {Schilke}, {Walmsley}, {Wienen}, {Tackenberg}, \& {Linz}}]{contreras2013}
{Contreras}, Y., {Schuller}, F., {Urquhart}, J.~S., {et~al.} 2013, \href{http://dx.doi.org/10.1051/0004-6361/201220155}{\color{magenta}\aap}, \href{https://ui.adsabs.harvard.edu/abs/2013A&A...549A..45C}{549, A45}

\bibitem[{{Cotton}(2008)}]{cotton2008}
{Cotton}, W.~D. 2008, \href{http://dx.doi.org/10.1086/586754}{\color{magenta}\pasp}, \href{http://adsabs.harvard.edu/abs/2008PASP..120..439C}{120, 439}

\bibitem[{{Csengeri} {et~al.}(2014){Csengeri}, {Urquhart}, {Schuller}, {Motte}, {Bontemps}, {Wyrowski}, {Menten}, {Bronfman}, {Beuther}, {Henning}, {Testi}, {Zavagno}, \& {Walmsley}}]{csengeri2014}
{Csengeri}, T., {Urquhart}, J.~S., {Schuller}, F., {et~al.} 2014, \href{http://dx.doi.org/10.1051/0004-6361/201322434}{\color{magenta}\aap}, \href{https://ui.adsabs.harvard.edu/abs/2014A&A...565A..75C}{565, A75}

\bibitem[{{Dokara} {et~al.}(2021){Dokara}, {Brunthaler}, {Menten}, {Dzib}, {Reich}, {Cotton}, {Anderson}, {Chen}, {Gong}, {Medina}, {Ortiz-Le{\'o}n}, {Rugel}, {Urquhart}, {Wyrowski}, {Yang}, {Beuther}, {Billington}, {Csengeri}, {Carrasco-Gonz{\'a}lez}, \& {Roy}}]{Dokara2021}
{Dokara}, R., {Brunthaler}, A., {Menten}, K.~M., {et~al.} 2021, \href{http://dx.doi.org/10.1051/0004-6361/202039873}{\color{magenta}\aap}, \href{https://ui.adsabs.harvard.edu/abs/2021A&A...651A..86D}{651, A86}

\bibitem[{{Dokara} {et~al.}(2023){Dokara}, {Gong}, {Reich}, {Rugel}, {Brunthaler}, {Menten}, {Cotton}, {Dzib}, {Khan}, {Medina}, {Nguyen}, {Ortiz-Le{\'o}n}, {Urquhart}, {Wyrowski}, {Yang}, {Anderson}, {Beuther}, {Csengeri}, {M{\"u}ller}, {Ott}, {Pandian}, \& {Roy}}]{Dokara2023}
{Dokara}, R., {Gong}, Y., {Reich}, W., {et~al.} 2023, \href{http://dx.doi.org/10.1051/0004-6361/202245339}{\color{magenta}\aap}, \href{https://ui.adsabs.harvard.edu/abs/2023A&A...671A.145D}{671, A145}

\bibitem[{{Dougherty} {et~al.}(2003){Dougherty}, {Pittard}, {Kasian}, {Coker}, {Williams}, \& {Lloyd}}]{doug2003}
{Dougherty}, S.~M., {Pittard}, J.~M., {Kasian}, L., {et~al.} 2003, \href{http://dx.doi.org/10.1051/0004-6361:20031048}{\color{magenta}\aap}, \href{https://ui.adsabs.harvard.edu/abs/2003A&A...409..217D}{409, 217}

\bibitem[{{Dulk}(1985)}]{dulk1985}
{Dulk}, G.~A. 1985, \href{http://dx.doi.org/10.1146/annurev.aa.23.090185.001125}{\color{magenta}\araa}, \href{https://ui.adsabs.harvard.edu/abs/1985ARA&A..23..169D}{23, 169}

\bibitem[{{Dzib} {et~al.}(2021){Dzib}, {Forbrich}, {Reid}, \& {Menten}}]{dzib2021}
{Dzib}, S.~A., {Forbrich}, J., {Reid}, M.~J., \& {Menten}, K.~M. 2021, \href{http://dx.doi.org/10.3847/1538-4357/abc68f}{\color{magenta}\apj}, \href{https://ui.adsabs.harvard.edu/abs/2021ApJ...906...24D}{906, 24}

\bibitem[{{Dzib} {et~al.}(2015){Dzib}, {Loinard}, {Rodr{\'\i}guez}, {Mioduszewski}, {Ortiz-Le{\'o}n}, {Kounkel}, {Pech}, {Rivera}, {Torres}, {Boden}, {Hartmann}, {Evans}, {Brice{\~n}o}, \& {Tobin}}]{dzib2015}
{Dzib}, S.~A., {Loinard}, L., {Rodr{\'\i}guez}, L.~F., {et~al.} 2015, \href{http://dx.doi.org/10.1088/0004-637X/801/2/91}{\color{magenta}\apj}, \href{https://ui.adsabs.harvard.edu/abs/2015ApJ...801...91D}{801, 91}

\bibitem[{{Dzib} {et~al.}(2013){Dzib}, {Rodr{\'\i}guez}, {Loinard}, {Mioduszewski}, {Ortiz-Le{\'o}n}, \& {Araudo}}]{dzib2013}
{Dzib}, S.~A., {Rodr{\'\i}guez}, L.~F., {Loinard}, L., {et~al.} 2013, \href{http://dx.doi.org/10.1088/0004-637X/763/2/139}{\color{magenta}\apj}, \href{https://ui.adsabs.harvard.edu/abs/2013ApJ...763..139D}{763, 139}

\bibitem[{{Dzib} {et~al.}(2023){Dzib}, {Yang}, {Urquhart}, {Medina}, {Brunthaler}, {Menten}, {Wyrowski}, {Cotton}, {Dokara}, {Ortiz-Le{\'o}n}, {Rugel}, {Nguyen}, {Gong}, {Chakraborty}, {Beuther}, {Billington}, {Carrasco-Gonzalez}, {Csengeri}, {Hofner}, {Ott}, {Pandian}, {Roy}, \& {Yanza}}]{dzib2023}
{Dzib}, S.~A., {Yang}, A.~Y., {Urquhart}, J.~S., {et~al.} 2023, \href{http://dx.doi.org/10.1051/0004-6361/202143019}{\color{magenta}\aap}, \href{https://ui.adsabs.harvard.edu/abs/2023A&A...670A...9D}{670, A9}

\bibitem[{{Elia} {et~al.}(2021){Elia}, {Merello}, {Molinari}, {Schisano}, {Zavagno}, {Russeil}, {M{\`e}ge}, {Martin}, {Olmi}, {Pestalozzi}, {Plume}, {Ragan}, {Benedettini}, {Eden}, {Moore}, {Noriega-Crespo}, {Paladini}, {Palmeirim}, {Pezzuto}, {Pilbratt}, {Rygl}, {Schilke}, {Strafella}, {Tan}, {Traficante}, {Baldeschi}, {Bally}, {di Giorgio}, {Fiorellino}, {Liu}, {Piazzo}, \& {Polychroni}}]{elia2021}
{Elia}, D., {Merello}, M., {Molinari}, S., {et~al.} 2021, \href{http://dx.doi.org/10.1093/mnras/stab1038}{\color{magenta}\mnras}, \href{https://ui.adsabs.harvard.edu/abs/2021MNRAS.504.2742E}{504, 2742}

\bibitem[{{Elia} {et~al.}(2017){Elia}, {Molinari}, {Schisano}, {Pestalozzi}, {Pezzuto}, {Merello}, {Noriega-Crespo}, {Moore}, {Russeil}, {Mottram}, {Paladini}, {Strafella}, {Benedettini}, {Bernard}, {Di Giorgio}, {Eden}, {Fukui}, {Plume}, {Bally}, {Martin}, {Ragan}, {Jaffa}, {Motte}, {Olmi}, {Schneider}, {Testi}, {Wyrowski}, {Zavagno}, {Calzoletti}, {Faustini}, {Natoli}, {Palmeirim}, {Piacentini}, {Piazzo}, {Pilbratt}, {Polychroni}, {Baldeschi}, {Beltr{\'a}n}, {Billot}, {Cambr{\'e}sy}, {Cesaroni}, {Garc{\'{\i}}a-Lario}, {Hoare}, {Huang}, {Joncas}, {Liu}, {Maiolo}, {Marsh}, {Maruccia}, {M{\`e}ge}, {Peretto}, {Rygl}, {Schilke}, {Thompson}, {Traficante}, {Umana}, {Veneziani}, {Ward-Thompson}, {Whitworth}, {Arab}, {Bandieramonte}, {Becciani}, {Brescia}, {Buemi}, {Bufano}, {Butora}, {Cavuoti}, {Costa}, {Fiorellino}, {Hajnal}, {Hayakawa}, {Kacsuk}, {Leto}, {Li Causi}, {Marchili}, {Martinavarro-Armengol}, {Mercurio}, {Molinaro}, {Riccio}, {Sano}, {Sciacca}, {Tachihara}, {Torii}, {Trigilio}, {Vitello}, \&
  {Yamamoto}}]{elia2017}
{Elia}, D., {Molinari}, S., {Schisano}, E., {et~al.} 2017, \href{http://dx.doi.org/10.1093/mnras/stx1357}{\color{magenta}\mnras}, \href{http://adsabs.harvard.edu/abs/2017MNRAS.471..100E}{471, 100}

\bibitem[{{Fazio} {et~al.}(2004){Fazio}, {Hora}, {Allen}, {Ashby}, {Barmby}, {Deutsch}, {Huang}, {Kleiner}, {Marengo}, {Megeath}, {Melnick}, {Pahre}, {Patten}, {Polizotti}, {Smith}, {Taylor}, {Wang}, {Willner}, {Hoffmann}, {Pipher}, {Forrest}, {McMurty}, {McCreight}, {McKelvey}, {McMurray}, {Koch}, {Moseley}, {Arendt}, {Mentzell}, {Marx}, {Losch}, {Mayman}, {Eichhorn}, {Krebs}, {Jhabvala}, {Gezari}, {Fixsen}, {Flores}, {Shakoorzadeh}, {Jungo}, {Hakun}, {Workman}, {Karpati}, {Kichak}, {Whitley}, {Mann}, {Tollestrup}, {Eisenhardt}, {Stern}, {Gorjian}, {Bhattacharya}, {Carey}, {Nelson}, {Glaccum}, {Lacy}, {Lowrance}, {Laine}, {Reach}, {Stauffer}, {Surace}, {Wilson}, {Wright}, {Hoffman}, {Domingo}, \& {Cohen}}]{fazio2004}
{Fazio}, G.~G., {Hora}, J.~L., {Allen}, L.~E., {et~al.} 2004, \href{http://dx.doi.org/10.1086/422843}{\color{magenta}\apjs}, \href{https://ui.adsabs.harvard.edu/abs/2004ApJS..154...10F}{154, 10}

\bibitem[{{Fomalont} {et~al.}(1991){Fomalont}, {Windhorst}, {Kristian}, \& {Kellerman}}]{fomalont1991}
{Fomalont}, E.~B., {Windhorst}, R.~A., {Kristian}, J.~A., \& {Kellerman}, K.~I. 1991, \href{http://dx.doi.org/10.1086/115952}{\color{magenta}\aj}, \href{https://ui.adsabs.harvard.edu/abs/1991AJ....102.1258F}{102, 1258}

\bibitem[{{Forbrich} {et~al.}(2021){Forbrich}, {Dzib}, {Reid}, \& {Menten}}]{forbrich2021}
{Forbrich}, J., {Dzib}, S.~A., {Reid}, M.~J., \& {Menten}, K.~M. 2021, \href{http://dx.doi.org/10.3847/1538-4357/abc68e}{\color{magenta}\apj}, \href{https://ui.adsabs.harvard.edu/abs/2021ApJ...906...23F}{906, 23}

\bibitem[{{Ghosh} {et~al.}(1989){Ghosh}, {Iyengar}, {Rengarajan}, {Tandon}, {Verma}, {Daniel}, \& {Ho}}]{ghosh1989}
{Ghosh}, S.~K., {Iyengar}, K.~V.~K., {Rengarajan}, T.~N., {et~al.} 1989, \href{http://dx.doi.org/10.1086/168122}{\color{magenta}\apj}, \href{https://ui.adsabs.harvard.edu/abs/1989ApJ...347..338G}{347, 338}

\bibitem[{{Ginsburg} {et~al.}(2017){Ginsburg}, {Goddi}, {Kruijssen}, {Bally}, {Smith}, {Galv{\'a}n-Madrid}, {Mills}, {Wang}, {Dale}, {Darling}, {Rosolowsky}, {Loughnane}, {Testi}, \& {Bastian}}]{gingsburg2017}
{Ginsburg}, A., {Goddi}, C., {Kruijssen}, J.~M.~D., {et~al.} 2017, \href{http://dx.doi.org/10.3847/1538-4357/aa6bfa}{\color{magenta}\apj}, \href{https://ui.adsabs.harvard.edu/abs/2017ApJ...842...92G}{842, 92}

\bibitem[{{Gong} {et~al.}(2023){Gong}, {Ortiz-Le{\'o}n}, {Rugel}, {Menten}, {Brunthaler}, {Wyrowski}, {Henkel}, {Beuther}, {Dzib}, {Urquhart}, {Yang}, {Pandian}, {Dokara}, {Veena}, {Nguyen}, {Medina}, {Cotton}, {Reich}, {Winkel}, {M{\"u}ller}, {Skretas}, {Csengeri}, {Khan}, \& {Cheema}}]{gong2023}
{Gong}, Y., {Ortiz-Le{\'o}n}, G.~N., {Rugel}, M.~R., {et~al.} 2023, \href{http://dx.doi.org/10.1051/0004-6361/202346102}{\color{magenta}\aap}, \href{https://ui.adsabs.harvard.edu/abs/2023A&A...678A.130G}{678, A130}

\bibitem[{{Green}(2019)}]{green2019}
{Green}, D.~A. 2019, \href{http://dx.doi.org/10.1007/s12036-019-9601-6}{\color{magenta}Journal of Astrophysics and Astronomy}, \href{https://ui.adsabs.harvard.edu/abs/2019JApA...40...36G}{40, 36}

\bibitem[{{Green} {et~al.}(2009){Green}, {Caswell}, {Fuller}, {Avison}, {Breen}, {Brooks}, {Burton}, {Chrysostomou}, {Cox}, {Diamond}, {Ellingsen}, {Gray}, {Hoare}, {Masheder}, {McClure-Griffiths}, {Pestalozzi}, {Phillips}, {Quinn}, {Thompson}, {Voronkov}, {Walsh}, {Ward-Thompson}, {Wong-McSweeney}, {Yates}, \& {Cohen}}]{green2009}
{Green}, J.~A., {Caswell}, J.~L., {Fuller}, G.~A., {et~al.} 2009, \href{http://dx.doi.org/10.1111/j.1365-2966.2008.14091.x}{\color{magenta}\mnras}, \href{https://ui.adsabs.harvard.edu/abs/2009MNRAS.392..783G}{392, 783}

\bibitem[{{Greisen}(2003)}]{greisen2003}
{Greisen}, E.~W. 2003, in Astrophysics and Space Science Library, Vol. 285, Information Handling in Astronomy - Historical Vistas, ed. A.~{Heck}, 109

\bibitem[{{Hales} {et~al.}(2012){Hales}, {Murphy}, {Curran}, {Middelberg}, {Gaensler}, \& {Norris}}]{hales2012}
{Hales}, C.~A., {Murphy}, T., {Curran}, J.~R., {et~al.} 2012, \href{http://dx.doi.org/10.1111/j.1365-2966.2012.21373.x}{\color{magenta}\mnras}, \href{http://adsabs.harvard.edu/abs/2012MNRAS.425..979H}{425, 979}

\bibitem[{{Helfand} {et~al.}(2006){Helfand}, {Becker}, {White}, {Fallon}, \& {Tuttle}}]{helfand2006}
{Helfand}, D.~J., {Becker}, R.~H., {White}, R.~L., {Fallon}, A., \& {Tuttle}, S. 2006, \href{http://dx.doi.org/10.1086/503253}{\color{magenta}\aj}, \href{http://adsabs.harvard.edu/abs/2006AJ....131.2525H}{131, 2525}

\bibitem[{{Hoare} {et~al.}(2005){Hoare}, {Lumsden}, {Oudmaijer}, {Urquhart}, {Busfield}, {Sheret}, {Clarke}, {Moore}, {Allsopp}, {Burton}, {Purcell}, {Jiang}, \& {Wang}}]{hoare2005}
{Hoare}, M.~G., {Lumsden}, S.~L., {Oudmaijer}, R.~D., {et~al.} 2005, in Massive Star Birth: A Crossroads of Astrophysics, ed. R.~{Cesaroni}, M.~{Felli}, E.~{Churchwell}, \& M.~{Walmsley}, Vol. 227, 370--375

\bibitem[{{Hoare} {et~al.}(2012){Hoare}, {Purcell}, {Churchwell}, {Diamond}, {Cotton}, {Chandler}, {Smethurst}, {Kurtz}, {Mundy}, {Dougherty}, {Fender}, {Fuller}, {Jackson}, {Garrington}, {Gledhill}, {Goldsmith}, {Lumsden}, {Mart{\'{\i}}}, {Moore}, {Muxlow}, {Oudmaijer}, {Pandian}, {Paredes}, {Shepherd}, {Spencer}, {Thompson}, {Umana}, {Urquhart}, \& {Zijlstra}}]{hoare2012}
{Hoare}, M.~G., {Purcell}, C.~R., {Churchwell}, E.~B., {et~al.} 2012, \href{http://dx.doi.org/10.1086/668058}{\color{magenta}\pasp}, \href{http://adsabs.harvard.edu/abs/2012PASP..124..939H}{124, 939}

\bibitem[{{Holwerda}(2005)}]{holwerda2005}
{Holwerda}, B.~W. 2005, ArXiv Astrophysics e-prints \href{http://adsabs.harvard.edu/abs/2005astro.ph.12139H}{[\eprint{astro-ph/0512139}]}

\bibitem[{{Irabor} {et~al.}(2018){Irabor}, {Hoare}, {Oudmaijer}, {Urquhart}, {Kurtz}, {Lumsden}, {Purcell}, {Zijlstra}, \& {Umana}}]{irabor2018}
{Irabor}, T., {Hoare}, M.~G., {Oudmaijer}, R.~D., {et~al.} 2018, \href{http://dx.doi.org/10.1093/mnras/sty1929}{\color{magenta}\mnras}, \href{http://adsabs.harvard.edu/abs/2018MNRAS.480.2423I}{480, 2423}

\bibitem[{{Kerton} {et~al.}(2013){Kerton}, {Arvidsson}, \& {Alexander}}]{kerton2013}
{Kerton}, C.~R., {Arvidsson}, K., \& {Alexander}, M.~J. 2013, \href{http://dx.doi.org/10.1088/0004-6256/145/3/78}{\color{magenta}\aj}, \href{https://ui.adsabs.harvard.edu/abs/2013AJ....145...78K}{145, 78}

\bibitem[{{Khan} {et~al.}(2024){Khan}, {Rugel}, {Brunthaler}, {Menten}, {Wyrowski}, {Urquhart}, {Gong}, {Yang}, {Nguyen}, {Dokara}, {Dzib}, {Medina}, {Ortiz-Le{\'o}n}, {Pandian}, {Beuther}, {Veena}, {Neupane}, {Cheema}, {Reich}, \& {Roy}}]{khan2024}
{Khan}, S., {Rugel}, M.~R., {Brunthaler}, A., {et~al.} 2024, \href{https://ui.adsabs.harvard.edu/abs/2024arXiv240705770K}{arXiv e-prints, arXiv:2407.05770}

\bibitem[{{Lucas} {et~al.}(2008){Lucas}, {Hoare}, {Longmore}, {Schr{\"o}der}, {Davis}, {Adamson}, {Bandyopadhyay}, {de Grijs}, {Smith}, {Gosling}, {Mitchison}, {G{\'a}sp{\'a}r}, {Coe}, {Tamura}, {Parker}, {Irwin}, {Hambly}, {Bryant}, {Collins}, {Cross}, {Evans}, {Gonzalez-Solares}, {Hodgkin}, {Lewis}, {Read}, {Riello}, {Sutorius}, {Lawrence}, {Drew}, {Dye}, \& {Thompson}}]{lucas2008}
{Lucas}, P.~W., {Hoare}, M.~G., {Longmore}, A., {et~al.} 2008, \href{http://dx.doi.org/10.1111/j.1365-2966.2008.13924.x}{\color{magenta}\mnras}, \href{https://ui.adsabs.harvard.edu/abs/2008MNRAS.391..136L}{391, 136}

\bibitem[{{Lumsden} {et~al.}(2013){Lumsden}, {Hoare}, {Urquhart}, {Oudmaijer}, {Davies}, {Mottram}, {Cooper}, \& {Moore}}]{lumsden2013}
{Lumsden}, S.~L., {Hoare}, M.~G., {Urquhart}, J.~S., {et~al.} 2013, \href{http://dx.doi.org/10.1088/0067-0049/208/1/11}{\color{magenta}\apjs}, \href{https://ui.adsabs.harvard.edu/abs/2013ApJS..208...11L}{208, 11}

\bibitem[{{Marleau} {et~al.}(2008){Marleau}, {Noriega-Crespo}, {Paladini}, {Clancy}, {Carey}, {Shenoy}, {Kraemer}, {Kuchar}, {Mizuno}, \& {Price}}]{marleau2008}
{Marleau}, F.~R., {Noriega-Crespo}, A., {Paladini}, R., {et~al.} 2008, \href{http://dx.doi.org/10.1088/0004-6256/136/2/662}{\color{magenta}\aj}, \href{https://ui.adsabs.harvard.edu/abs/2008AJ....136..662M}{136, 662}

\bibitem[{{Medina} {et~al.}(2018){Medina}, {Dzib}, {Tapia}, {Rodr{\'\i}guez}, \& {Loinard}}]{medina2018}
{Medina}, S. N.~X., {Dzib}, S.~A., {Tapia}, M., {Rodr{\'\i}guez}, L.~F., \& {Loinard}, L. 2018, \href{http://dx.doi.org/10.1051/0004-6361/201731774}{\color{magenta}\aap}, \href{https://ui.adsabs.harvard.edu/abs/2018A&A...610A..27M}{610, A27}

\bibitem[{{Medina} {et~al.}(2019){Medina}, {Urquhart}, {Dzib}, {Brunthaler}, {Cotton}, {Menten}, {Wyrowski}, {Beuther}, {Billington}, {Carrasco-Gonzalez}, {Csengeri}, {Gong}, {Hofner}, {Nguyen}, {Ortiz-Le{\'o}n}, {Ott}, {Pandian}, {Roy}, {Sarkar}, {Wang}, \& {Winkel}}]{Medina2019}
{Medina}, S. N.~X., {Urquhart}, J.~S., {Dzib}, S.~A., {et~al.} 2019, \href{http://dx.doi.org/10.1051/0004-6361/201935249}{\color{magenta}\aap}, \href{https://ui.adsabs.harvard.edu/abs/2019A&A...627A.175M}{627, A175}

\bibitem[{{Messineo} {et~al.}(2011){Messineo}, {Davies}, {Figer}, {Kudritzki}, {Valenti}, {Trombley}, {Najarro}, \& {Rich}}]{messineo2011}
{Messineo}, M., {Davies}, B., {Figer}, D.~F., {et~al.} 2011, \href{http://dx.doi.org/10.1088/0004-637X/733/1/41}{\color{magenta}\apj}, \href{https://ui.adsabs.harvard.edu/abs/2011ApJ...733...41M}{733, 41}

\bibitem[{{Mohan} {et~al.}(2022){Mohan}, {Vig}, \& {Mandal}}]{mohan2022}
{Mohan}, S., {Vig}, S., \& {Mandal}, S. 2022, \href{http://dx.doi.org/10.1093/mnras/stac1159}{\color{magenta}\mnras}, \href{https://ui.adsabs.harvard.edu/abs/2022MNRAS.514.3709M}{514, 3709}

\bibitem[{{Molinari} {et~al.}(2016){Molinari}, {Schisano}, {Elia}, {Pestalozzi}, {Traficante}, {Pezzuto}, {Swinyard}, {Noriega-Crespo}, {Bally}, {Moore}, {Plume}, {Zavagno}, {di Giorgio}, {Liu}, {Pilbratt}, {Mottram}, {Russeil}, {Piazzo}, {Veneziani}, {Benedettini}, {Calzoletti}, {Faustini}, {Natoli}, {Piacentini}, {Merello}, {Palmese}, {Del Grande}, {Polychroni}, {Rygl}, {Polenta}, {Barlow}, {Bernard}, {Martin}, {Testi}, {Ali}, {Andr{\'e}}, {Beltr{\'a}n}, {Billot}, {Carey}, {Cesaroni}, {Compi{\`e}gne}, {Eden}, {Fukui}, {Garcia-Lario}, {Hoare}, {Huang}, {Joncas}, {Lim}, {Lord}, {Martinavarro-Armengol}, {Motte}, {Paladini}, {Paradis}, {Peretto}, {Robitaille}, {Schilke}, {Schneider}, {Schulz}, {Sibthorpe}, {Strafella}, {Thompson}, {Umana}, {Ward-Thompson}, \& {Wyrowski}}]{Molinari2016}
{Molinari}, S., {Schisano}, E., {Elia}, D., {et~al.} 2016, \href{http://dx.doi.org/10.1051/0004-6361/201526380}{\color{magenta}\aap}, \href{https://ui.adsabs.harvard.edu/abs/2016A&A...591A.149M}{591, A149}

\bibitem[{{Molinari} {et~al.}(2010){Molinari}, {Swinyard}, {Bally}, {Barlow}, {Bernard}, {Martin}, {Moore}, {Noriega-Crespo}, {Plume}, {Testi}, {Zavagno}, {Abergel}, {Ali}, {Anderson}, {Andr{\'e}}, {Baluteau}, {Battersby}, {Beltr{\'a}n}, {Benedettini}, {Billot}, {Blommaert}, {Bontemps}, {Boulanger}, {Brand}, {Brunt}, {Burton}, {Calzoletti}, {Carey}, {Caselli}, {Cesaroni}, {Cernicharo}, {Chakrabarti}, {Chrysostomou}, {Cohen}, {Compiegne}, {de Bernardis}, {de Gasperis}, {di Giorgio}, {Elia}, {Faustini}, {Flagey}, {Fukui}, {Fuller}, {Ganga}, {Garcia-Lario}, {Glenn}, {Goldsmith}, {Griffin}, {Hoare}, {Huang}, {Ikhenaode}, {Joblin}, {Joncas}, {Juvela}, {Kirk}, {Lagache}, {Li}, {Lim}, {Lord}, {Marengo}, {Marshall}, {Masi}, {Massi}, {Matsuura}, {Minier}, {Miville-Desch{\^e}nes}, {Montier}, {Morgan}, {Motte}, {Mottram}, {M{\"u}ller}, {Natoli}, {Neves}, {Olmi}, {Paladini}, {Paradis}, {Parsons}, {Peretto}, {Pestalozzi}, {Pezzuto}, {Piacentini}, {Piazzo}, {Polychroni}, {Pomar{\`e}s}, {Popescu}, {Reach}, {Ristorcelli},
  {Robitaille}, {Robitaille}, {Rod{\'o}n}, {Roy}, {Royer}, {Russeil}, {Saraceno}, {Sauvage}, {Schilke}, {Schisano}, {Schneider}, {Schuller}, {Schulz}, {Sibthorpe}, {Smith}, {Smith}, {Spinoglio}, {Stamatellos}, {Strafella}, {Stringfellow}, {Sturm}, {Taylor}, {Thompson}, {Traficante}, {Tuffs}, {Umana}, {Valenziano}, {Vavrek}, {Veneziani}, {Viti}, {Waelkens}, {Ward-Thompson}, {White}, {Wilcock}, {Wyrowski}, {Yorke}, \& {Zhang}}]{molinari2010}
{Molinari}, S., {Swinyard}, B., {Bally}, J., {et~al.} 2010, \href{http://dx.doi.org/10.1051/0004-6361/201014659}{\color{magenta}\aap}, \href{https://ui.adsabs.harvard.edu/abs/2010A&A...518L.100M}{518, L100}

\bibitem[{{Motte} {et~al.}(2018){Motte}, {Bontemps}, \& {Louvet}}]{Motte2018}
{Motte}, F., {Bontemps}, S., \& {Louvet}, F. 2018, \href{http://dx.doi.org/10.1146/annurev-astro-091916-055235}{\color{magenta}\araa}, \href{https://ui.adsabs.harvard.edu/abs/2018ARA&A..56...41M}{56, 41}

\bibitem[{{Nguyen} {et~al.}(2021){Nguyen}, {Rugel}, {Menten}, {Brunthaler}, {Dzib}, {Yang}, {Kauffmann}, {Pillai}, {Nandakumar}, {Schultheis}, {Urquhart}, {Dokara}, {Gong}, {Medina}, {Ortiz-Le{\'o}n}, {Reich}, {Wyrowski}, {Beuther}, {Cotton}, {Csengeri}, {Pandian}, \& {Roy}}]{Nguyen2021}
{Nguyen}, H., {Rugel}, M.~R., {Menten}, K.~M., {et~al.} 2021, \href{http://dx.doi.org/10.1051/0004-6361/202140802}{\color{magenta}\aap}, 651, A88

\bibitem[{{Nguyen} {et~al.}(2022){Nguyen}, {Rugel}, {Murugeshan}, {Menten}, {Brunthaler}, {Urquhart}, {Dokara}, {Dzib}, {Gong}, {Khan}, {Medina}, {Ortiz-Le{\'o}n}, {Reich}, {Wyrowski}, {Yang}, {Beuther}, {Cotton}, \& {Pandian}}]{Nguyen2022}
{Nguyen}, H., {Rugel}, M.~R., {Murugeshan}, C., {et~al.} 2022, \href{http://dx.doi.org/10.1051/0004-6361/202244115}{\color{magenta}\aap}, \href{https://ui.adsabs.harvard.edu/abs/2022A&A...666A..59N}{666, A59}

\bibitem[{{Oliveira}(2008)}]{oliveira2008}
{Oliveira}, J.~M. 2008, in Handbook of Star Forming Regions, Volume II, ed. B.~{Reipurth}, Vol.~5, 599

\bibitem[{{Ortiz-Le{\'o}n} {et~al.}(2021){Ortiz-Le{\'o}n}, {Menten}, {Brunthaler}, {Csengeri}, {Urquhart}, {Wyrowski}, {Gong}, {Rugel}, {Dzib}, {Yang}, {Nguyen}, {Cotton}, {Medina}, {Dokara}, {K{\"o}nig}, {Beuther}, {Pandian}, {Reich}, \& {Roy}}]{Ortiz2021}
{Ortiz-Le{\'o}n}, G.~N., {Menten}, K.~M., {Brunthaler}, A., {et~al.} 2021, \href{http://dx.doi.org/10.1051/0004-6361/202140817}{\color{magenta}\aap}, 651, A87

\bibitem[{{Purcell} {et~al.}(2013){Purcell}, {Hoare}, {Cotton}, {Lumsden}, {Urquhart}, {Chandler}, {Churchwell}, {Diamond}, {Dougherty}, {Fender}, {Fuller}, {Garrington}, {Gledhill}, {Goldsmith}, {Hindson}, {Jackson}, {Kurtz}, {Mart{\'{\i}}}, {Moore}, {Mundy}, {Muxlow}, {Oudmaijer}, {Pandian}, {Paredes}, {Shepherd}, {Smethurst}, {Spencer}, {Thompson}, {Umana}, \& {Zijlstra}}]{purcell2013}
{Purcell}, C.~R., {Hoare}, M.~G., {Cotton}, W.~D., {et~al.} 2013, \href{http://dx.doi.org/10.1088/0067-0049/205/1/1}{\color{magenta}\apjs}, \href{http://adsabs.harvard.edu/abs/2013ApJS..205....1P}{205, 1}

\bibitem[{{Rashid} {et~al.}(2024){Rashid}, {Roy}, {Pandian}, {Dutta}, {Dokara}, {Vig}, \& {Menten}}]{rashid2024}
{Rashid}, M., {Roy}, N., {Pandian}, J.~D., {et~al.} 2024, \href{https://ui.adsabs.harvard.edu/abs/2024arXiv240518978R}{\href{http://dx.doi.org/10.48550/arXiv.2405.18978}{\color{magenta}arXiv e-prints}, arXiv:2405.18978}

\bibitem[{{Reid} {et~al.}(1995){Reid}, {Argon}, {Masson}, {Menten}, \& {Moran}}]{Reid1995}
{Reid}, M.~J., {Argon}, A.~L., {Masson}, C.~R., {Menten}, K.~M., \& {Moran}, J.~M. 1995, \href{http://dx.doi.org/10.1086/175518}{\color{magenta}\apj}, \href{https://ui.adsabs.harvard.edu/abs/1995ApJ...443..238R}{443, 238}

\bibitem[{{Schuller} {et~al.}(2009){Schuller}, {Menten}, {Contreras}, {Wyrowski}, \& {Schilke}}]{schuller2009}
{Schuller}, F., {Menten}, K.~M., {Contreras}, Y., {Wyrowski}, F., \& {Schilke}, e.~a. 2009, \href{http://dx.doi.org/10.1051/0004-6361/200811568}{\color{magenta}\aap}, \href{http://adsabs.harvard.edu/abs/2009A%26A...504..415S}{504, 415}

\bibitem[{{Urquhart} {et~al.}(2007){Urquhart}, {Busfield}, {Hoare}, {Lumsden}, {Clarke}, {Moore}, {Mottram}, \& {Oudmaijer}}]{urquhart_radio_south}
{Urquhart}, J.~S., {Busfield}, A.~L., {Hoare}, M.~G., {et~al.} 2007, \href{http://dx.doi.org/10.1051/0004-6361:20065837}{\color{magenta}\aap}, \href{http://adsabs.harvard.edu/abs/2007A%26A...461...11U}{461, 11}

\bibitem[{{Urquhart} {et~al.}(2014){Urquhart}, {Csengeri}, {Wyrowski}, {Schuller}, {Bontemps}, {Bronfman}, {Menten}, {Walmsley}, {Contreras}, {Beuther}, {Wienen}, \& {Linz}}]{urquhart2014c}
{Urquhart}, J.~S., {Csengeri}, T., {Wyrowski}, F., {et~al.} 2014, \href{http://dx.doi.org/10.1051/0004-6361/201424126}{\color{magenta}\aap}, \href{http://adsabs.harvard.edu/abs/2014A%26A...568A..41U}{568, A41}

\bibitem[{{Urquhart} {et~al.}(2009){Urquhart}, {Hoare}, {Purcell}, {Lumsden}, {Oudmaijer}, {Moore}, {Busfield}, {Mottram}, \& {Davies}}]{urquhart2009}
{Urquhart}, J.~S., {Hoare}, M.~G., {Purcell}, C.~R., {et~al.} 2009, \href{http://dx.doi.org/10.1051/0004-6361/200912108}{\color{magenta}\aap}, \href{https://ui.adsabs.harvard.edu/abs/2009A&A...501..539U}{501, 539}

\bibitem[{{Urquhart} {et~al.}(2018){Urquhart}, {K{\"o}nig}, {Giannetti}, {Leurini}, {Moore}, {Eden}, {Pillai}, {Thompson}, {Braiding}, {Burton}, {Csengeri}, {Dempsey}, {Figura}, {Froebrich}, {Menten}, {Schuller}, {Smith}, \& {Wyrowski}}]{urquhart2018_csc}
{Urquhart}, J.~S., {K{\"o}nig}, C., {Giannetti}, A., {et~al.} 2018, \href{http://dx.doi.org/10.1093/mnras/stx2258}{\color{magenta}\mnras}, \href{http://adsabs.harvard.edu/abs/2018MNRAS.473.1059U}{473, 1059}

\bibitem[{{Urquhart} {et~al.}(2013){Urquhart}, {Thompson}, {Moore}, {Purcell}, {Hoare}, {Schuller}, {Wyrowski}, {Csengeri}, {Menten}, {Lumsden}, {Kurtz}, {Walmsley}, {Bronfman}, {Morgan}, {Eden}, \& {Russeil}}]{urquhart2013}
{Urquhart}, J.~S., {Thompson}, M.~A., {Moore}, T.~J.~T., {et~al.} 2013, \href{http://dx.doi.org/10.1093/mnras/stt1310}{\color{magenta}\mnras}, \href{https://ui.adsabs.harvard.edu/abs/2013MNRAS.435..400U}{435, 400}

\bibitem[{{Urquhart} {et~al.}(2022){Urquhart}, {Wells}, {Pillai}, {Leurini}, {Giannetti}, {Moore}, {Thompson}, {Figura}, {Colombo}, {Yang}, {K{\"o}nig}, {Wyrowski}, {Menten}, {Rigby}, {Eden}, \& {Ragan}}]{urquhart2022}
{Urquhart}, J.~S., {Wells}, M.~R.~A., {Pillai}, T., {et~al.} 2022, \href{http://dx.doi.org/10.1093/mnras/stab3511}{\color{magenta}\mnras}, \href{https://ui.adsabs.harvard.edu/abs/2022MNRAS.510.3389U}{510, 3389}

\bibitem[{{Wang} {et~al.}(2020){Wang}, {Beuther}, {Rugel}, {Soler}, {Stil}, {Ott}, {Bihr}, {McClure-Griffiths}, {Anderson}, {Klessen}, {Goldsmith}, {Roy}, {Glover}, {Urquhart}, {Heyer}, {Linz}, {Smith}, {Bigiel}, {Dempsey}, \& {Henning}}]{wang2020}
{Wang}, Y., {Beuther}, H., {Rugel}, M.~R., {et~al.} 2020, \href{http://dx.doi.org/10.1051/0004-6361/201937095}{\color{magenta}\aap}, \href{https://ui.adsabs.harvard.edu/abs/2020A&A...634A..83W}{634, A83}

\bibitem[{{Wang} {et~al.}(2018){Wang}, {Bihr}, {Rugel}, {Beuther}, {Johnston}, {Ott}, {Soler}, {Brunthaler}, {Anderson}, {Urquhart}, {Klessen}, {Linz}, {McClure-Griffiths}, {Glover}, {Menten}, {Bigiel}, {Hoare}, \& {Longmore}}]{wang2018}
{Wang}, Y., {Bihr}, S., {Rugel}, M., {et~al.} 2018, \href{http://dx.doi.org/10.1051/0004-6361/201833642}{\color{magenta}\aap}, \href{https://ui.adsabs.harvard.edu/\#abs/2018A&A...619A.124W}{619, A124}

\bibitem[{{Wilner} {et~al.}(1999){Wilner}, {Reid}, \& {Menten}}]{wilner1999}
{Wilner}, D.~J., {Reid}, M.~J., \& {Menten}, K.~M. 1999, \href{http://dx.doi.org/10.1086/306907}{\color{magenta}\apj}, \href{https://ui.adsabs.harvard.edu/abs/1999ApJ...513..775W}{513, 775}

\bibitem[{{Wright} {et~al.}(2010){Wright}, {Eisenhardt}, {Mainzer}, {Ressler}, {Cutri}, {Jarrett}, {Kirkpatrick}, {Padgett}, {McMillan}, {Skrutskie}, {Stanford}, {Cohen}, {Walker}, {Mather}, {Leisawitz}, {Gautier}, {McLean}, {Benford}, {Lonsdale}, {Blain}, {Mendez}, {Irace}, {Duval}, {Liu}, {Royer}, {Heinrichsen}, {Howard}, {Shannon}, {Kendall}, {Walsh}, {Larsen}, {Cardon}, {Schick}, {Schwalm}, {Abid}, {Fabinsky}, {Naes}, \& {Tsai}}]{Wright2010}
{Wright}, E.~L., {Eisenhardt}, P. R.~M., {Mainzer}, A.~K., {et~al.} 2010, \href{http://dx.doi.org/10.1088/0004-6256/140/6/1868}{\color{magenta}\aj}, \href{https://ui.adsabs.harvard.edu/abs/2010AJ....140.1868W}{140, 1868}

\bibitem[{{Yang} {et~al.}(2023){Yang}, {Dzib}, {Urquhart}, {Brunthaler}, {Medina}, {Menten}, {Wyrowski}, {Ortiz-Le{\'o}n}, {Cotton}, {Gong}, {Dokara}, {Rugel}, {Beuther}, {Pandian}, {Csengeri}, {Veena}, {Roy}, {Nguyen}, {Winkel}, {Ott}, {Carrasco-Gonzalez}, {Khan}, \& {Cheema}}]{yang2023}
{Yang}, A.~Y., {Dzib}, S.~A., {Urquhart}, J.~S., {et~al.} 2023, \href{http://dx.doi.org/10.1051/0004-6361/202347563}{\color{magenta}\aap}, \href{https://ui.adsabs.harvard.edu/abs/2023A&A...680A..92Y}{680, A92}

\bibitem[{{Yang} {et~al.}(2021){Yang}, {Urquhart}, {Thompson}, {Menten}, {Wyrowski}, {Brunthaler}, {Tian}, {Rugel}, {Yang}, {Yao}, \& {Mutale}}]{yang2021}
{Yang}, A.~Y., {Urquhart}, J.~S., {Thompson}, M.~A., {et~al.} 2021, \href{http://dx.doi.org/10.1051/0004-6361/202038608}{\color{magenta}\aap}, \href{https://ui.adsabs.harvard.edu/abs/2021A&A...645A.110Y}{645, A110}

\bibitem[{{Yanza} {et~al.}(2024, MNRAS, submitted){Yanza}, {Dzib}, {Palau}, {Rodriguez}, {Masque}, \& et~al.}]{yanza2024}
{Yanza}, V., {Dzib}, S.~A., {Palau}, A., {et~al.} 2024, MNRAS, submitted

\bibitem[{{Yanza} {et~al.}(2022){Yanza}, {Masqu{\'e}}, {Dzib}, {Rodr{\'\i}guez}, {Medina}, {Kurtz}, {Loinard}, {Trinidad}, {Menten}, \& {Rodr{\'\i}guez-Rico}}]{yanza2022}
{Yanza}, V., {Masqu{\'e}}, J.~M., {Dzib}, S.~A., {et~al.} 2022, \href{http://dx.doi.org/10.3847/1538-3881/ac67ec}{\color{magenta}\aj}, \href{https://ui.adsabs.harvard.edu/abs/2022AJ....163..276Y}{163, 276}

\bibitem[{{Zhang} {et~al.}(2019){Zhang}, {Reid}, {Zhang}, {Wu}, {Hu}, {Sakai}, {Menten}, {Zheng}, {Brunthaler}, {Dame}, \& {Xu}}]{zhang2019}
{Zhang}, B., {Reid}, M.~J., {Zhang}, L., {et~al.} 2019, \href{http://dx.doi.org/10.3847/1538-3881/ab141d}{\color{magenta}\aj}, \href{https://ui.adsabs.harvard.edu/abs/2019AJ....157..200Z}{157, 200}

\end{thebibliography}

\end{document}